\documentclass[12pt]{article}
\usepackage{amsmath}
\usepackage{graphicx}
\usepackage{array}
\usepackage[top= 2.5cm,bottom= 2.5 cm, left=2cm,right=2cm]{geometry}

\begin{document}

\title{The New Symmetries Beyond the Standard Model \\ (The Body-centred Cubic
Periodic Symmetries \\ in Particle Physics)}
\author{Jiao-Lin Xu\\
$\qquad$$\qquad$$\qquad$$\qquad$$\qquad$$\qquad$\\
The Center for Simulational Physics,\\
The Department of Physics and Astronomy\\
University of Georgia,\\
Athens, GA 30602, USA\\
\texttt{jxu@hal.physast.uga.edu}\\
$\qquad$$\qquad$$\qquad$$\qquad$$\qquad$$\qquad$\\
Institute of Theoretical Physics,\\
 Chinese Academy of Sciences, P.
O. Bax 2735, Beijing 100080, China.\\
}

\date{\today}
\maketitle

\begin{abstract}
This paper proposes new symmetries (the body-centred cubic periodic
symmetries) beyond the standard model. Using a free particle
expanded Schrodinger equation with the body-centred cubic periodic
symmetry condition, the paper deduces a full baryon spectrum
(including mass M, I, S, C, B, Q, J and P) of all 116 observed
baryons. All quantum numbers of all deduced baryons are completely
consistent with the corresponding experimental results. The deduced
masses of all 116 baryons agree with (more than average 98 percent)
the experimental baryon masses using only four constant parameters.
The body-centred cubic periodic symmetries with a periodic constant
``a'' about $10^{-23}$m play a crucial rule. The results strongly
suggest that the new symmetries really exist. This paper predicts
some kind of ``Zeeman effect'' of baryons, for example: one
experimental baryon N(1720)$\frac{3}{2}^{+}$ with $ \Gamma$ = 200
Mev is composed of two N baryons [(N(1659)$\frac{3}{2}^{+}$ +
N(1839)$\frac{3}{2}^{+}$] = $\overline{N(1749)}$$\frac{3}{2}^{+}$
with $\Gamma$ = 1839-1659 = 180 Mev.
\end{abstract}
PACS numbers 11.30.-j \,\,\,\, 12.40.Yx\\
key words: new symmetry, beyond standard model, baryon spectrum,
phenomenology.

\section{Introduction}
$\qquad$The standard model of particle physics has been enormously
successful in predicting a wide range of phenomena. M. K. Gaillard,
P. D. Grannis and F. J. Sciulli \cite{Standard} point out. ``And,
just as ordinary quantum mechanics fails in the relativistic limit,
we do not expect the standard model to be valid at arbitrarily short
distances. However, its remarkable success strongly suggests that
the standard model will remain an excellent approximation to nature
at distance scales as small as $10^{-18}$ m." How about the nature
at distance scales smaller than $10^{-18}$ m?

\emph{The Particle Physics Roadmap} \cite{Roadmap}, in the``Beyond
the Standard Model" section, proposes nine terms ``as a mission
statement for the field. 1) \textbf{Are there undiscovered
principles of nature: new symmetries, new physical laws?} 2) How can
we solve the mystery of dark energy? 3) Are there extra dimensions
of space? $\ldots$  9) What happened to the antimatter ?" The first
mission of all is to discover \textbf{new symmetries}.

Where we can find the undiscovered symmetries? How do we find the
new symmetries? Professor \textbf{Wilczek} \cite{Wilczek} points
out: ``\textbf{that empty space--the vacuum}--is in reality a richly
structured, highly symmetrical, medium. $\ldots$. Because the vacuum
is a complicated material governed by locality and symmetry, one can
learn how to analyze it by studying other such materials--that is,
condensed matter." Wilczek not only pointed out where we can find
the symmetries--``that empty space--the vacuum", but also provided a
very practical and efficient way for the study--``learning from
studying condensed matter."

Since the simplest and common symmetries of condensed matter is the
body centred cubic periodic symmetries, this paper suggests that
there are new symmetries beyond the standard model: the local
body-centred cubic periodic symmetries with a periodic constant ``a"
$\sim$ $10^{-23}m$ around baryons. Using the symmetries and a free
particle expanded Schrodinger equation, the paper can deduce the
full baryon spectrum (including mass M, isospin I, strange number S,
charmed number C, bottom number B, electric charge Q, total angular
momentum J and parity P) with only four constant parameters. The
deduced quantum numbers (I, S, C, B, Q, J and P) of all baryons (116
baryons) are completely the same as the 116 corresponding
experimental baryons. The deduced masses of all 116 baryons agree
with the corresponding experimental baryons with more than average
98 percent accuracy. The symmetries play a crucial rule in the
deduction. Thus the paper will show that the body-centred cubic
periodic symmetries really exist in nature at distance scales
smaller as $ 10^{-23}$ m by deduction of the full baryon spectrum
that agrees with the experimental results.

The great success and exploration of the standard model, has
prepared the conditions for deduction of the full baryon spectrum.
Many excellent experimental physicists have searched for free quarks
with as high as possible energies (accelerators and cosmic ray) to
``unglue" quarks for more than 40 years, yet no one individually
free quark has been found \cite{Free Quark}. This has been a great
exploration. The result strongly shows that the binding energy of
the quarks in the baryons is extremely high. From that fact, we
infer that the low mass spectrum of the baryons might not have
originated from the internal quark motions inside the baryons since
the energy levels of the internal quark motions might be much higher
than the observable mass spectrum of the baryons. Such as the energy
levels of nuclei with much high binding energy are much higher than
the energy levels of molecules with much lower binding energy. The
fact is really so, ``the experimentally observed excitation spectrum
of the nucleon (N and $\Delta$ resonances) is compared to the
results of a typical quark model calculation. Many more states are
predicted than observed but on the other hand states with certain
quantum numbers appear in the spectrum  at excitation energies much
lower than predicted." \cite{Much lower} Many outstanding
theoretical physicists have deduced the mass spectrum of the baryons
using the internal motions of the quarks inside baryons for more
than 40 years, but a completely satisfying united theoretical mass
spectrum has still not arisen \cite{Quark calculation}. The quark
model does not give out the precise quark masses in theory and the
quark masses cannot be measured directly by experiments, so we do
not know the precise quark masses. \textbf{Trying to deduce the
baryon masses from the unknown precise masses of the inside quarks
is very difficult work. Therefore, we try to deduce the mass
spectrum of the baryons from outside symmetries.} Then, using the
qqq baryon model, we deduce the quark masses from the deduced
($\simeq$ measured) masses of the most important baryons.

In the low-energy phenomena, the baryons usually act as single
particles. They are just like the atoms in chemical reactions in
that the atoms look like elementary particles without internal
structure. All baryons automatically decay into the ground baryon
(proton and neutron) in a very short time \cite{Baryon}.
\textbf{These facts make us think that the baryons might be
different excited states of only one kind of particles. The simplest
model that really works is a single free particle expanded
Schrodinger equation with the body centered cubic periodic (BCCP)
condition.}

In order to deduce the full spectrum of baryons using the BCCP
symmetries, we will propose three postulates to set up a single free
particle expanded Schrodinger equation and a body-centred cubic
periodic (BCCP) condition in Section 2. We will deduce energy bands
from the expanded Schrodinger equation with BCCP condition in
Section 3. We will find phenomenological formulae about quantum
numbers of the energy bands in Section 4. We will deduce the baryons
using the phenomenological formulae and the energy bands in Section
5. We will compare the deduced baryons with the experimental results
in Section 6. We will deduce the masses and the mass parameters of
the five quarks from the masses of the most important deduced
baryons in Section 7. Finally we will discuss some important
problems (in Section 8) and will give the conclusions (in Section 9)
as well as will give some predictions (in Section 10).

\section{The Expanded Schrodinger Equation and \\
the body-centred cubic Periodic Condition}

$\qquad$All the experiments of particle physics show that (1) the
baryons always are born and act as whole particles (not individual
quarks), and (2) the baryons always automatically decay into the
ground state baryons (protons and neutrons) in a very short time.
From these very important experiments, we infer a single free
particle approximation of the baryons (neglecting the internal
structure and motions) to deduce the masses and the quantum numbers
(I, S, C, B, L, J and P) of the baryons.

\subsection{The first postulate (perfect SU(N) symmetry groups)}

The quark model \cite{Quark Model} assumes five quarks (u, d, s ,c
and b) with special quantum numbers (see Table 14.1 of \cite
{Quarks}) as the components of baryons and mesons to explain the
observable quantum numbers of baryons and mesons with sum laws. The
five quarks satisfy SU(5)$_f$ symmetries. The existence of baryons
with t-quark is very unlikely due to the short lifetime of the top
quark. In order to explain the mass differences of baryons, it also
assumes that the five (four or three) quarks satisfy SU(5)$_f$
(SU(4)$_f$ or SU(3)$_f$) approximate symmetries and the symmetries
have been broken by the mass differences of the quarks.

Since the quark model was born more than forty years ago, many new
particles have been discovered. Especially the surprising
experimental facts--the quark confinement \cite{Free Quark} was
discovered which is entirely unexpected. These new experimental
discoveries should be adopted to estimate the quark masses, and to
further check whether the $SU(5)_f$ symmetry would be broken due to
the differences in quark masses.

The experimental facts of the quark confinement and the stability of
proton (proton p with lifetime $\tau$ about $10^{30}$ years)
inevitably result in huge binding energies of the three quarks in
baryons. According to the E = M$C^{2}$ formula, the huge binding
energies and the small experimental masses of baryons inevitably
lead to huge quark masses. From qqq baryon model of the quark model,
for the five most important baryons (proton, neutron,
$\Lambda(1116)$, $\Lambda_c(2286)$ and $\Lambda_b(5624)$), there are
five equations:
\begin{equation}
m_u + m_u + m_d - |E_{bin}(uud)| = 938 (Mev),  \label{m938}
\end{equation}
\begin{equation}
m_u + m_d + m_d - |E_{bin}(udd)| = 940 (Mev),  \label{m940}
\end{equation}
\begin{equation}
m_u + m_d +m_s - |E_{bin}(uds)| = 1116 (Mev),  \label{m1116}
\end{equation}
\begin{equation}
m_u + m_d +m_c - |E_{bin}(udc)| = 2286 (Mev),  \label{m2286}
\end{equation}
\begin{equation}
m_u + m_d +m_b - |E_{bin}(udb)| = 5624 (Mev).  \label{m5624}
\end{equation}
In order to find the maximum possible values of the quark mass
differences, we assume that the mass difference of the five baryons
are completely from the mass differences of the five quarks, the
binding energy are completely the same,
\begin{equation}
|E_{bin}(uud)|=|E_{bin}(udd)|=|E_{bin}(uds)|=|E_{bin}(udc)|
=|E_{bin}(udd)|=\,\,|E_{Bind}|. \label{Same BindE}
\end{equation}
From (\ref{Same BindE}), (\ref{m938}) and (\ref{m940})  we have
\begin{equation}
m_d\,\,-\,\,m_u =  2. \label{Eu-d}
\end{equation} \\
Putting (\ref{Eu-d}) into (\ref{m938}) and (\ref{m940}), from
(\ref{Same BindE}),we get

\begin{equation}
m_u\,\, =\, \frac{1}{3}[|E_{Bind}|]\,\,+\,\,312,\label{mu}
\end{equation}
\begin{equation}
m_d\,\, =\, \frac{1}{3}[|E_{Bind}|]\,\,+\,\,314.\label{md}
\end{equation}
Similarly, we can find:
\begin{equation}
m_s\,\, =\, \frac{1}{3}[|E_{Bind}|]\,\,+\,\,490,\label{ms}
\end{equation}
\begin{equation}
m_c\,\, =\, \frac{1}{3}[|E_{Bind}|]\,\,+\,\,1660,\label{mc}
\end{equation}
\begin{equation}
m_b\,\, =\, \frac{1}{3}[|E_{Bind}|]\,\,+\,\,4998.\label{mb}
\end{equation}

Thus, from (\ref{mu}), (\ref{md}), (\ref{ms}), (\ref{mc}) and
(\ref{mb}), we have the masses of the quarks for the binding energy
= zero, 300 Gev, 3000 Gev and 30000 Gev as shown in Table1 1:
$\newline$\\

\begin{tabular}{l}
\ \ \ \ Table 1.\ The masses of the quarks \\
\begin{tabular}{|l|l|l|l|l|l|l|l|}
\hline E$_{bind}$(uud) & m$_u$  &  m$_d$ & m$_s$ & m$_c$ & m$_b$  \\
\hline 0 Gev & 312  & 314 & 490 &  1660 & 4998\\
\hline 300 Gev & 100312  &100314 & 100490 & 101660 & 104998\\
\hline 3000 Gev & 1000312 & 1000314 & 1000490 & 1001660 & 1004998\\
\hline 30000 Gev & 10000312 &10000314 & 10000490&10001660&10004998\\
\hline
\end{tabular}%
\end{tabular}%
\newline

The huge quark masses and the very small differences of the baryon
masses naturally bring about that the very small mass differences of
the quarks can be neglected as shown in Table 1. And the $SU(3)_f$,
$SU(4)_f$ and $SU(5)_f$ symmetries give out completely correct
quantum numbers of all baryons and all mesons. These facts force us
to think that the $SU(3)_f$, $SU(4)_f$ and $SU(5)_f$ symmetries not
only really exists, but also are perfect symmetries. The observable
mass differences of baryons are not from the mass difference of the
quarks (while \textbf{they are from outside symmetry as shown in the
following sections}). Thus the experimental facts now have already
shown that the SU(N) (N = 3, 4 and 5) symmetry groups are perfect
since the five quark masses are the same essentially. Therefore we
propose the first postulate:

Postulate I: \textbf{All five quarks have the same bare mass m$_q$
in the vacuum (but different physical masses in the observable
baryons) and they satisfy perfect SU(3)$_f$ (for u,d and s),
SU(4)$_f$(for u, d, s and c) and SU(5)$_f$ (for u, d, s, c and b)
symmetries. In other words, the flavor SU(N)$_f$ (N = 3, 4 and 5)
symmetry groups are not broken.}

According to Postulate I, the five quarks have the same unknown
large bare masses. Since SU(5)$_f$ symmetry group and SU(3)$_c$
symmetry group are both perfect, the interaction energies of the
three quarks inside various baryons are the same. Thus we have a
corollary:

Corollary: \textbf{The baryons composed of any three quarks will
have the same unknown large bare mass M$_b$ in the vacuum.}

The observable physical masses of the various baryons with the same
bare mass M$_b$ but different quantum numbers, however, are
different since they are in different energy excited states (from
the vacuum).

The experimentally observed baryons almost all belong to SU(4)
multiplets of baryons made of u, d, s and c. The quark model has
already given the quantum numbers of the baryons of the 20-plet with
an SU(3) Octet as appearing in Table 2 \cite{20-plet}:
\newline
\newline
\
\begin{tabular}{l}
$\qquad$Table 2.\ The quantum numbers of baryons with s = 1/2 \\
\begin{tabular}{|l|l|l|l|l|l|l|l|}
\hline Baryons & I & $I_Z$ & S & C & Q & Bare M & Color \\
\hline N & $\frac{1}{2}$ & $\frac{1}{2}$,-$\frac{1}{2}$ & 0
& 0 & +1, 0 & M$_b$ & 0 \\
\hline $\Sigma$ & 1 & 1,0,-1 & -1 & 0 & 1, 0, -1 & M$_b$ & 0 \\
\hline $\Lambda$ & 0 & 0 & -1 & 0 & 0 & M$_b$ & 0 \\
\hline $\Xi$ & $\frac{1}{2}$ & $\frac{1}{2}$,-$\frac{1}{2}$ &
-2 & 0 & 0,-1 & M$_b$ & 0 \\
\hline $\Sigma$$_c$ & 1 & 1, 0, -1 & 0 & +1 &
+2, +1, 0 & M$_b$ & 0 \\
\hline $\Lambda_c$ & 0 & 0 & 0 & +1 & +1 & M$_b$ & 0 \\
\hline $\Xi_c$ & 1/2 & $\frac{1}{2}$,-$\frac{1}{2}$ & -1 & +1 &
1, 0 & M$_b$ & 0 \\
\hline $\Xi_c$ & 1/2 & $\frac{1}{2}$,-$\frac{1}{2}$ & -1 & +1 &
 1, 0 & M$_b$ & 0 \\
\hline $\Omega$$_c$ & 0 & 0 & -2 & +1 & 0 & M$_b$ & 0 \\
\hline $\Xi$$_{cc}$ & 1/2 & $\frac{1}{2}$,-$\frac{1}{2}$ & 0 & +2 &
+2, +1 & M$_b$ & 0 \\
\hline $\Omega$$_{cc}$ & 0 & 0 & -1 & +2 & +1
& M$_b$ & 0 \\
\hline
\end{tabular}%
\end{tabular}%
\newline\\
The quark model has already given the quantum numbers of baryons in
the 20-plet of SU(4)$_f$ with an SU(3) Decuplet as appearing in
Table 3 \cite{20-plet}:
\newline

\begin{tabular}{l}
$\qquad$Table 3.\ The quantum numbers of baryons with s = 3/2 \\
\begin{tabular}{|l|l|l|l|l|l|l|l|l}
\hline Baryons & I & I$_z$ & S & C & Q & Bare M & Color \\
\hline $\Delta$ & $\frac{3}{2}$ & $\frac{3}{2}$,$\frac{1}{2}$,
-$\frac{1}{2}$,-$\frac{3}{2}$ & 0 & 0 & 2,1,0,-1 & M$_b$ & 0 \\
\hline $\Sigma$ & 1 & 1,0,-1 & -1 & 0 & 1, 0, -1 & M$_b$ & 0 \\
\hline $\Xi$ & $\frac{1}{2}$ & $\frac{1}{2}$,-$\frac{1}{2}$ & -2 & 0
& 0,-1 & M$_b$ & 0 \\
\hline $\Omega$ & 0 & 0 & -3 & 0 & -1 & M$_b$ & 0 \\
\hline $\Sigma_c$ & 1 & 1, 0, -1 & 0 & +1 & 2, 1, 0 & M$_b$ & 0 \\
\hline $\Xi$$_c$ & $\frac{1}{2}$ & $\frac{1}{2}$,-$\frac{1}{2}$ & -1
& +1 & 1, 0 & M$_b$ & 0 \\
\hline $\Omega_c$ & 0 & 0 & -2 & +1 & 0 & M$_b$ & 0  \\
\hline $\Xi_{cc}$ & $\frac{1}{2}$ &
$\frac{1}{2}$,-$\frac{1}{2}$ & 0 & +2 & +2, +1 & M$_b$ & 0  \\
\hline $\Omega$$_{cc}$ & 0 & 0 & -1 & +2 & +1 & M$_b$ & 0  \\
\hline $\Omega$$_{ccc}$ & 0 & 0 & 0 & +3 & +2 & M$_b$ & 0  \\
\hline
\end{tabular}%
\end{tabular}%
\newline\\
According to particle and anti-particle symmetry, there always
exists an anti-quark for each quark of the quarks u, d, s, c and b
in the vacuum, and there always exists an anti-baryon for each
baryon of Table 2 or Table 3 in the vacuum. We can easily find the
quantum numbers of the anti-quarks from the quarks in Table 14.1 of
\cite {Quarks} and anti-baryons from Table 2 and Table 3. Thus we do
not list the quantum numbers of the anti-quarks and the
anti-baryons.

For convenience, we make an idea baryon with a unknown large bare
mass $M_b$ and colorless. We call the idea particle b-particle. In
other words, for all the baryons, we give the common name
``b-particle." A b-particle represents any baryon with the same
large bare mass $M_b$ and other common properties (colorless) in the
vacuum.

\subsection{The second postulate (the free b-particle expanded
Schrodinger equation and the BCCP symmetry condition)}

$\qquad$A big macroscopic body can affect the space around it. For
example, it is well known that under the effect of the sun, the
space around the sun is curved. The small microscopic particles look
to affect the space around it also. The micro-particles have
wave-particle duality. The wave-particle duality might hint that
there are some periodic symmetries in the space. The periodic
symmetries cause the wave-particle duality. The symmetry of the
apparent empty space is a very complex problem. Frank Wilczek
\cite{Wilczek} point out: ``that empty space--the vacuum--is in
reality a richly structured, highly symmetrical, medium." $\ldots$,
one can learn how to analyze it by studying other such
materials--that is, condensed matter." The most natural symmetry of
condensed matter with the highest density and the simplest structure
is the body-centred cubic periodic symmetries. After many years
research, we have discovered that a body-centred cubic periodic
condition is the best physical and mathematical simplification for
the problem and this condition will gives a full baryon spectrum
that consistent with experimental results. Thus we propose the
second postulate:\newline

Postulate II: \textbf{After a pair of b-particle and anti b-particle
is excited from the vacuum, the excited b-particle gets a baryon
number $\mathcal{B}$ = +1, the excited anti b-particle gets a baryon
number $\mathcal{B}$ = -1. They all get at least an energy $M_0C^2$.
At the same time, as a single particle, the b-particle is freely
moving within a local weak body-centred cubic periodic (BCCP)
symmetry space around it. The b-particle with large bare mass $M_b$
obeys a free particle expanded Schrodinger equation with a body
center cubic periodic condition. The periodic constant ``a" $\sim$
$10^{-23}$ m.}\newline

The following sections will show that it can give the correct baryon
spectrum. According to Postulate II, when a b-particle (or an anti
b-particle) is in the vacuum state, its baryon number $\mathcal{B}$
= 0. If, however, a pair of b-particle and anti b-particle is
excited out from the vacuum state, they become an observable baryon
and an anti-baryon with the baryon numbers
\begin{equation}
\mathcal{B} = +1,\,\,\, for\,\,\, b-particler;\,\,\, \mathcal{B} = -
1,\,\,\, for\,\,\, anti\,\,b-particle. \label{Baryon Number}
\end{equation}

The anti b-particle well annihilate with a existing b-particle
imminently.

The excited b-particle will obey a free b-particle expanded
Schrodinger equation
\begin{equation}
\frac{\hbar ^2}{2M_{b}}\nabla ^{2}\Psi + \varepsilon \Psi = 0,
\label{b-particle}
\end{equation}
where $M_b$ is the unknown very large bare mass of the b-particle
(and anti b-baryon) with baryon number $\mathcal{B}$ = +1 (or an
anti-baryon number $\mathcal{B}$ = -1). This Schrodinger equation is
an expanded Schrodinger equation beyond the standard model. It is
different from the ordinary Schrodinger equation. (1) In the
ordinary equation the mass of the particle is an observable mass of
the particle, but in this equation the mass is the bare mass of the
b-particle. (2) The ordinary equation is working in smooth and
straight space, but this equation is working in a BCCP symmetry
space. (3) This equation is available to the distance scale $\sim$
$10^{-23}$m beyond the standard model. It calculates the effect of
the vacuum material around the excited b-particle. We simplify the
effect into a body-centred cubic periodic condition and a bare mass
in the expanded Schrodinger equation phenomenologically. (4) This
Schrodinger equation can replace the Dirac equation, since the bare
mass \cite{Bare Mass} of the b-particle (or anti b-particle) $M_{b}$
is much larger than the observable masses of the baryons. The
ultimate test of the validity of any physical theory must be whether
it is consistent with observations and measurements of physical
phenomena. Thus the ultimate test of these approximations is the
experiments. The following sections will show that the expansion and
the approximations are very good.

The body-centred cubic periodic symmetry (BCCP) condition is

\begin{equation}
\Psi(\vec{r}+\vec{S}) = \Psi(\vec{r}) \\
,  \label{Boundary}
\end{equation}
where the $\vec{S}$ is
\begin{equation}
\vec{S}= \mathbf{A}\mathbf{j}\label{S} \\
\end{equation}
the matrix $\mathbf{A}$ is the A matrix of a body-centred cubic
lattice. It is
\[
\mathbf{A} =  \left[
\begin{array}{ccc}
-a/2 & a/2 & a/2 \\
a/2 & -a/2 & a/2 \\
a/2 & a/2 & -a/2 %
\end{array}%
\right].
\]%
The constant ``a" is the periodic constant of the body-centred cubic
periodic symmetry; a $\sim$ $10 ^{-23}$ m. The vector $\vec{j}$ is
\[
\vec{j} =  \left[
\begin{array}{ccc}
j_1  \\
j_2  \\
j_3  \\
\end{array}%
\right]
\]%
\label{J vactor} where $j_1$, $j_2$ and $j_3$ are positive or negative
integers or zero.

From (\ref{b-particle}) and (\ref{Boundary}), we can see that the
equation and the condition are independent from spin, isospin,
strange number, charm number, bottom number, electric charge and
color that the b-particle might have. They are dependent only on the
bare mass $M_b$ of the b-particle. From the Corollary of Postulate
I, all baryons have the same bare mass $M_B$. Thus the equation and
condition describe all baryons. In mathematics this equation and its
condition are the same with the free electron limit in a body center
cubic lattice of the energy band theory. In the free electron limit
case \cite{Free Particle}, the periodic potential of the lattice
becomes arbitrarily weak while the symmetry properties (the BCC
lattice) of the wave functions are preserved. The mathematical form
of the free limit case is that a free particle Schrodinger equation
(V($\vec{r}$) = 0) and the wave functions satisfy the symmetry of
the body-centred cubic lattice. According to the energy band theory
\cite{Energy Band}, an excited (from the vacuum) b-particle in the
BCCP symmetry space will be in an energy band with BCCP symmetry. In
order to find the observable baryons, we propose the third
postulate.

\subsection{The third postulate (the observable baryons)}

$\qquad$According to the energy band theory \cite{Energy Band}, with
the expanded Schrodinger equation and BCCP condition, we can find
energy bands. Then using phenomenological formula, we can find the
intrinsic quantum numbers (I, S, C, B and Q) of an energy band from
its degeneracy, the rotational fold of symmetry axis and the index
number $\vec{n}$ of the energy band that are deduced from the
expanded Schrodinger equation with BCCP boundary condition. In order
to recognize the baryons from the energy bands, we propose the third
postulate:

Postulate III: \textbf{The intrinsic quantum numbers (I, S, C, B and
Q) of baryons are given by the perfect SU(N) symmetry in the vacuum.
The intrinsic quantum numbers of the energy bands are deduced from
the expanded Schrodinger equation and the BCCP condition as well as
phenomenological formulae of the quantum numbers. The baryons with
certain quantum numbers (as appearing in Table 2 and Table 3) will
be excited only into the energy bands with the same quantum numbers
as the baryon with in the vacuum. The lowest energy (the excited
energy from vacuum and the excited energy in a band) of this energy
band is the observable mass of the observable baryon}.

Thus the intrinsic quantum numbers (I, S, C, B and Q) of the
observable baryons are mainly determined by the SU(N) symmetries. We
will deduce the energy bands from the expanded Schrodinger equation
and the BCCP condition in Section 3.

\section{The Energy Bands and Wave Functions of the B-particle}

$\qquad$The expanded Schrodinger equation of the b-particle
(\ref{b-particle}) descriibes a b-particle with unknown very large
bare mass $M_b$ motion inside the vacuum material; the wave
functions satisfy the body-centred cubic periodic symmetries of the
local space around the b-particle. In mathematical terms, this
equation and its condition are the same as the free electron limit
in a body center cubic lattice of the energy band theory \cite{Free
Particle}. For the equation and condition, there are already
solutions in energy band theory \cite{Energy Band}. We will use the
mathematical results. This, however, is not a solid state problem.
The periodic constant ``a" is about $10^{-23}$m. It is much smaller
than the standard atomic lattice constant (about 2-6 $\times$
$10^{-10}m$). This is a particle work--deducing the baryon spectrum
using free particle limit approximation methods of energy band
theory.

\subsection{The solutions of the free b-particle expanded Schrodinger
equation with the BCCP condition}

$\qquad$According to the energy band theory, the solution of the
equation (\ref{b-particle}) and the BCCP condition (\ref{Boundary})
is the Bloch wave function \cite{Bloch}:

\begin{equation}
\Psi_{\vec{k}}(\vec{r}) = e^{i\vec{k}.\vec{r}}u(\vec{r}),
\label{Bloch}
\end{equation}
where $\vec{k}$ is a wave vector and u($\vec{r}$) is a periodic
function with body-centred cubic periodic symmetries
\begin{equation}
u(\vec{r}+\vec{S}) = u(\vec{r}),  \label{u(r)}
\end{equation}
here the $\vec{S}$ is shown in formula (\ref{S}). If we put the
Bloch function ($\ref{Bloch}$) into the Schrodinger equation
($\ref{b-particle}$), we get an equation of the u($\vec{r}$):
\begin{equation}
\nabla ^{2}u + 2i\vec{k}\cdot\nabla u +
\frac{2M_{b}}{\hbar^2}(\varepsilon - \frac{\hbar^2 k^2}{2M_{b}})u =
0.  \label{eq. of u}
\end{equation}
A periodic solution [(3.91) of \cite{Energy Band}] of the above
equation is
\begin{equation}
u_{\vec{l}}(\vec{r})= e^{-i\vec{l}\vec{B}.\vec{r}}.  \label{u}
\end{equation}
The corresponding energy value [(3.92) of \cite{Energy Band}] is

\begin{equation}
\varepsilon_k =\frac{\hbar^2}{2M_{b}}(\vec{k}-\vec{l}.\vec{B})^2
\label{energy}
\end{equation}%
where the vector $\vec{l}$ is
\begin{equation}
\vec{l} = (l_1,\, l_2,\, l_3);  \label{L value}
\end{equation}
$l_1$, $l_2$ and $l_3$ are positive or negative integers or zero;
and the matrix $\mathbf{B}$ is the B matrix of the body centred
cubic lattice

\[
\mathbf{B} =  \left[
\begin{array}{ccc}
0 & 2\pi/a & 2\pi/a \\
2\pi/a & 0 & 2\pi/a \\
2\pi/a & 2\pi/a & 0 \label{Matrix B}%
\end{array}%
\right],
\]
where a ($\sim$ $10^{-23}$m) is the periodic constant of the body
center cubic lattice.

Since the wave function satisfies the body-centred cubic periodic
condition, the wave vector $\vec{k}$ satisfies the reciprocal
(lattice) periodic symmetry of the body-centred cubic (lattice). If
we take
\begin{equation}
\vec{k} =\frac{2\pi}{a}(\xi,\, \eta,\, \zeta) ,  \label{k-vector}
\end{equation}%
using $\vec{l}$ value(\ref{L value}) and B-matrix value, we can get
\begin{equation}
(\vec{k}-\vec{l}.\vec{B})^2 = (\frac{2\pi}{a})^2
[(\xi-n_1)^2+(\eta-n_2)^2+(\zeta-n_3)^2],  \label{E(kn)}
\end{equation}
where
\begin{equation}
\begin{tabular}{l}
$n_{1}$= $l_2$+$l_3$, \\
$n_{2}$= $l_3$+$l_1$, \\
$n_{3}$= $l_1$+$l_2$; \label{n1n2n3}%
\end{tabular}
\end{equation}%
and
\begin{equation}
\begin{tabular}{l}
$l_{1}$ =1/2( -n$_{1}$+n$_{2}$+n$_{3}$), \\
$l_{2}$ =1/2( +n$_{1}$-n$_{2}$+n$_{3}$), \\
$l_{3}$ =1/2( +n$_{1}$+n$_{2}$-n$_{3}$).%
\end{tabular}
\label{N Conditions} \\
\end{equation}%
Condition (\ref{N Conditions}) implies that the vector $\vec{n}$ =
(n$_{1}$, n$_{2}$, n$_{3}$) can only take certain values that make
the $\vec{l}$ to be an integral vector ($l_1$, $l_2$, $l_3$ are
positive or negative integers or zero). For example, the $\vec{n}$
cannot take (0,0,1) or ( 1,1,-1), but can take (0,0,2) and (1,-1,2)
since the wave functions must satisfy the symmetries of the BCCP
symmetry. From (\ref{energy}) and (\ref{E(kn)}), we get
\begin{equation}
\varepsilon_k = \frac{h^2}{2M_{b}a^2} [(\xi-n_1)^2+(\eta-n_2)^2+
(\zeta-n_3)^2] .  \label{Energy}
\end{equation}%
If we assume
\begin{equation}
\alpha =\frac{h^2}{2M_{b}a^2}  \label{alpha}
\end{equation}
and
\begin{equation}
E(\vec{k},\vec{n}) = (n_{1}-\xi)^{2} +(n_{2}-\eta)^{2}
+(n_{3}-\zeta)^2) \label{Energy}
\end{equation}%
we get the energy formula
\begin{equation}
\varepsilon(\vec{k},\vec{n}) =\alpha E(\vec{k},\vec{n}).
\label{mass}
\end{equation}

From (\ref{u}), (\ref{Bloch}), (\ref{k-vector}), (\ref{L value}),
(\ref{n1n2n3}) and the expression of the \textbf{B} matrix, we can
find the total wave function
\begin{equation}
\Psi_{\vec{k}}(\vec{r}) = exp(- \frac{i2\pi}{a})[(n_{1}-\xi)x+(n_{2}-%
\eta)y+(n_{3}-\zeta)z].  \label{Plain Wave}
\end{equation}

\subsection{The energy bands and wave function of the b-particle}

$\qquad$According to the energy band theory, the first Brillouin
zone of the body-centred cubic lattice appears in Fig. 1. [depicted
from (Fig. 1) of \cite{Callaway} and (FIGURE 8.10) of \cite{Joshi}
as well as Fig. 42 of \cite{Energy Band}: $\qquad $ $\qquad $
$\qquad $ $\qquad $ $\qquad $ $\qquad $ $\qquad $ $\qquad $ $\qquad
$ \ \

\includegraphics[scale=0.8]{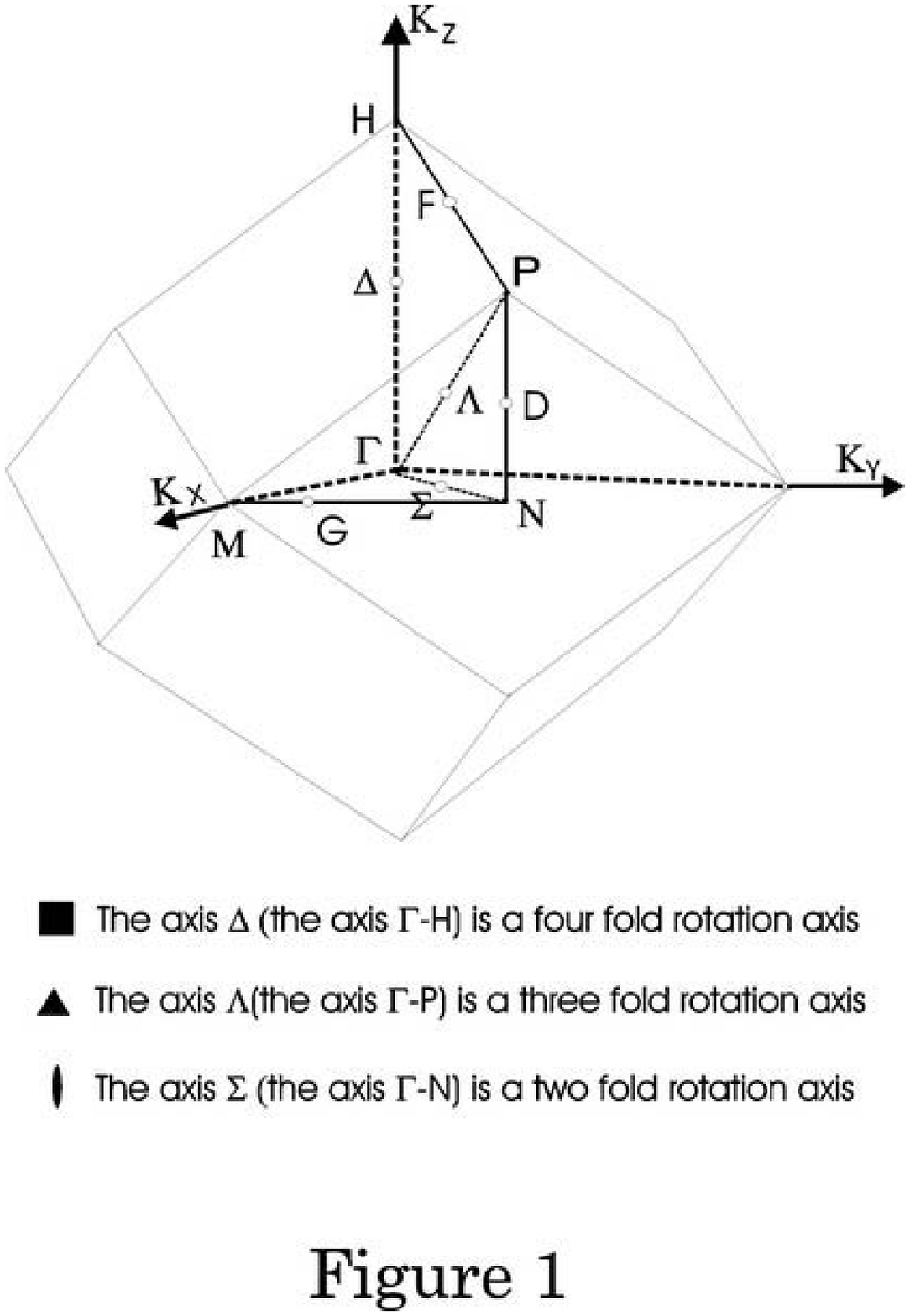}

Fig. 1. \ The first Brillouin zone of the body-centred cubic
lattice. The symmetry points and symmetry axes are showing. The
$\Delta$-axis is a four-fold rotary axis, the strange number S = 0,
the baryon family $\Delta$ ($\Delta^{++},$ $\Delta^{+},$
$\Delta^{0},$ $\Delta^{-}$) and N ($N^{+}, N^{0}$ ) will be born on
the axis. The $\Lambda$-axes and F-axis are three-fold rotary axes,
the strange number S =\ -1, the baryon family $\Sigma$
($\Sigma^{+}$,$\Sigma^{0}$,$\Sigma^{-}$) and $\Lambda$ will be born
on the axes. The $\Sigma$-axes and G-axis are two-fold rotary axes,
the strange number S = -2, the baryon family $\Xi$ ($\Xi^{0}$,
$\Xi^{-}$) will appear on the axes. The D-axis is parallel to the
axis $\Delta$, S = 0. The D-axis is a two-fold rotary axis, the
baryon family N (N$^{+}$, N$^{0}$) will appear on the axis.

The energy band theory \cite{Energy Band} has already given out the
symmetry groups of the symmetry points ($\Gamma$, H, M, P and N) and
the symmetry axes [$\Delta$ ($\Gamma$-H), $\Lambda$ ($\Gamma$-P),
$\Sigma$ ($\Gamma$-N), D (P-N), F (P-H) and G (M-n)].

At the points $\Gamma$, H and M, the wave functions have the
symmetries of the O$_h$ group. The irreducible representations and
basis function (or symmetry type) of the O$_h$ group appear in Table
A1.\newline

At the point P, the wave functions have the symmetries of $T_d$
group. The irreducible representations, characters and basis
function (or symmetry type) of the group appear in the Table
A2.\newline

At the points N, the group of the wave vector N is of order 8. Each
operation forms a class by itself, so that there are eight
irreducible representations, and all the states are not degenerate.
The wave functions have the symmetry of D$_{2h}$ group. The
irreducible representations, characters and basis function (or
symmetry type) of the group appear in Table A3.\newline

On the $\Delta$-axis, the wave vectors obey the symmetry group
$C_{4v}$. The irreducible representations, characters and basis
function (or symmetry type) of the group $C_{4v}$ appear in Table
A4.

On the $\Lambda$-axis, the wave vectors obey the symmetry group
$C_{3v}$. The irreducible representations, characters and basis
function (or symmetry type) of the group $C_{3v}$ appear in Table
A5.

On the $\Sigma$-axis, D-axis and G-axis, the wave vectors obey the
symmetry group $C_{2v}$. The irreducible representations, characters
and basis function (or symmetry type) of the group $C_{2v}$ appear
in Table A6.

The compatibility relations between states along symmetry axes and
the states at the end-points are in the following tables:

The compatibility relations of the $\Gamma $-$\Delta $, $\Gamma
$-$\Lambda$ and $\Gamma $-$\Sigma $ appear in Table A7.

The compatibility relations of the H-$\Delta$ and H-F appear in
Table A8.

The compatibility relations of the M-G appear in Table A9.

The compatibility relations of the P-D, P-F and P-$\Lambda$ appear
in Table A10.

The compatibility relations of the N-D, N-G and N-$\Sigma$ appear in
Table A11. $\qquad \qquad \qquad \qquad \qquad \qquad \qquad \qquad
$\newline

The coordinates of the symmetry points ($\Gamma $, H, P, N and M) are:

\begin{equation}  \label{Coordinats of Point}
\Gamma = (0, 0, 0), H = (0, 0, 1), P = (1/2, 1/2, 1/2), \\
N = (1/2, 1/2, 0), M = (1, 0, 0). \\
\end{equation}%

The coordinates ($\xi $, $\eta $, $\zeta $) of the symmetry axes are:

\begin{eqnarray}
\Delta = (0, 0, \zeta), 0 < \zeta < 1; \ D = (1/2, 1/2, \xi),
0 < \xi < 1/2;
\nonumber \\
\Lambda = (\xi, \xi , \xi), 0 < \xi < 1/2; \ F = (\xi, \xi, 1-\xi),
0 < \xi
< 1/2;  \nonumber \\
\Sigma = (\xi, \xi, 0), 0 < \xi < 1/2; \ G = (\xi, 1-\xi, 0),
1/2 < \xi < 1.
\ \   \label{Coordinats of Axes}
\end{eqnarray}
For any valid value of the vector $\vec{n}$, substituting the ($\xi
$, $\eta $, $\zeta $) coordinates of the symmetry points or the
symmetry axes into Eq.(\ref{Energy}) and Eq.(\ref{Plain Wave}), we
can get the E($\vec{k}$, $\vec{n}$) values and the wave functions at
the symmetry points and on the symmetry axes. In order to show how
to calculate the energy bands, we give the calculation of some low
energy bands in the symmetry axis $\Delta $ as an example (the
results appear in Fig. 2).

First, from (\ref{Energy}) and (\ref{Plain Wave}), we find the
formulae for the E($\vec{k}$, $\vec{n}$) values and the wave
functions at the end points $ \Gamma $ and H of the symmetry axis
$\Delta $, as well as on the symmetry axis $\Delta $ itself:
\begin{equation}
E_{\Gamma } = n_{1}^{2} + n_{2}^{2} + n_{3}^{2},  \label{gammae}
\end{equation}%
\begin{equation}
\Psi _{\Gamma }= exp(-\frac{i2\pi}{a}[n_{1}x+n _{2}y+n_{3}z]);
\label{gammaksi}
\end{equation}%
\begin{equation}
E_{H} = n_{1}^{2} + n_{2}^{2} + (n_{3}-1)^{2},  \label{he}
\end{equation}%
\begin{equation}
\Psi _{H} = exp (-\frac{i2\pi}{a}[n_{1}x+n_{2}y+(n_{3}-1)z]);
\label{hksi}
\end{equation}%
\begin{equation}
E_{\Delta } = n_{1}^{2} + n_{2}^{2} + (n_{3} -\zeta)^{2},
\label{deltae}
\end{equation}%
\begin{equation}
\Psi_{\Delta} = exp(-\frac{i2\pi}{a}[n_{1}x+n_{2}y+(n_{3}-\zeta)z]).
\label{deltaksi}
\end{equation}%
Then, using (\ref{gammae})--(\ref{deltaksi}), and beginning from the
lowest possible energy, we can obtain the corresponding integer
vectors $\vec{n}$ = (n$_{1}$, n$_{2}$, n$_{3}$) (satisfying
condition (\ref{N Conditions})) and the wave functions.

The lowest E($\vec{k}$, $\vec{n}$) is at ($\xi $, $\eta $, $\zeta $)
= 0 (the point $\Gamma $) and with only one value of $\vec{n}$ =
(0,0,0) \ \ (see (\ref{gammae}) and (\ref{gammaksi})):
\begin{equation}
\vec{n} = (0, 0, 0), \ \ \ \ \ E_{\Gamma } = 0, \ \ \ \ \Gamma_1 : \ \ \ \ \
\Psi _{\Gamma } = 1.  \label{GROUND E-W}
\end{equation}
This wave function is a $\Gamma_1$ symmetry type from Table A1.\newline

Starting from E$_\Gamma$ = 0, along the axis $\Delta $, there is one energy
band (the lowest energy band) E$_{\Delta }$ = $\zeta ^{2}$, with n$_{1}$ = n$%
_{2}$ = n$_{3}$ = 0 (see(\ref{deltae}) and (\ref{deltaksi})) ending at the
point E$_{H}$=1:
\begin{equation}
\vec{n} = (0, 0, 0), \ \ \ \  \\
E_{\Gamma } = 0,\, E_{\Delta } = \zeta ^{2},\, E_{H}= 1, \ \ \Delta_1 : \\
\Psi_{\Delta}= exp [i\frac{2\pi}{a}\zeta z]. \ \ \ \ \ \ \ \ \
\end{equation}%
\newline
This wave function is a $\Delta_1$ symmetry type from Table A4.\newline

At the point H of the energy band E$_{\Gamma }$ =0 $\rightarrow \ $E$%
_{\Delta }$= $\zeta ^{2}$ $\rightarrow $ \ \ E$_{H}$=1 (the end
point). Also at point H, $E_H$ = 1 when $\vec{n}$ =$(\pm 1,0,1)$,
$(0,\pm 1,1)$ and (0,0,2) (see (\ref{he}) and (\ref{hksi})):
\begin{equation}
E_{H} = 1, \ \ \Psi _{H} = e^{[i\frac{2\pi}{a}(\pm x)]}, e^{[i\frac{2\pi}{a}
(\pm y)]}, e^{[i\frac{2\pi}{a}(\pm z)]}.
\end{equation}
From Table A1, the symmetrized wave functions are \newline
\begin{equation}
H_1: \Psi =cos\frac{2\pi}{a}x+cos\frac{2\pi}{a}y+cos\frac{2\pi}{a}z,
\end{equation}
\label{H1}\newline
\begin{equation}
H_{12}: \Psi ={cos\frac{2\pi}{a}z-\frac{1}{2}(cos\frac{2\pi}{a}x+ cos\frac{%
2\pi}{a}y), cos\frac{2\pi}{a}x-cos\frac{2\pi}{a}y}
\end{equation}
\label{H1}\newline
\begin{equation}
H_{15}: \Psi = (sin\frac{2\pi}{a}x, sin\frac{2\pi}{a}y, cos\frac{2\pi}{a}z)
\end{equation}
\label{H15}\newline

Starting from $E_{H}=1$, along the axis $\Delta $, there are three energy
bands ending at the points $E_{\Gamma }=0$, $E_{\Gamma }=2$ and $E_{\Gamma
}=4$, respectively:
\begin{equation}
\vec{n}=(0,0,0), \\
E_{H}=1\rightarrow E_{\Delta }=\zeta ^{2}\rightarrow E_{\Gamma }=0, \ \
\Delta_1 : \Psi _{\Delta }=exp [i\frac{2\pi}{a}(\zeta z)].
\end{equation}%
This is a $\Delta_1$ symmetry type function from Table A4 on the
$\Delta$ axis.
\begin{equation}
\vec{n}=(0,0,2), \\
E_{H}=1\rightarrow E_{\Delta }= (2-\zeta)^{2}\rightarrow
E_{\Gamma }=4, \\
\Psi _{\Delta }= exp [i\frac{2\pi}{a}(2-\zeta )z]. \end{equation}
This is a $\Delta_1$ symmetry type function from Table A4 on the
$\Delta$ axis.
\begin{equation}
\vec{n}=(\pm 1,0,1)(0,\pm1,1), \\
E_{H}=1\rightarrow E_{\Delta }= 1+(1-\zeta)^{2}\rightarrow
E_{\Gamma}=2,
\end{equation}
\begin{equation}
\Psi _{\Delta }=e^{(-i\frac{2\pi}{a})[\pm x+(1-\zeta )z]},
e^{(-i\frac{2\pi}{a})[\pm y+(1-\zeta )z]}.
\end{equation}

The energy bands with four sets of values $\vec{n}$ $\ (\vec{n}=(\pm
$ 1,0,1$ ),$ $($0,$\pm $1,1$))$ are ending at $E_{\Gamma }=2$. The
four wave functions can be composed of $\Delta_1$,
$\Delta_2^{\prime}$ and $\Delta_5$ symmetry type functions:\newline
\begin{equation}
\\
\Delta_{1}:
e^{i\frac{2\pi}{a}(\zeta-1)z}(cos\frac{2\pi}{a}x+cos\frac{2\pi}{a
}y),
\end{equation}
\begin{equation}
\\
\Delta'_{2}: e^{i\frac{2\pi}{a}(\zeta-1)z}(cos\frac{2\pi}{a}x-cos
\frac{2\pi}{a}y),
\end{equation}
\begin{equation}
\\
\Delta_{5}: [e^{i\frac{2\pi}{a}(\zeta-1)z}sin\frac{2\pi}{a}x,
e^{i\frac{2\pi }{a}(\zeta-1)z}sin\frac{2\pi}{a}y].
\end{equation}
From (\ref{gammae}), $E_{\Gamma }=2$ also when $\vec{n}$ takes
another eight sets of values: $\vec{n} =(1,\pm 1,0)$, $(-1,\pm
1,0)$, and $(\pm 1,0,-1)$, $ (0,\pm 1,-1)$. Putting the $12$ sets of
$\vec{n}$ values into Eq. (\ref{gammaksi}), we can obtain 12 plane
wave functions:
\begin{equation}
\Psi _{\Gamma }= exp(-\frac{i2\pi}{a}(\pm x+z)), \Psi _{\Gamma }=
exp(-\frac{ i2\pi}{a}(\pm y+z)),
\end{equation}
\begin{equation}
\Psi _{\Gamma }= exp(-\frac{i2\pi}{a}[x \pm y]), \Psi _{\Gamma }=
exp(-\frac{i2\pi}{a}(-x \pm y)),
\end{equation}
\begin{equation}
\Psi _{\Gamma }= exp(-\frac{i2\pi}{a}(\pm x-z)), \Psi _{\Gamma }=
exp(-\frac{i2\pi}{a}(\pm y -z)).
\end{equation}
Using the 12 plane wave functions, we can compose $\Gamma_1$,
$\Gamma_{12}$, $\Gamma_{25}^{\prime}$, $\Gamma_{15}$ and
$\Gamma_{25}$ symmetry types at the point $E_{\Gamma}$ = 2. Since
this paper does not need the complex wave functions, we omit them.

Starting from $E_{\Gamma }=2$, along the axis $\Delta $, there are
three energy bands ended at the points $E_{H}=1$, $E_{H}=3$ and
$E_{H}=5$, respectively:
\begin{equation}
\vec{n} = (\pm1,0,1), (0,\pm 1,1);E_{\Gamma } = 2\rightarrow
E_{\Delta} = 1+ (1-\zeta) ^{2}\rightarrow E_H = 1,
\end{equation}
\begin{equation}
\vec{n}= (1,\pm 1, 0),(-1,\pm 1,0); E_{\Gamma } = 2\rightarrow
E_{\Delta }= 2+\zeta ^{2}\rightarrow E_H = 3,
\end{equation}
\begin{equation}
\vec{n}= (\pm 1, 0, -1),(0,\pm 1, -1); E_{\Gamma } = 2\rightarrow
E_{\Delta}=1+(\zeta +1)^{2}\rightarrow E_H=5.
\end{equation}

Continuing the process, we can find all low energy bands and the
corresponding wave functions as well as the irreducible
representations with symmetry types. The wave functions are not
necessary for deduction of quantum numbers and masses of baryons, so
we only show the energy bands with their index numbers $\vec{n}$
values and the irreducible representations at symmetry points and on
symmetry axes in Fig.2--Fig.8. There are six figures (Fig.2--Fig.7);
each of them shows the energy bands for one of the six axes in Fig.
1. For convenience, we only deduce the E($\vec{k}$, $\vec{n}$)
values in \ref{Energy}. Using (\ref{mass}), the energy of the energy
band $\varepsilon(\vec{k},\vec{n})$ = $\alpha E(\vec{k},\vec{n})$.
\newline

According to Postulate II, when a b-particle excites from the
vacuum, it will get an energy $M_{0}C^2$ at least; at the same time
it will go into an energy band with energy $\epsilon$. From
Postulate III, the observable mass $M_f$ of the baryons is
\begin{equation}
M_{Fig}\,\,\, = \,\,\, M_{0}\,\,+\,\, Minimum(\epsilon).
\label{Mass}
\end{equation} \\
From (\ref{Mass}), (\ref{Energy}) and (\ref{mass}), the masses
$M_{Fig}$ of the baryons (appearing in the Fig.2--Fig.7) can change
to

\begin{equation}
M_{Fig} = M_0 + minimum(\alpha E(\vec{k},\vec{n})).  \label{TMass}
\end{equation}

From (\ref{TMass}) and the experimental mass spectrum of baryons
\cite{Baryon}, we find that the lowest mass of all baryons is
[minimum($\alpha$ E($\vec{k}$,$\vec{n}$)) = 0]
\begin{equation}
M_{Fig} = M_0 =(1/2)(M_{p} + M_{n}) =939\,\,(Mev).  \label{M0 value}
\end{equation}
Comparing the mass spectrum of the baryons \cite{Baryon} and the
mass of the energy bands of the b-particle (\ref{TMass}), we find
\begin{equation}
\alpha = 360\,\, (Mev).  \label{360 Mev}
\end{equation}
From (\ref{M0 value}), (\ref{360 Mev}), (\ref{TMass}) and
(\ref{Energy}), we get the mass formula of the baryons that is
\begin{equation}
M_{Fig} = 939 + 360\,\times\,
minimum[(n_1-\xi)^2+(n_2-\eta)^2+(n_3-\zeta)^2].  \label{M0+360}
\end{equation}

Using (\ref{M0+360}), we can find the masses of the energy band
shown in Fig.2--Fig.8. These figures are schematic ones where the
straight lines of the energy bands should be parabolic curves. The
energy band indexes [$\vec{n}$= ($n_1, n_2, n_3$),
$\vec{n^{\prime}}$ = ($n^{\prime}_1, n^{\prime}_2, n^{\prime}_3$),
$\ldots$] appear above the lines of the energy bands; the
irreducible representations at symmetry points appear near the
points; the lowest energies (baryon masses) and highest energy of
the energy bands are marked near by the two end points of the energy
bands; and the irreducible representations on symmetry axes appear
above (or under) the lines of the energy band also. The index
numbers of energy bands appearing in the figures [such as
(101,011,-101,0-11)] are the simplified index numbers of (1, 0, 1;
0, 1, 1; -1, 0, 1; 0, -1, 1).

Both end points of an energy band (the intersections of the energy
band line and the two vertical lines) show the highest and lowest
energies (in Mev) of the energy bands in Figure 2--8. We can easily
find the lowest energy of the energy bands from Fig.2--Fig.8. It is
the mass of the baryon as shown in Fig.2--Fig.8.

Fig.2. The energy bands on the $\Delta-$axis. The energy
939+360$E_{\Gamma}$ (Mev) is the energy at $\Gamma$ point (one end)
of the energy band, and the energy 939+360$E_{H}$ (Mev) is the
energy at H point (another end) of the energy band.

Fig.3. The energy bands on the $\Lambda$-axis. The energy
939+360$E_{\Gamma}$ (Mev) is the energy at $\Gamma$ point (one end)
of the energy band, and the energy 939+360$E_{P}$ (Mev) is the
energy at P point (another end) of the energy band.

Fig.4. The energy bands on the $\Sigma$-axis. The energy
939+360$E_{\Gamma}$ (Mev) is the energy at $\Gamma$ point (one end)
of the energy band, and the energy 939+360$E_{N}$ (Mev) is the
energy at N point (another end) of the energy band.

Fig.5. The energy bands on the D-axis. The energy 939+360$E_P$ (Mev)
is the energy at P point (one end) of the energy band, and the
energy 939+360$E_{N}$ (Mev) is the energy at N point (another end)
of the energy band.

Fig.6. The energy bands on the F-axis. The energy 939+360$E_P$ (Mev)
is the energy at P point (one end) of the energy band, and the
energy 939+360$E_{H}$ (Mev) is the energy at H point (another end)
of the energy band.

Fig.7. The energy bands on the G-axis. The energy 939+360$E_N$ is
the energy at N point (one end) of the energy band, and the energy
939+360$E_{H}$ is the energy at H point (another end) of the energy
band.

Fig.8. The single energy bands on the $\Delta$-axis. The energy
939+360$E_{\Gamma}$ (Mev) is the energy at $\Gamma$ point (one end)
of the energy band, and the energy 939+360$E_{H}$ (Mev) is the
energy at H point (another end) of the energy band. \\

\includegraphics[scale=0.6,angle=90]{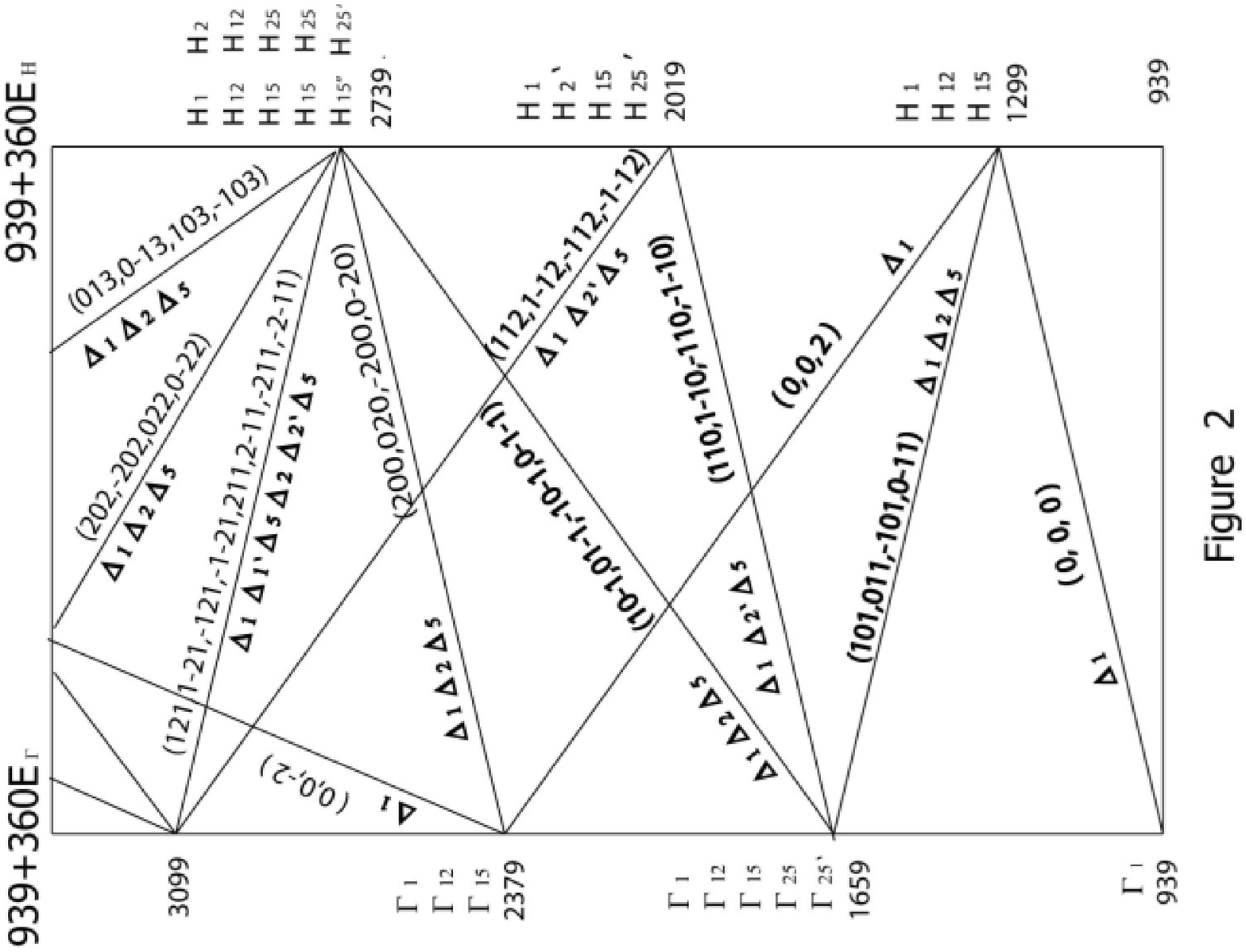}\ \ \ \ \ \newline
\includegraphics[scale=0.6,angle=90]{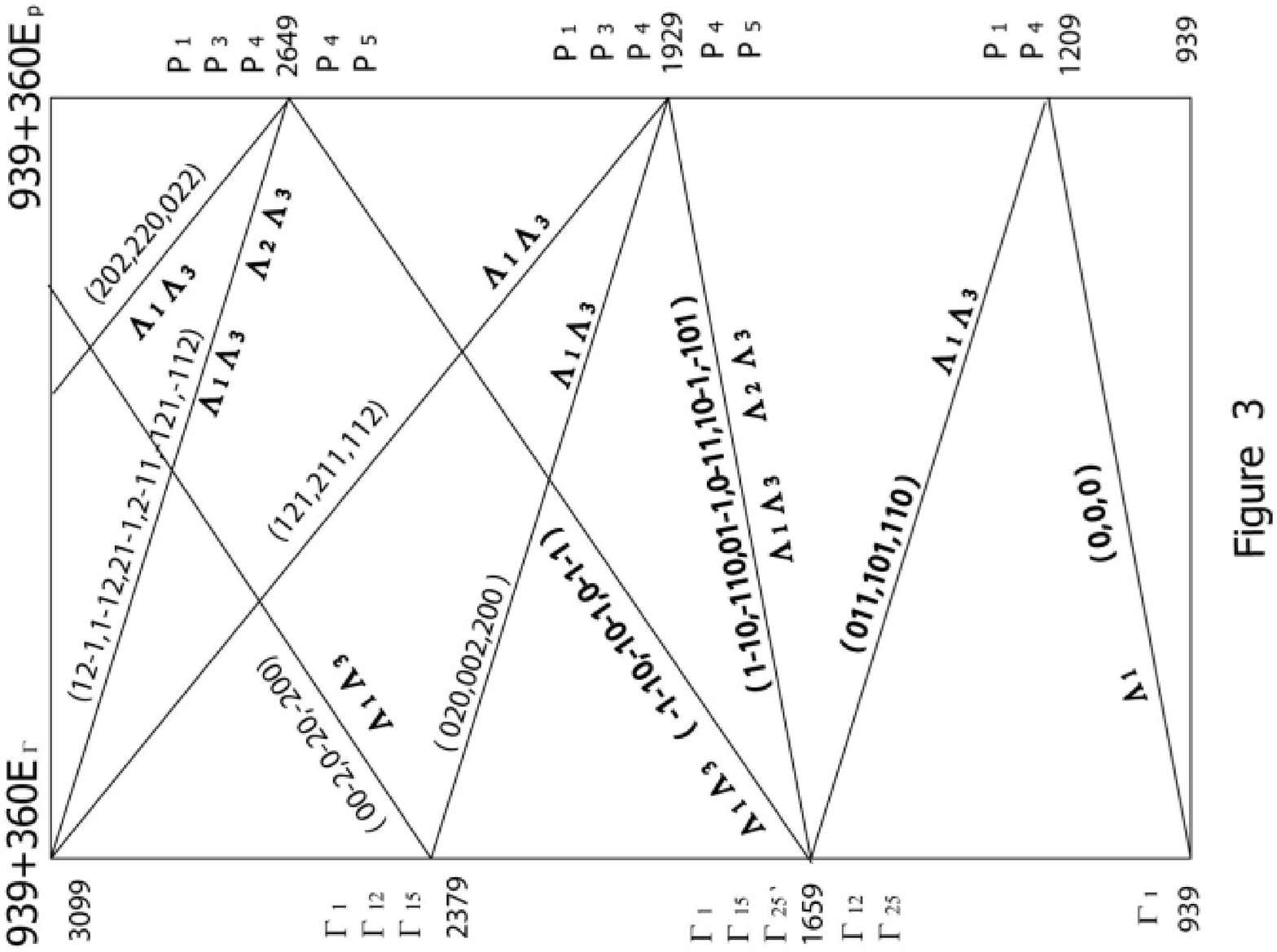}\ \ \ \ \ \newline
,\, $\qquad \qquad \qquad \qquad \qquad \qquad \qquad \qquad
$\newline
\includegraphics[scale=0.6,angle=90]{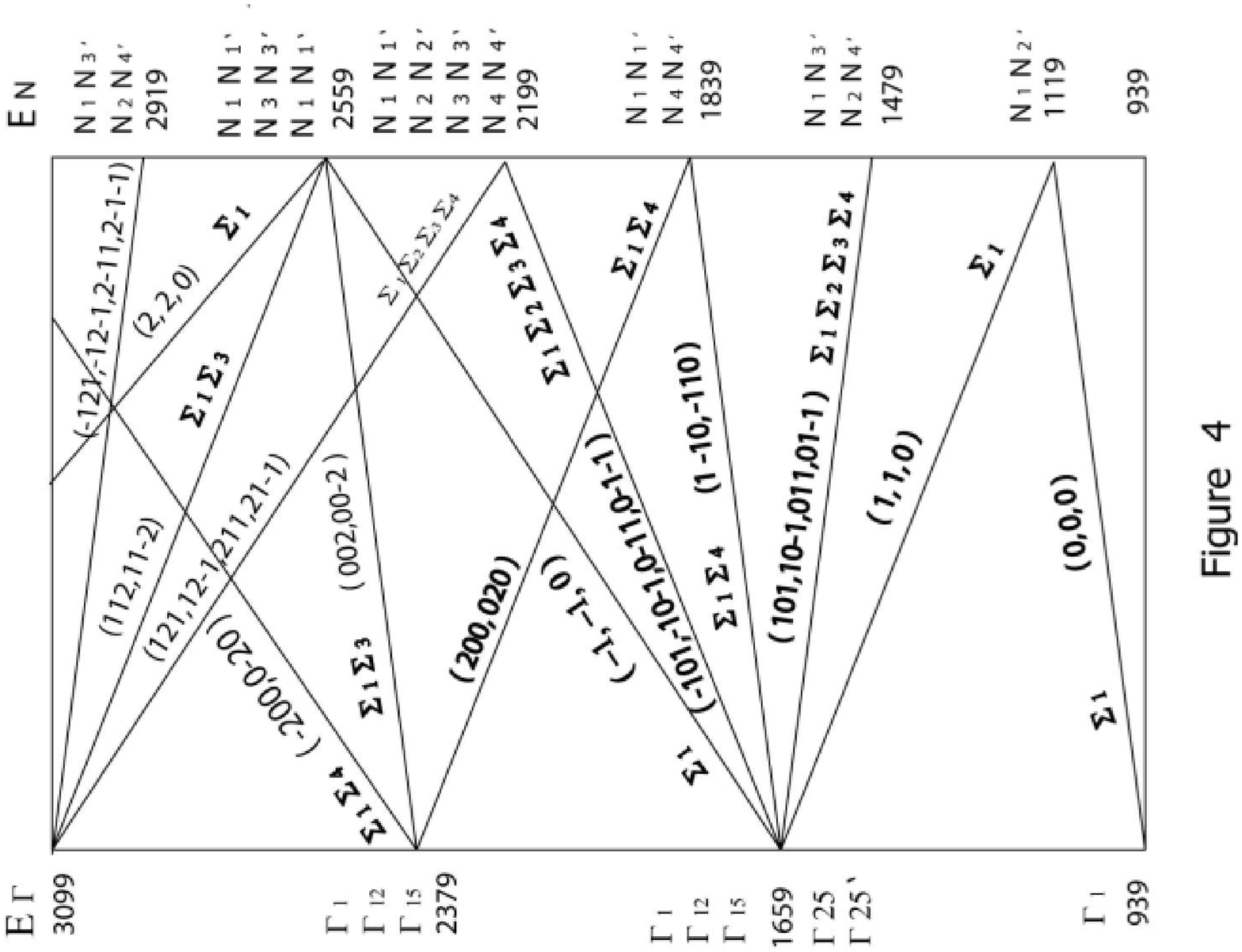}\ \ \ \ \ \newline
\includegraphics[scale=0.6,angle=90]{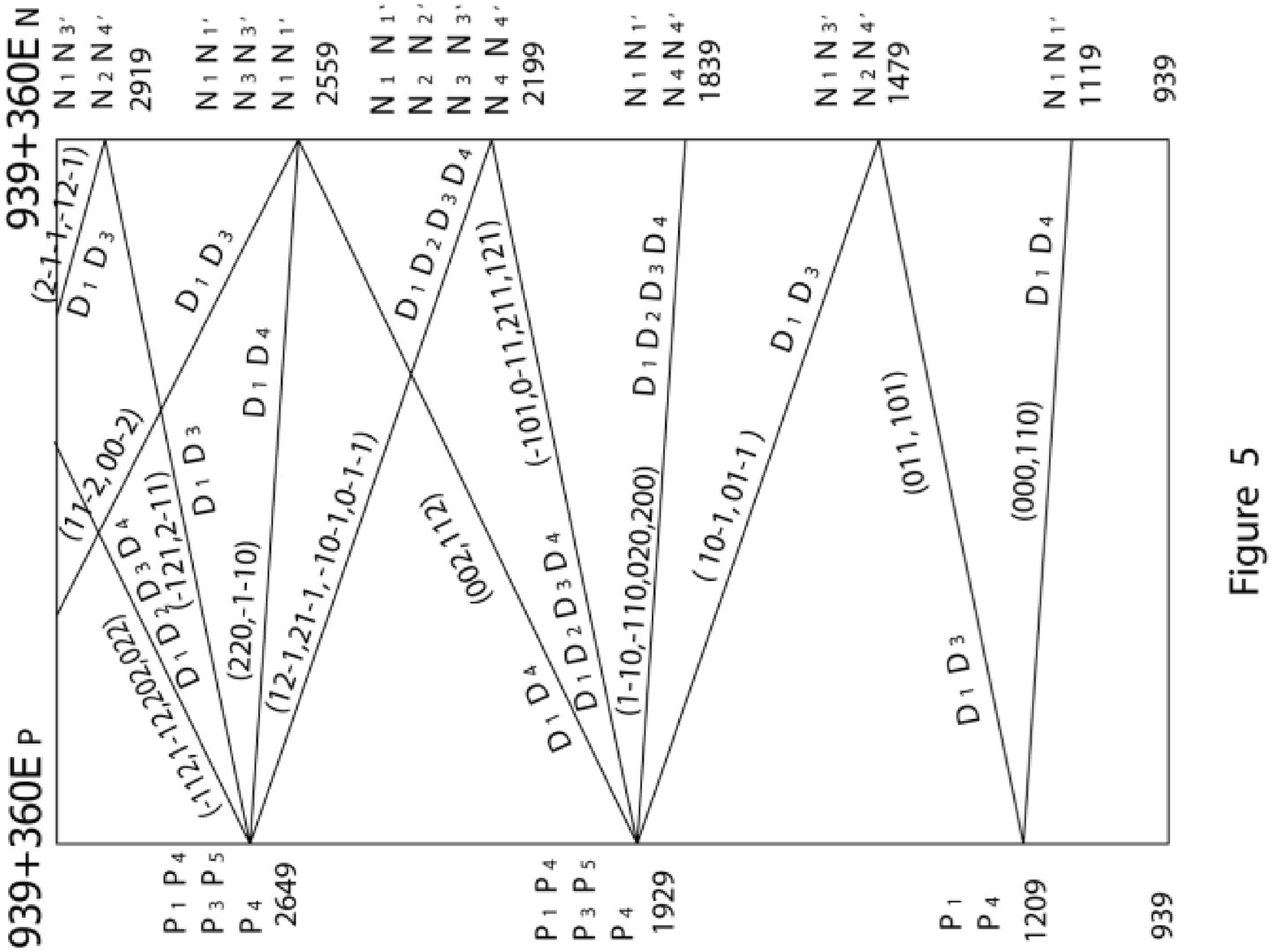} \ \ \ \ \newline
,\, $\qquad \qquad \qquad qquad \qquad \qquad \qquad \qquad $\newline
\includegraphics[scale=0.6,angle=90]{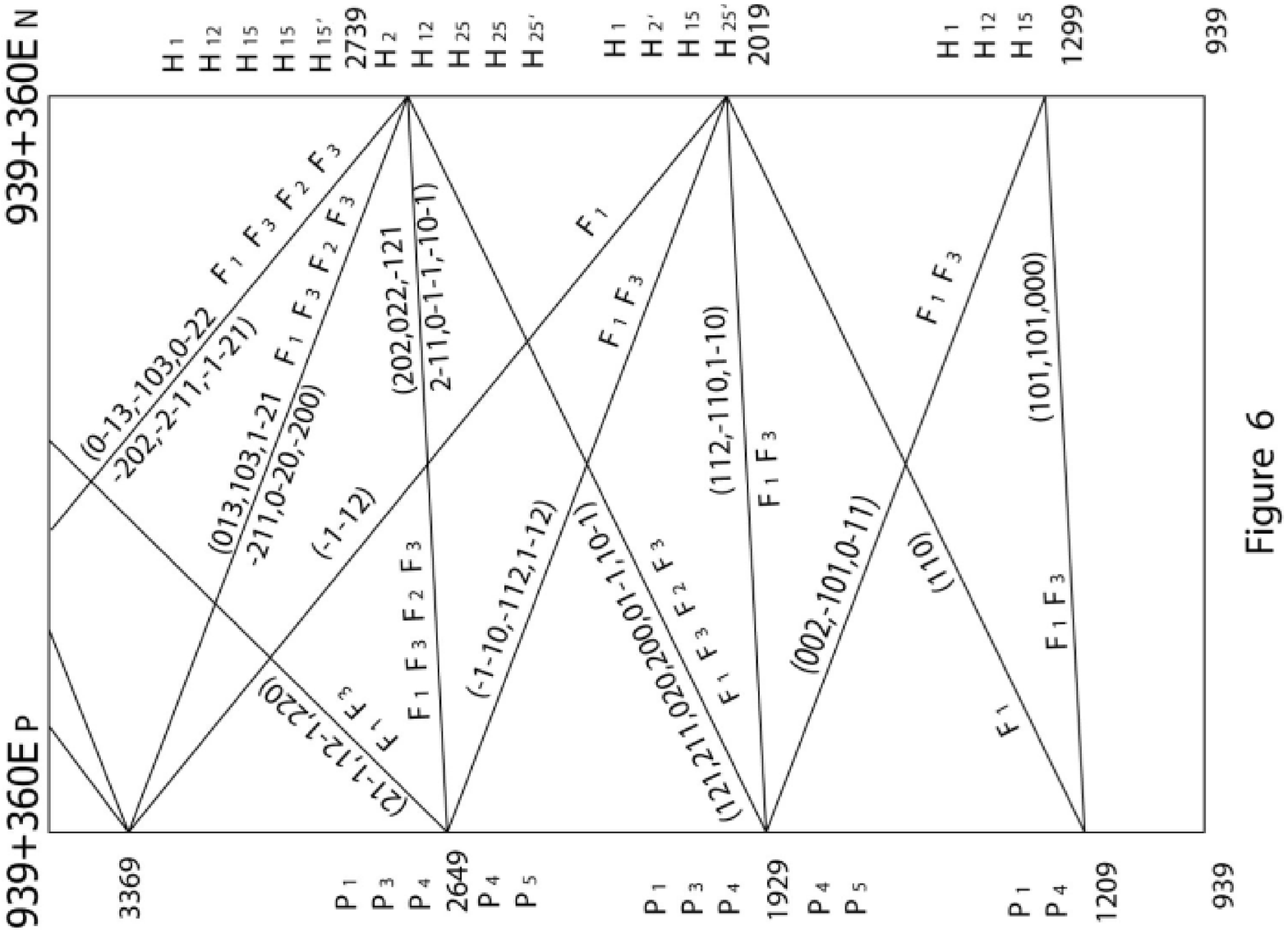} \ \ \ \ \newline
,\,
\includegraphics[scale=0.6,angle=90]{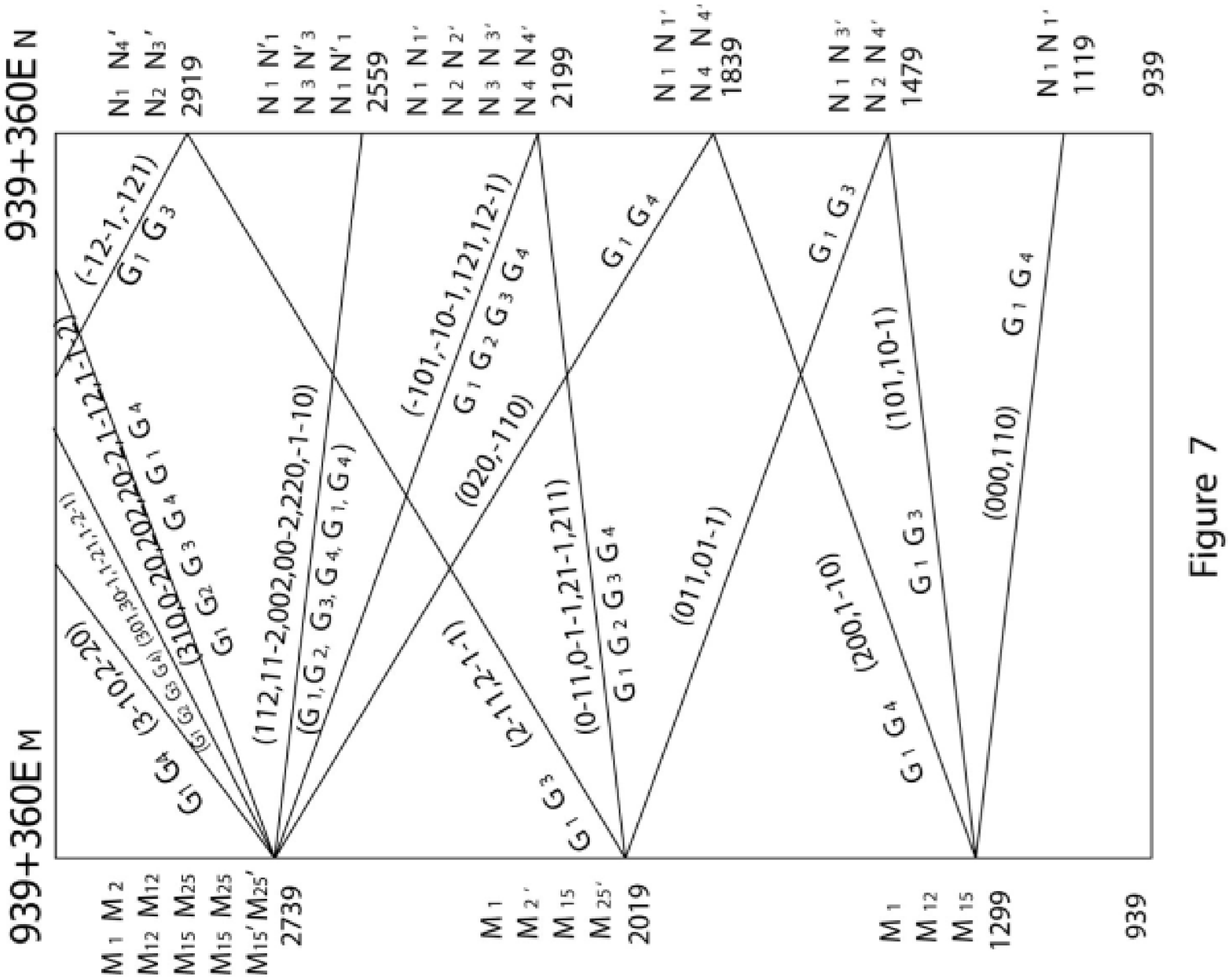} \ \ \ \ \newline
,\,$\qquad \qquad \qquad \qquad \qquad \qquad \qquad \qquad
$\newline
\includegraphics[scale=0.6,angle=90]{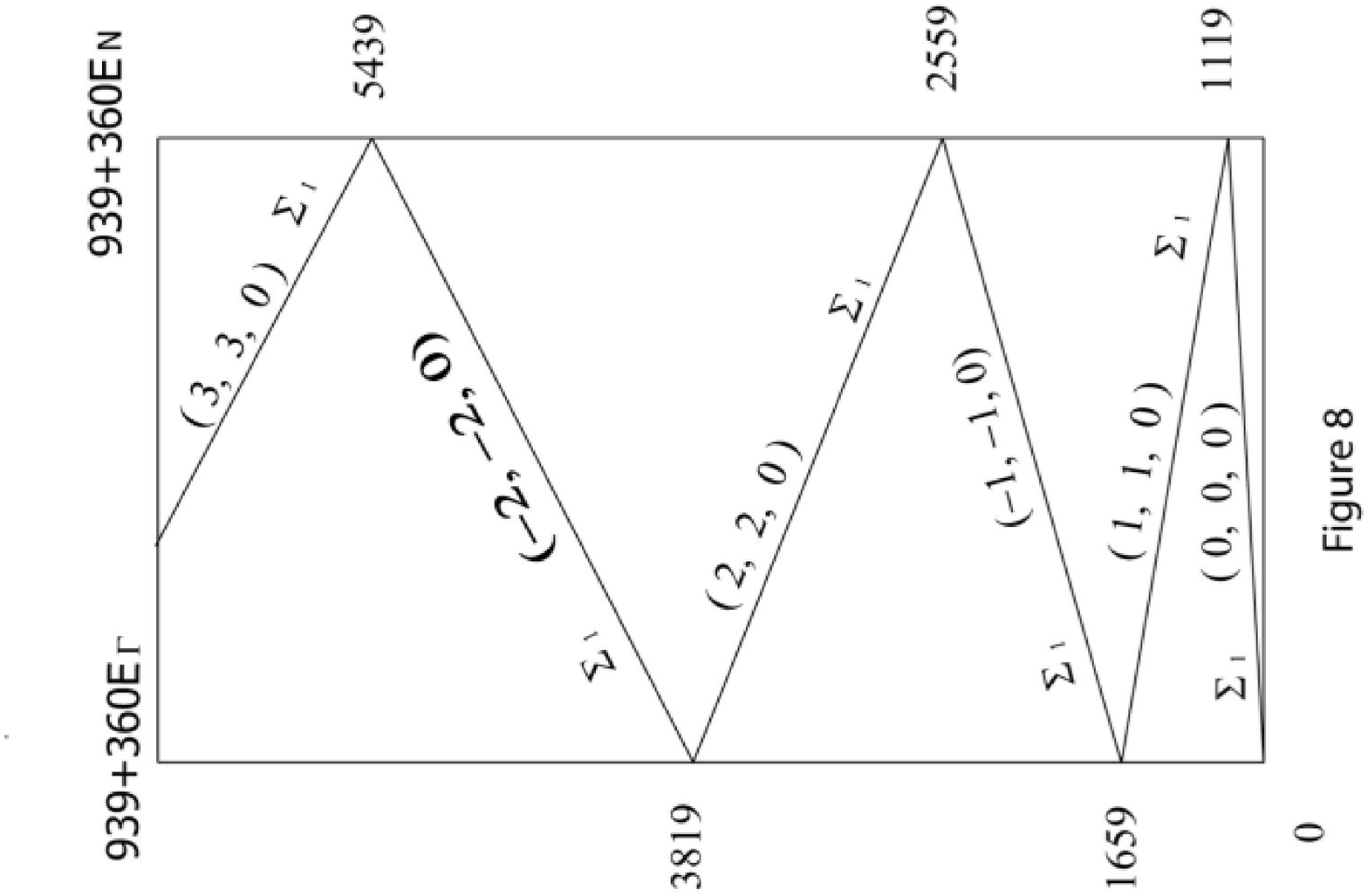} \ \ \ \ \newline
,\,
\section{The Phenomenological Formulae for the Quantum Numbers of
the Energy bands}

$\qquad$We have deduced the energy bands of the b-particle in the
above section (Fig.2--Fig.8). From postulates II and III, these band
exited states of the b-particle are baryons. Their masses $M_{Fig}$
have already appeared in Fig.2--Fig.8. In order to recognize the
baryons, we need phenomenological formulae of the baryon quantum
numbers (I, S, C, B, Q, J and P). Using these formulae, we can find
the quantum numbers for each energy band. Then comparing the quantum
numbers of the energy bands with the quantum numbers of the
experimental baryons \cite{Baryon} and quark model in Table 2 and
Table 3, we can recognize the baryons of the energy bands. The
formulae follow: \newline %

1. Isospin number I: I results from the energy band degeneracy d
using a formula
\begin{equation}
d + \delta(|n|) =2I_m +1,  \label{IsoSpin}
\end{equation}
\newline
where
\begin{equation}
|n| = \left\vert n_{1}\right\vert +\left\vert n_{2}\right\vert
+\left\vert n_{3}\right\vert,  \label{n value}
\end{equation}
and $\delta(|n|)$ is a Dirac function; for $|n|$ = 0, $\delta(|n|)$
= 1; for $|n|$ $\neq$ 0, $\delta(|n|)$ = 0.

If the d is larger than the rotary fold R of the symmetry axis,%
\begin{equation}
d > R, \label{deg > R}
\end{equation}%
the d will be divided $\gamma$ sub-degeneracies first,
\begin{equation}
\gamma = \frac{d}{R};  \label{Subdeg}
\end{equation}%
then, using (\ref{IsoSpin}), we can find the isospin values for each
sub-energy band. Using the $I_m$ value, we can find the 2$I_m$ +1
components: $I_{z}$ = $I_m$, $I_m$-1, $I_m$ -2, ..., -$I_m$. At the
same time, there is another I value for $I_m$ = 1 and $\frac{3}{2}$:
\begin{equation}
I = I_m - 1.  \label{I-2}
\end{equation}
Thus for $I_m$ = 1 ($\Sigma$ baryons with I = 1 and S = -1) there is
another baryon ($\Lambda$ baryons with I = 0 and S = -1); and also
for $I_m$ = $ \frac{3}{2}$ ($\Delta$ baryons with I = $\frac{3}{2}$
and S = 0), there is another baryon (N-baryons with I =
$\frac{1}{2}$ and S = 0).

2. Strange number S: For an energy band of a symmetry axis with a
symmetry rotary fold R, its strange number S results from the rotary
fold R of the symmetry axis with
\begin{equation}
S_{axis} = \mathcal{B}(1-\delta(|n|))(R-4), \label{Strange Number}
\end{equation}
where $\mathcal{B}$ = 1 for baryons (or = -1 for anti baryons) from
Postulate II and the number 4 is the highest possible rotary fold
number of the symmetry axes in the body-centred cubic periodic
symmetry. The $|n|$ = $|n_1|$ + $|n_2|$ + $|n_3|$, $\delta(|n|)$ is
a Dirac function: if $|n|$ = 0 $\delta(|n|)$ = 1, if $|n|$ $\ne$ 0
$\delta(|n|)$ = 0. From (\ref{Strange Number}), for the ground
energy band with
$\vec{n}$ = (0, 0, 0),\\
\begin{equation}
the\,\,\,\, S_{ground} = 0. \label{Groud}
\end{equation}\\
Using (\ref{Strange Number}), we can find that the strange numbers
of the bands on the $\Delta$-axis (R = 4) $S_{\Delta}$ = 0, the
strange number of the energy bands on the $\Lambda$-axis (R = 3)
$S_{\Lambda}$ = -1, and the strange number of the energy bands on
the $\Sigma$-axis (R = 2) $S_{\Sigma}$ = -2:
\begin{equation}
S_{\Delta} = 0,\,\,\, S_{\Lambda} = -1\,\,\, and\,\,\, S_{\Sigma} =
-2. \label {S = 0,-1,-2}
\end{equation}
Using (\ref{Strange Number}) we can also find the opposite strange
numbers for the anti-baryon.

The formula (\ref{Strange Number}) is working well for the three
axes (the $\Delta$-axis, the $\Lambda$-axis and the $\Sigma$-axis)
inside the first Brillouin zone. How to find the strange numbers for
the three surface axes though? We think that the strange numbers of
the surface axes are determined by an inside axis having the same
direction with it. The D-axis being parallel to the $\Delta$-axis,
from (\ref{S = 0,-1,-2}), the strange number of the bands on the
D-axis is
\begin{equation}
S_{D} = S_{\Delta}\,\,=\,\, 0.  \label{S-D}
\end{equation}

The F-axis being parallel to the $\Lambda$-equivalent-axis and R =
3, from (\ref{S = 0,-1,-2}), the strange number of the bands on the
F-axis is
\begin{equation}
S_{F} = S_{\Lambda}\,\,=\,\, -1.  \label{S-F}
\end{equation}

The G-axis being parallel to the $\Sigma$-equivalent-axis and R = 2,
from (\ref{S = 0,-1,-2}), the strange number of the bands on the
G-axis is
\begin{equation}
S_{G} = S_{\Sigma}\,\,=\,\, -2.  \label{S-G}
\end{equation}

We have deduced all strange numbers of all energy bands with d $\ge$
R on all symmetry axes using the formula (\ref{Strange Number}). For
the energy bands with d $\ge$ R, the formula (\ref{Strange Number})
is working well. For the energy bands with d less than R, the
formula might need to adjust since these energy bands have not the
R-fold rotary symmetries of the symmetry axis.

While to R - d = 2 cases, for first case R = 4 and d = 2, R - d = 2.
The R = 4 $\to$ $\Delta$-axis with $S_{\Delta}$ = 0 from
($\Delta$-baryon), d = 2 $\to$ I = $\frac{1}{2}$ from
(\ref{IsoSpin}). The band with I = $\frac{1}{2}$ and $S_{\Delta}$ =
0 might be a N-baryon, in this case the formula might not need to
adjust. The second case R = 3 and d = 1 R - d = 2. The R = 3 $\to$
the $\Lambda$-axis with $S_{\Lambda}$ = -1 and I = 1 the
$\Sigma$-baryon. The d = 1 I = 0 from (\ref{IsoSpin}). The band with
S = -1 and I = 0 might be a $\Lambda$-baryon. In this case the
formula might be not need to adjust. Thus for
\begin{equation}
R - d = 2 \label{R-d=2}
\end{equation}
the formula (\ref{Strange Number}) does not need to adjust, the
energy band has S = $S_{ax}$. For other cases,
\begin{equation}
d < R \,\,\,\, and \,\,\,\,R-d \neq 2,  \label{dlessR}
\end{equation}%
\newline
we have to find a revised formula of the strange number.

3. Electric charge Q: Using the baryon number $\mathcal{B}$, the
isospin I and the strange number S (C and b) of the energy band,
from the Gell-Mann--Nishijima relation \cite{GMN}, we can deduce the
electric charge Q of the energy band:
\begin{equation}
Q=I_{z}+\frac{1}{2}(\mathcal{B}+S+C+B).  \label{GMN}
\end{equation}

We must emphasize that (\ref{GMN}) can give the correct electric
charges based on the quark model. According to Postulate III, the
charges of baryons really come from the charges of the quarks.
Without the quarks, we cannot have solid physical foundation to
deduce the correct charges of the baryons with this formula.

In the above paragraphs (about strange number paragraphs), we stated
that ``for an energy band with the degeneracy d less than R and R-d
$\neq$ 2, the formula (\ref{Strange Number}) need to adjust. For
example, a single band on $\Delta$ axis (see Fig. 2), S = 0 from
(\ref{Strange Number}), I = 0 from(\ref{IsoSpin}). Using
Gell-Mann--Nishijima relation (\ref{GMN}), we get a fractional
charge:
\begin{equation}
Q=0+ \frac{1}{2}(1 + 0 + 0 + 0)=\frac{1}{2}.
\end{equation}
The experiments show that the charges of the baryons are always
integers or zero and never have fractional charge. Thus there is a
mistake. We believe that the Gell-Mann--Nishijima relation, the
baryon number ($\mathcal{B}$=1) and the isospin (I = 0) are all
correct, but the strange number (S = R - 4 = 0) is not correct since
the energy band (with d less than R and R-d $\neq$ 2) has already
broken the R fold rotation symmetry of the symmetry axis.

4. In order to find the correct formula of the strange number for
the energy bands with d $\textless$ R and R - d $\neq$ 2, we compare
the important single baryons (I = 0) with the possible corresponding
single bands. For the $\Sigma$-axis, from Fig. 4, we have:
\\$\newline$
\begin{tabular}{l}
Table 4.\ The S of baryons and the n-value of the single bands of
$\Sigma$-axis \\
\begin{tabular}{|l|l|l|l|l|l|l|l|l}
\hline Baryon or Band & Isospin & M & S or $S_{\Sigma}$ & $\Delta$S
&
n-value & S \\
\hline $\Lambda(1116)$ & 0 & 1116 & S=\,\,-1 &  &  & -1 \\
\hline $E_{N}$=1/2 & 0 & 1119 & $S_{\Sigma}$=\,\,-2 & $\Delta$S=+1 &
(1,1,0) & -1 \\
\hline $\Omega$(1672) & 0 & 1672 & S=\,\,-3 &  &  & -3  \\
\hline $E_\Gamma$=2 & 0 & 1659 & $S_{\Sigma}$=\,\,-2 & $\Delta$S=-1
&
(-1,-1,0) & -3 \\
\hline
\end{tabular}%
\end{tabular}%
\\ $\newline$

In Table 4, the I, M and S of $\Lambda$(1116) and $\Omega$(1672) are
the experimental values \cite{Baryon}; the I values of the two bands
are from (\ref{IsoSpin}), the M values of the two bands are from
Fig. 4 and the $S_{ax} $ values of the two bands are from
(\ref{Strange Number}).\newline

Since the two bands are born on the $\Sigma$-axis, the strange
number of the two bands will near the $S_\Sigma$\,= - 2. From Table
4, we can see that if we assume the strange number equals $S_\Sigma$
plus a $\Delta$S
\begin{equation}
S = S_{\Sigma} + \Delta S  \label{s+Ds}
\end{equation}
and
\begin{equation}
\Delta S = \mathcal{B}\,Sign(n_{1}+n_{2}+n_{3})\time 1  \label{Ds}
\end{equation}
we will get correct strange numbers. The sign function
\begin{equation}
Sign(n_{1}+n_{2}+n_{3})\,=\,\,+1,\,\,for\,\,(n_{1}+n_{2}+n_{3}) > 0, \\
\end{equation}%
\newline
\begin{equation}
Sign(n_{1}+n_{2}+n_{3})\,=\,\,-1,\,\, for\,\,(n_{1}+n_{2}+n_{3}) < 0, \\
\end{equation}%
\newline
\begin{equation}
Sign(n_{1}+n_{2}+n_{3})\, =\,\,0,\,\, for\,\,(n_{1}+n_{2}+n_{3}) = 0. \\
\end{equation}%
\newline

For single bands (I = 0) of the $\Delta$-axis, from Fig. 2, we have:
$\qquad$ \newline

\begin{tabular}{l}
Table 5.\ The S of baryons and the n-value of the single bands
on the $\Delta$-axis \\
\begin{tabular}{|l|l|l|l|l|l|l|l|l}
\hline Baryon or Band & I & M & S or $S_{\Delta}$ & $\Delta$S &
n-value
& S or C \\
\hline $\Lambda(1405)$ & 0 & 1405 & S=\,\,-1 &  &  & -1 \\
\hline $E_{H}$=1 & 0 & 1299 & $S_{\Delta}$=\,\,0 & $\Delta$S=\, \,-1
&
(0,0,2) & -1 \\
\hline $\Lambda_c$(2286) & 0 & 2286 & C=\,\,+1 &  &  & +1  \\
\hline $E_\Gamma$=4 & 0 & 2379 & $S_{\Delta}$=\,\,0 &
$\Delta$S=\,\,+1 &
(0,0,-2) & +1 \\
\hline
\end{tabular}%
\end{tabular}%
\newline
$\qquad \qquad \qquad \qquad \qquad \qquad \qquad \qquad $\newline

From Table 5 we can see that the formula (\ref{Ds}) is not working
simply for the single bands on the $\Delta$-axis. For the
$\Delta$-axis, the band with n =(0, 0, 2), $\Delta$ S = +1 from
(\ref{Ds}), but it needs $\Delta$ S = -1 as appearing in Table 5.
Thus the correct $\Delta$ S formulae, for the energy bands with d
less than R and R - d $\neq$ 2, is
\begin{equation}
\Delta S = \mathcal{B}[(1 - 2\delta({S_{ax}})]
Sign(n_{1}+n_{2}+n_{3}), \label{DS}
\end{equation}
\begin{equation}
S = S_{axis} + \Delta S.  \label{S+DS}
\end{equation}
From Table 4 and Table 5, we can see that the correct strange
numbers of the bands with d less than R and R - d $\neq$ 2 are
(\ref{S+DS}). The strange numbers oscillate around the strange
number value of the axis $S_{axis}$($\ref{Strange Number}$). This
situation shows that the fluctuation exists also in the particle
world.

5. Charmed number C and bottom number B: the strange number of the
single bands on the $\Delta$-{axis}, when $\vec{n}$= (0, 0, -2), S =
$S_{\Delta}$ + $\Delta$ S = +1. In order to compare it with the
experimental results, we would like to give the baryon of the energy
band a new name. The new name will be the charmed baryon $\Lambda_c$
with charmed number C = +1 . If an energy band (baryon) with a
$\Delta$ S = +1 (\ref{DS}) and its mass $\geq$ mass of
$\Lambda_c(2286)$, then we call it the charmed baryon with a charmed
number
\begin{equation}
C = +1.  \label{Charmed}
\end{equation}%
\newline
If an energy band with a $\Delta$ S = +1 (\ref{DS}) and S = -1 as
well as its energy being near even large the mass of the $\Lambda_b$
= 5624 Mev, we call it (the baryon of the energy band) the bottom
baryon with
\begin{equation}
B\,= \,-1.  \label{Bottom}
\end{equation}

6. The spins of the baryons: The spins of the baryons are determined
by the quark model. Such as in Table 2 (baryons with s =
$\frac{1}{2}$) and Table 3 (baryons with s = $\frac{3}{2}$). Since
the quarks all have spin s = $\frac{1}{2}$ and the baryons all
composed of three quarks, there are only two possible spin values: s
= $\frac{1}{2}$ and s = $\frac{3}{2}$. For the lowest energy band at
each symmetry points ($\Gamma$, H, M, N and P) and the lowest mass
baryons of each kind of baryons ($\Sigma$, $\Xi$, $\Sigma_c$,
$\Xi_c$, $\Omega_c$, $\Xi_{cc}$, $\Omega_{cc}$ and $\Lambda_b$) see
Table 2, the spin s = $\frac{1}{2}$ since the energies are not
enough to excited the spin to higher value s = $\frac{3}{2}$ ;
otherwise the s = $\frac{3}{2}$. Thus the corresponding energy bands
of the varied ground baryons [N(939), $\Lambda$(1116),
$\Sigma$(1193), $\Xi(1318)$, $\Lambda_c(2286)$, $\Sigma_c(2455)$,
$\Xi_c(2470)$, $\Xi_c(2577)$, $\Omega_c(2698)$, and
$\Lambda_b(5624)$.] have the spin
\begin{equation}
s \,\,= \,\, \frac{1}{2}.  \label{s = 1/2}
\end{equation}
The other energy bands (baryons) have spin s
\begin{equation}
s = \frac{1}{2}\,\,\, and\,\,\, \frac{3}{2}.  \label{Excited s}
\end{equation}

7. The orbit angular momentum L: the wave functions of the energy
bands are not eigenfunctions of the orbit angular momenta since they
satisfy the body-centred cubic periodic symmetries. Thus the energy
bands of the b-particle have short lifetimes and might be
``strange.'' There are, however, equivalent orbit angular momenta.
The equivalent orbit angular momenta are determined by the symmetry
types of the irreducible representations. The lowest equivalent
orbit angular momentums (L) of the irreducible representations have
appeared in Table A1--Table A6.

The single energy bands of the b-particle are usually longer
lifetime baryons (p, n, $\Lambda$, $\Omega$, $\Lambda_c$ and
$\Lambda_b$). Their orbit angular momenta are determined by the
irreducible representations of the symmetry group of the symmetry
axis. Such as for $\Delta_1$ of the $\Delta$ -axis, $\Lambda_1$ of
the $\Lambda$-axis, $\Sigma_1$ of the $\Sigma$-axis, the orbit
angular momenta $L_{axis}$ are

\begin{equation}
L_{axis}\,\,\,\,=\,\,\,\,0.  \label{Laxis}
\end{equation}

The energy band theory \cite{Callaway} points out that the basis
functions make possible an approximate correspondence (see Table
A1), for examples, $\Gamma_1$ $\to$ s wave (L = 0); $\Gamma_{15}$
$\to$ p wave (L = 1); $\Gamma_{12},\Gamma^{\prime}_{25} \,\to \,\,$
d wave (L = 2); $ \Gamma^{\prime}_{2}, \Gamma_{25}$, $\to$ f wave (L
= 3), $ \ldots$ (see Table A1).\newline

The energy band theory \cite{Callaway} further more points out that
these correspondences are only approximate in that if a wave
function, belonging to $\Gamma^{\prime}_{25}$, for instance, is
expanded in spherical harmonics (the eigenfunctions of angular
momentum), terms involving L = 4 and L = 6, etc., might be present
in addition to those of L = 2. Thus, when the energy higher the
possible L values of a representation are higher as following:
\begin{equation}
L\,\,=\,\,L_{low}\,\,+\,\, 2 \,\,+\,\ 2 \,\,\ldots, \label{L+2}
\end{equation}
the $L_{low}$ is the lowest possible L value which shown in Table
A1--A6 for each representation. The highest L value is given in
formula (\ref{Max J}).

The case can been seen from Table 6:

\begin{tabular}{l}
\\
Table 6A. \ Full rotation group compatibility table for the group O\\
\ \ \ \ \ \
\begin{tabular}{|l|l|l|l|}
\hline $D_{0}^{\pm}$ & $\Gamma_1$ \\
\hline $D_{1}^{\pm}$ & $\Gamma_4$   \\
\hline $D_{2}^{\pm}$ & $\Gamma_3$ + $\Gamma_5$  \\
\hline $D_{3}^{\pm}$ & $\Gamma_2$ + $\Gamma_4$ + $\Gamma_5$  \\
\hline $D_{4}^{\pm}$ & $\Gamma_1$ + $\Gamma_3$ + $\Gamma_4$ +
$\Gamma_5$  \\
\hline $D_{5}^{\pm}$ & $\Gamma_2$ + 2$\Gamma_4$ + $\Gamma_5$ \\
\hline $D_{6}^{\pm}$ & $\Gamma_1$ + $\Gamma_2$ + $\Gamma_3$ +
$\Gamma_4$ + 2$\Gamma_5$ \\
\hline $\ldots$ \\
\hline
\end{tabular}%
\end{tabular}%
\newline
The point group $O_h$ = O $\otimes$ $C_1$. The $O_h$ group is the
point group of the points $\Gamma$, point H and point M. $\qquad$
$\qquad $ $\qquad $ $\qquad $ $\qquad $ $\qquad$ $\qquad$\newline

8. The total angular momenta of the energy bands (baryons) are found
by
\begin{equation}
\vec{J} = \vec{s} + \vec{L}.  \label{J}
\end{equation}\\
Generally a particle with higher angular momenta has higher energy.
The experimental baryon spectrum \cite{Baryon} shows that if the
masses of the baryons are higher, the angular momenta of the baryons
are higher also. In order to simplify, we make a phenomenological
formulae for the highest total angular momentum of a baryon. The
maximum $J_{Max}$ values of a baryon will reach but do not exceed
$J_{max}$:
\begin{equation}
J_{max}\,\,\, \leq \frac{M-939}{360}\,\,+\,\, \frac{1}{2}[1 +
\delta(S)\frac{M-939}{939}] , \label{Max J}
\end{equation}%
where M is the mass of the baryon and the S is the strange number of
the baryon. The $\delta(S)$ is a Dirac function: if S = 0,
$\delta(S)$ = 1; if S $\ne$ 0, $\delta(S)$ = 0. Since the values of
J are quantized: $\frac{1}{2}$, $\frac{3}{2}$, $\frac{5}{2}$,
$\frac{7}{2}$, $\ldots$, if (\ref{Max J}) gives out a value that is
different from these quantized values, for example $\frac{13}{4}$,
the $J_{max}$ will take the nearest quantized value $\frac{7}{2}$.
If (\ref{Max J}) gives values that have two nearest quantized J
values, for examples $\frac{4}{2}$, $\frac{6}{2}$, $\frac{8}{2}$,
$\ldots$, the $J_{max}$ will take the larger ones:
\begin{equation}
 \frac{4}{2} \to \frac{5}{2},\,\,\,\frac{6}{2} \to \frac{7}{2},
 \,\,\,\frac{8}{2} \to \frac{9}{2}, \ldots . \label{even J}
\end{equation}

Using (\ref{Max J}) and (\ref{even J}), we can find the $J_{Max}$ of
baryons as appearing in Table 6B: $\newline$

\begin{tabular}{l}
$\qquad$$\qquad$$\qquad$$\qquad$$\qquad$Table 6B.\ The $J_{Max}$ of
baryons with mass M\\
\begin{tabular}{|l|l|l|l|l|l|l|l|l|l|l}
\hline M & 1299 & 1479& 1659 & 1814 & 1839 & 1929 & 1945& 2019 \\
\hline $J_{Max}$ \,\,of \,\,
$\Lambda\,\,or\,\,\Sigma$\,\,or\,\,$\Xi$ &$\frac{3}{2}$&
$\frac{5}{2}$& $\frac{5}{2}$ & $\frac{5}{2}$ & $\frac{7}{2}$ &
$\frac{7}{2}$ & $\frac{7}{2}$ &
$\frac{7}{2}$\\
\hline $J_{Max}$\,\,of \,\,$\Delta$\,\,or\,\, N\,\, &$\frac{3}{2}$ &
$\frac{5}{2}$ & $\frac{5}{2}$ & $\frac{7}{2}$& $\frac{7}{2}$ &
$\frac{7}{2}$ & $\frac{7}{2}$ & $\frac{9}{2}$ \\
\hline M &2199 &2379 &2444 &2559 & 2649 & 2739 & 2919 &3099 \\
\hline $J_{Max}$\,\,of \,\, $\Lambda\,\,or\,\,\Sigma\,\,or\,\,\Xi $&
$\frac{9}{2}$ & $\frac{9}{2}$ & $\frac{9}{2}$ & $\frac{11}{2}$ &
$\frac{11}{2}$ & $\frac{11}{2}$&
$\frac{13}{2}$ &  $\frac{13}{2}$\\
\hline $J_{Max}$\,\,of\,\, $\Delta$\,\,or\,\, N& $\frac{9}{2}$ &
$\frac{11}{2}$ & $\frac{11}{2}$ & $\frac{11}{2}$&$\frac{13}{2}$
& $\frac{13}{2}$ & $\frac{15}{2}$ & $\frac{17}{2}$\\
\hline
\end{tabular}%
\end{tabular}%
$\newline$\\

Table A1--Table A6 give out the lowest possible L value of the
representations. For higher mass baryons, the formulae (\ref{Max J})
and (\ref{even J}) and Table 6B will give out correct J values of
baryons.

9. Parity P: If the orbit angular momentum of the baryon is L, for a
ground state of a kind of baryons, the parity P is
\begin{equation}
P = (-1)^{Laxis},  \label{Ground P}
\end{equation}
where the Laxis is the orbit angular momentum from symmetry axes
(\ref{Laxis}).

For an excited baryon, the parity P is
\begin{equation}
P = - (-1)^{Lpoint},  \label{Excited P}
\end{equation}
where Lpoint is the orbit angular momentum from Table A1--A6.

10. An adjusted mass for some special baryons with $\Delta$ S $\neq$
0: For the energy bands with d less than R and d $\neq$ R - 2, there
is a $\Delta$S $ \neq$ 0. At the same time, there will be an
adjusted energy (mass) de. Thus for these baryons, the total mass
will be $M_{Fig}$ (\ref{TMass}) plus the adjusted mass de:
\begin{equation}
M\,\,= M_{Fig} + de.  \label{Total Mass}
\end{equation}
For simplification we assume that the adjusted energy de is:
\begin{equation}
de = - \beta[\delta(S_{ax} + f)(C-S)+ B + fS-n_{c}f]\Delta S,
\label{de}
\end{equation}
where $\beta$ = 115 (Mev); $\delta$($S_{ax}$ + f) is a Dirac
function, for $S_{ax}$ + f = 0, it = 1; otherwise it = 0. The
$S_{ax}$ is the strange number of the symmetry axis. The S is
strange number of the baryon. The C is the charmed number of the
baryon. For the three inside axes ($\Delta$-axis, $\Lambda$-axis and
$\Sigma$-axis), f = 0; while for the three surface axes (the d-axis,
the F-axis and the G-axis), f = 1. The $n_{c}$ = 0, 1, 2,$\ldots$,
is the order number of the charmed baryons. For one kind of charmed
baryon in a symmetrical axis, to the energy band (charmed baryon)
with the same energy and the same parity, the order number $n_r$
from the lowest to higher J values, the order number $n_{c}$ = 0,
$n_{c}$ = 1, $n_{c}$ = 2, $\ldots$. The adjusted energy formula de
can be simplified to each symmetrical axes as appearing in Table 7.

$\newline$ \\
\begin{tabular}{l}
Table 7.\ The adjusted energy formulae of the symmetry axes\\
\begin{tabular}{|l|l|l|l|l|l|l|l|l}
\hline $\Delta$-axis, $S_{\Delta}$ = 0, f = 0 &
de = - 115(C-S)$\times$$\Delta$ S \\
\hline $\Lambda$-axis, $S_{\Lambda}$ = -1, f = B = 0 &
de = 0 \\
\hline $\Sigma$-axis, $S_{\Sigma}$ = -2, f = 0 &
de = - 115B$\times$ $\Delta$S\\
\hline D-axis, $S_{D}$= 0, f = 1, B = 0 & de =
115($n_c$\,\,-\,\,S)$\times$$\Delta$ S,\,\,\,
$n_{c}$ = 0, 1, 2,$\ldots$ \\
\hline F-axis, $S_{F}$ = -1, f = 1,B = 0 & de = 115($n_c$
\,\,-\,\,C)$\times$$\Delta$S,\,\,\,$n_{c}$ = 0, 1, 2,$\ldots$\\
\hline G-axis, $S_{G}$ = -2, f = 1,B = 0 & de =
115($n_c$\,\,-\,\,S)$\times$ $\Delta$S,\,\,\, $n_{c}$ = 0, 1,
2,$\ldots$\\
\hline
\end{tabular}%
\end{tabular}%
$\newline$\\

\section{Deduction of the Baryons Spectrum
(Including M, I, S, C, B, Q, J and P of Baryons)}

$\qquad$The energy bands of the b-particle have been deduced in
section 3 (as appearing in Fig.2--Fig.8). Using above
phenomenological formulae (\ref{IsoSpin})--(\ref{de}), we can deduce
the quantum numbers of the energy bands from Fig.2--Fig.8. According
to Postulate III, for each energy band of the b-particle, there is
only one kind of baryons that has the same quantum numbers with the
energy band. This baryon can been excited (from the vacuum) into the
energy band to form an observable physical baryon. The mass of the
baryon is the lowest energy (of the energy band) that can be deduced
using (\ref{Total Mass}) and Fig.2--Fig. 8 ($M_{Fig}$) as well as
Table 7 (de).

From Fig.2--Fig.8 and Table 7, we can get the quantum numbers and
masses of the energy bands of the b-particle. We can get a baryon
number $\mathcal{B}$ = 1 (-1) for each energy band of the b-particle
(or anti b-particle) from Postulate II. Thus each energy band always
has a baryon number $\mathcal{B}$ = + 1. We will directly cite this
result (each band has $\mathcal{B}$ = +1) without deduction again.

First we deduce the ground baryons with the lowest mass of all
baryons.
\subsection{The ground state baryons}

$\qquad$From Fig.2, Fig.3 and Fig.4, we can see there are three
lowest energy bands with $\vec{n}$ = (0, 0, 0) and d = 1 on the
$\Delta$-axis, $\Lambda$-axis and $\Sigma$-axis. From the Postulate
II and (\ref{Baryon Number}), for each band of the three bands, we
get
\begin{equation}
\mathcal{B}= 1.  \label{B = 1}
\end{equation}%
Since $\vec{n}$ = (0, 0, 0) and d = 1, from (\ref{IsoSpin}), we get
\begin{equation}
I = \frac{1}{2}.  \label{I = 1/2}
\end{equation}%
Since $\vec{n}$ = (0, 0, 0), from (\ref{Strange Number}),
(\ref{S+DS}), (\ref {DS}), (\ref{Charmed}) and (\ref{Bottom}) we get
\begin{equation}
S = C = B = 0 .  \label{S = C = b}
\end{equation}

The three energy bands with $\vec{n}$ = (0, 0, 0) [in fact they are
a single band in three dimensional reciprocal lattice space] have
the I = $\frac{1}{2}$, S = C = B = 0. For I = $\frac{1}{2}$ band,
there are two members with $I_z$= + $\frac{1}{2}$ and
-$\frac{1}{2}$. Using Gell-Mann--Nishijima relation (\ref{GMN}), we
can deduce their electric charge: for $I_z$ = $\frac{1}{2}$, Q =
$I_z$ + 1/2($\mathcal{B}$ + S + C + B)= +1; for $I_z$= -
$\frac{1}{2}$, Q = $I_z$ + 1/2($\mathcal{B}$ + S + C + B)= 0. Thus
the three bands (one band) have I = $\frac{1}{2}$, S = C = B = 0 and
Q = +1, 0. Comparing the quantum numbers (I, S, C, B and Q) of the
bands with the quantum numbers of the baryons in Table 2 and in
\cite{Baryon}, we find there is only the N-baryon having the same
quantum numbers as the band. According to Postulate III, the
N-baryon will excite from the vacuum into the bands to form the
observable N-baryons.

From Fig.2, Fig.3 and Fig.4, we get $M_{Fig}$ = 939 (Mev), and from
(\ref{de}) and (\ref{S = C = b}), de = 0. From (\ref{Total Mass}),
the mass M of the bind is
\begin{equation}
M = M_{Fig} + de  = 939\,\, (Mev).  \label{939}
\end{equation}%

From Fig. 2, the energy band with $\vec{n}$ = (0, 0, 0) has
$\Delta_1$ type. Then from Table A4, $L_{axis}$ = 0 and P = +1 from
(\ref{Ground P}). From Fig. 3, the energy band with $\vec{n}$ = (0,
0, 0) has $\Lambda_1$ type. Then from Table A5, $L_{axis}$ = 0 and P
= +1 from (\ref{Ground P}). From Fig. 4, the energy band with
$\vec{n}$ = (0, 0, 0) has $\Sigma_1$ type. Then from Table A6,
$L_{ax}$ = 0 and P = +1 from (\ref{Ground P}). Thus the three bands
have L = 0, P = + and s = $\frac{1}{2}$ from (\ref{s = 1/2}). From
(\ref{J}), thus $J^{P}$ = $\frac{1}{2}^{+}$.

Therefore the three single energy bands with $\vec{n}$ = (0, 0, 0)
of the b-particle are the ground baryons N(939)(proton and neutron)
with $\mathcal{B}$ = +1; I = $\frac{1}{2}$ ; S = C = B = 0 and Q =
+1, 0; $J^{P}$ = $\frac{1}{2}^{+}$ and M = 939 (Mev):
\begin{equation}
\vec{n}\,\, = \,\,(0, 0, 0), \,\,\,N(939)\,\,\frac{1}{2}^{+}.
\label{N(939)}
\end{equation}

In Fig.4 and Fig.8, at $M_{Fig}$ = 1119, the single band with
representation $\Sigma_1$ and $\vec{n}$ = (1,1,0) on the
$\Sigma$-axis , has $\mathcal{B}$ = +1, I = 0 from (\ref{IsoSpin}).
Its $S_{\Sigma}$ = -2 from (\ref{Strange Number}) and $\Delta$S = +1
from (\ref{DS}), thus the band has S = $S_{\Sigma}$ + $\Delta$S = -
1. It also has Q = 0 from (\ref{GMN}). In total the band has I = 0,
S = -1 and Q = 0. Comparing the quantum numbers of the band with the
quantum numbers of the baryons in Table 2 and in \cite{Baryon}, we
find there is only the $\Lambda$-baryon having the same quantum
numbers as the band. According to Postulate III, the
$\Lambda$-baryon will excite from the vacuum into the band to form
an observable $\Lambda(1119)$-baryon. For the single band with
$\vec{n}$ = (1, 1, 0) and $\Sigma_1$, from (\ref{Laxis}), (\ref{J})
and (\ref{Ground P}), we have

\begin{equation}
\vec{n} = \,\,(1,\,\,1,\,\,0)\,\,\,\Lambda(1119)\,\,\,\frac{1}{2}^+
\label{1119}
\end{equation}\\

The band with $\vec{n}$ = (1,1,0) of the two-fold band with
$M_{Fig}$ = 1119 and $\vec{n}$ = (1,1,0; 0,0,0) in Fig.5 and Fig.7
has the $\mathcal{B}$ = +1. The two-fold asymmetry band with (1,1,0;
0,0,0) will divide into two single bands. The band with (0,0,0)
forms N(939) from (\ref{N(939)}). The another band with $\vec{n}$ =
(1,1,0) forms the $\Lambda(1119)$-baryon similarly to the one
mentioned above.

In Fig.6 at $M_{Fig}$ = 1209, the single band with $\vec{n}$ =
{1,1,0} might form a $\Lambda(1209)$-baryon as mentioned above.
Since these bands with $\vec{n}$ = (1,1,0) in various axes are the
same band in the receptive three dimension space, the lowest energy
of the three dimensional band is 1119 (Mev). Thus this band forms a
$\Lambda(1119)$-baryon.

Similarly to above, the band with $\vec{n}$ = (1,1,0) (of the
three-fold band with $M_{Fig}$ = 1209 and $\vec{n}$ =
(1,1,0;1,0,1;0,1,1) in Fig.3) form a $\Lambda(1119)$ also. Therefore
the energy bands with  $\vec{n}$ = (1,1,0) and $M_{Fig}$ = 1119 or
1209 will form a $\Lambda(1119)$-baryon as
\begin{equation}
the\,\, band\,\, with\,\, (110)\,\, and\,\, M_{Fig}\,\, =
\,\,1119\,\,or\,\, 1209\,\,\to \,\,\Lambda(1119).
\label{Lambda(1119)}
\end{equation}\\

\subsection{The baryons on the $\Delta$-axis}

$\qquad$The $\Delta $-axis is a four-fold rotary axis (see Fig.1), R
= 4. From Postulate II and (\ref{Strange Number}), we have
\begin{equation}
\mathcal{B} = 1\,\,\, and \,\,\, S_{\Delta} = 0. \label{Sd=0}
\end{equation}
Thus all energy bands of the $\Delta$-axis have $\mathcal{B}$ = 1
and $S_{\Delta}$ = 0.

For low energy levels, there are eight-fold degenerate and four-fold
degenerate energy bands as well as single bands on the axis in
Fig.2. The eight-fold degenerate band will divide into two four-fold
degenerate bands from (\ref{Subdeg}).

For four-fold degenerate bands, using (\ref{IsoSpin}), we get $I_m$
= 3/2, and $ I_{z}$ = 3/2, 1/2, -1/2, -3/2. From (\ref{GMN}), the
four-fold energy bands have Q = 2, 1, 0, -1. Thus, each four-fold
degenerate band of the b-particle has
\begin{equation}
\mathcal{B} = 1,\, S = 0,\, I_m = 3/2,\, Q = 2,\, 1,\, 0,\, -1.
\label{Delta}
\end{equation}%

Comparing the quantum numbers of the four-fold energy bands with the
quantum numbers of the baryons in Table 3 and the experimental
baryons in \cite{Baryon}, we find that there are only the
$\Delta$-baryon having the same quantum numbers as the energy band.
Thus, from Postulate III, the $\Delta$-baryons will excite from the
vacuum into the energy bands to form the observable
$\Delta$-baryons. From Fig. 2, we can find the mass $M_{Fig}$ of the
$\Delta$-baryon corresponding to the energy band as appearing in
Table 8. Since C = S = 0 of the $\Delta$-baryons, de = 0 from Table
7. Thus the $\Delta$-baryons have M = $M_{Fig}$ from (\ref{Total
Mass}).

At the same time, for each four-fold degenerate band, there is
always at least a two-fold band [sometimes, there are two two-fold
bands] with S = 0, I = $\frac{1}{2}$ from (\ref{I-2}), I$_{z} =
\frac{1}{2}, -\frac{1}{2}$ and Q = +1 and 0 from (\ref{GMN}).
Comparing the quantum numbers of the bands with the quantum numbers
of baryons deduced by the quark model in Table 2 and the
experimental baryons \cite{Baryon}, we find there is only the
N-baryons having the same quantum numbers as the band. According to
Postulate III, the N-baryons will excited from the vacuum into the
bands to form the observable N(M)-baryons. Similarly to the
$\Delta$-baryons, we can find the mass of the N-baryons
corresponding to the energy bands as appearing in Table 8.

At $M_f$ = 1659  $\vec{n}$ = (110,1-10,-110,-1-10), there are two
representations $\Gamma'_{25}$ and $\frac{1}{2}\Gamma_{12}$ that
they have the same L = 2. Thus there are two N(1659) with L = 2.
Similarly, at $M_f$ = 2739 and $\vec{n}$ = (211,2-11,-211,-2-11),
there are two N(2739) also shown in Table 8.

For the single bands, d = 1 I = 0 from (\ref{IsoSpin}),
$\mathcal{B}$ = 1 and $S_{\Delta}$ = 0.

At $M_{Fig}$ = 1299, the single band with $\vec{n}$ =(0, 0, 2) in
Fig. 2, from (\ref{S+DS}) and (\ref{DS}), has the strange number
\begin{equation}
S = S_{axis}+\Delta S = 0 - \,1 =\,-1 .  \label{D-single-S}
\end{equation}
The charge Q = 0 from (\ref{GMN}). The mass = $M_{Fig}$ + de =1299
+115 = 1414 from $(\ref{Total Mass})$ and Table 7. Thus the single
band has $\mathcal{B}$ = 1, I = 0, S = -1, Q = 0. Comparing the
quantum numbers with the quantum numbers of the baryons in Table 2,
we find there will be a $\Lambda(1414)$-baryon in the band.

At $M_{Fig}$ = 2379, the band with $\vec{n}$ = (0, 0, -2), has S =
+1 = C from (\ref{S+DS}) and (\ref{Charmed}):
\begin{equation}
S = S_{axis}+\Delta S = 0 + 1 = +1\,\,\to \,\,C. \label{C}
\end{equation}
Thus the band has $\mathcal{B}$ = 1, I = 0, C = +1 and Q = +1 from
(\ref{GMN}). Comparing the quantum numbers of the band with the
quantum numbers of the baryons in Table 2 and in \cite{Baryon}, we
find there is only the $\Lambda_c$-baryon having the same quantum
numbers as the band. According to Postulate III, the
$\Lambda_c$-baryon will excite from the vacuum into the band to form
the observable $\Lambda_c(M)$-baryon. Using Table 7, de = - 115
(Mev) from (\ref{C}) and Table 7. Thus the baryon mass M = $M_{Fig}$
+ de = 2379 -115 = 2264 (Mev) from (\ref{Total Mass}). For the
baryon $\Lambda_c(2264)$, $J^{P}$ = $\frac{1}{2}^{+}$, from
(\ref{Laxis}), (\ref{J}) and (\ref{Ground P}).

It is very important to pay attention to \textbf{the charmed baryon
$\Lambda_c(2264)$ born on the single energy band of the
$\Delta$-axis from $\Delta $S = +1} as appearing in Table 8.

Table B1 gives the possible $J^{P}$ values of the deduced baryons;
Table 6B gives the $J_{Max}$ values for the deduced baryon. The
correct $J^{P}$ appear in Table 8.
$\newline \\
\begin{tabular}{l}
$\qquad$$\qquad$$\qquad$$\qquad$$\qquad$ Table 8.\ The
baryons on the $\Delta$-axis \\
\begin{tabular}{|l|l|l|l|l|l|l|l|l|l|l|}
\hline $M_{Fig}$ & ($n_1,n_2,n_3$) & S & C & de & Baryon(M) &
J(1/2) & J(3/2) & P \\
\hline 939 & (0,\,\,0,\,\,0) & 0 & 0 & 0 & N(939) &
$\frac{1}{2}$ &  & +  \\
\hline 1299 &
\begin{tabular}{l}
(101,-101, \\
011,0-11)  %
\end{tabular}&
0 & 0 & 0 &
\begin{tabular}{l}
$\Delta$(1299) \\
N(1299)%
\end{tabular}
&
\begin{tabular}{l}
$\frac{3}{2}$ \\
$\frac{1}{2}$,$\frac{3}{2}$%
\end{tabular}  &  &  +  \\
\hline 1299 & (0,0,2) & -1 & 0 & 115 & $\Lambda$(1414) &
$\frac{1}{2}$ &  & - \\
\hline 1659 &
\begin{tabular}{l}
(110,1-10, \\
-110,-1-10)
\end{tabular}  & 0 & 0 & 0 &
\begin{tabular}{l}
$\Delta(1659)$ \\
N(1659) \\
N(1659)%
\end{tabular}&
\begin{tabular}{l}
$\frac{3}{2},\frac{5}{2}$  \\
$\frac{3}{2},\frac{5}{2}$  \\
$\frac{3}{2},\frac{5}{2}$  %
\end{tabular}&
\begin{tabular}{l}
$\frac{1}{2}$,$\frac{3}{2},\frac{5}{2}$ \\
$\frac{1}{2}$,$\frac{3}{2},\frac{5}{2}$ \\
$\frac{1}{2}$,$\frac{3}{2},\frac{5}{2}$ %
\end{tabular}& - \\
\hline 1659 &
\begin{tabular}{l}
(10-1,-10-1,  \\
01-1,0-1-1)   %
\end{tabular} & 0 & 0 & 0 &
\begin{tabular}{l}
$\Delta(1659)$ \\
N(1659)%
\end{tabular} & $\frac{1}{2},\frac{3}{2}$ &
$\frac{1}{2}$ $\frac{3}{2}$$\frac{5}{2}$ & + \\
\hline 2019 &
\begin{tabular}{l}
(112,1-12,  \\
-112,-1-12) %
\end{tabular} & 0 & 0 & 0 &
\begin{tabular}{l}
$\Delta(2019)$ \\
N(2019)%
\end{tabular} &
\begin{tabular}{l}
$\frac{1}{2}$,$\frac{3}{2}$; \\
$\frac{5}{2}$,$\frac{7}{2}$%
\end{tabular}
&
\begin{tabular}{l}
$\frac{1}{2}$,$\frac{3}{2}$,$\frac{5}{2}$ \\
$\frac{3}{2}$,$\frac{5}{2}$,$\frac{7}{2}$,$\frac{9}{2}$  %
\end{tabular} & +  \\
\hline 2379 &
\begin{tabular}{l}
(200,-200,   \\
020,0-20)    %
\end{tabular} & 0 & 0 & 0 &
\begin{tabular}{l}
$\Delta(2379)$ \\
N(2379)%
\end{tabular}
&
\begin{tabular}{l}
$\frac{3}{2}$,$\frac{5}{2}$ \\
$\frac{7}{2}$,$\frac{9}{2}$ \\
$\frac{1}{2}$,$\frac{3}{2}$ \\
$\frac{5}{2}$,$\frac{7}{2}$ \\
$\frac{9}{2}$,$\frac{11}{2}$ %
\end{tabular}
&
\begin{tabular}{l}
$\frac{1}{2}$,$\frac{3}{2}$,$\frac{5}{2}$,$\frac{7}{2}$ \\
$\frac{5}{2}$,$\frac{7}{2}$,$\frac{9}{2}$,$\frac{11}{2}$ \\
$\frac{1}{2}$,$\frac{3}{2}$,$\frac{5}{2}$ \\
$\frac{3}{2}$,$\frac{5}{2}$,$\frac{7}{2}$,$\frac{9}{2}$ \\
$\frac{7}{2}$,$\frac{9}{2}$,$\frac{11}{2}$  %
\end{tabular}
&
\begin{tabular}{l}
- \\
- \\
+ \\
+ \\
+ %
\end{tabular} \\
\hline 2379 & (0, 0,-2) & 0 & 1 & -115 & $\Lambda_{c}(2264)$ &
$\frac{1}{2}$ &  & +\\
\hline 2739 &
\begin{tabular}{l}
(121,1-21, \\
-121,-1-21, \\
211,2-11, \\
-211,-2-11)%
\end{tabular}
& 0 & 0 & 0 &  &  &  &  \\
 \hline 2739 &
\begin{tabular}{l}
(121,1-21, \\
-121,-1-21,%
\end{tabular}
& 0 & 0 & 0 &
\begin{tabular}{l}
$\Delta(2739) $ \\
N(2739)%
\end{tabular}
&
\begin{tabular}{l}
$\frac{1}{2}$,$\frac{3}{2}$ \\
$\frac{5}{2}$,$\frac{7}{2}$ \\
$\frac{9}{2}$,$\frac{11}{2}$%
\end{tabular}
&
\begin{tabular}{l}
$\frac{1}{2}$,$\frac{3}{2}$,$\frac{5}{2}$ \\
$\frac{3}{2}$,$\frac{5}{2}$,$\frac{7}{2}$,$\frac{9}{2}$ \\
$\frac{7}{2}$,$\frac{9}{2}$,$\frac{11}{2}$,$\frac{13}{2}$  %
\end{tabular}
& +   \\
\hline 2739 &
\begin{tabular}{l}
(211,2-11, \\
-211,-2-11)%
\end{tabular}
& 0 & 0 & 0 &
\begin{tabular}{l}
$\Delta(2739)$ \\
N(2739) \\
N(2739)%
\end{tabular}
&
\begin{tabular}{l}
$\frac{3}{2}$,$\frac{5}{2}$ \\
$\frac{7}{2}$,$\frac{9}{2}$ \\
$\frac{11}{2}$,$\frac{13}{2}$  %
\end{tabular}
&
\begin{tabular}{l}
$\frac{1}{2}$,$\frac{3}{2}$,$\frac{5}{2}$,$\frac{7}{2}$ \\
$\frac{5}{2}$,$\frac{7}{2}$,$\frac{9}{2}$,$\frac{11}{2}$ \\
$\frac{9}{2}$,$\frac{11}{2}$,$\frac{13}{2}$   %
\end{tabular} & -\\
\hline 2739 &
\begin{tabular}{l}
(202,-202, \\
022,0-22)%
\end{tabular}
& 0 & 0 & 0 &
\begin{tabular}{l}
$\Delta(2739)$ \\
N(2739)%
\end{tabular}
&
\begin{tabular}{l}
$\frac{5}{2}$,$\frac{7}{2}$ \\
$\frac{9}{2}$,$\frac{11}{2}$ %
\end{tabular}
&
\begin{tabular}{l}
$\frac{3}{2}$,$\frac{5}{2}$,$\frac{7}{2}$,$\frac{9}{2}$ \\
$\frac{7}{2}$,$\frac{9}{2}$,$\frac{11}{2}$,$\frac{13}{2}$   %
\end{tabular}
& + \\
\hline 2739 &
\begin{tabular}{l}
(013,0-13, \\
103,-103)%
\end{tabular}
& 0 & 0 & 0 &
\begin{tabular}{l}
$\Delta(2739)$ \\
N(2739)%
\end{tabular} &
\begin{tabular}{l}
$\frac{7}{2}$,$\frac{9}{2}$ \\
$\frac{11}{2}$,$\frac{13}{2}$ %
\end{tabular} &
\begin{tabular}{l}
$\frac{5}{2}$,$\frac{7}{2}$,$\frac{9}{2}$,$\frac{11}{2}$ \\
$\frac{9}{2}$,$\frac{11}{2}$,$\frac{13}{2}$ %
\end{tabular}
& - \\

\hline 3099 &
\begin{tabular}{l}
(12-1,1-2-1, \\
-121,-1-2-1)%
\end{tabular}
& 0 & 0 & 0 &
\begin{tabular}{l}
$\Delta(3099)$ \\
N(3099)%
\end{tabular} &
\begin{tabular}{l}
$\frac{3}{2}$,$\frac{5}{2}$ \\
$\frac{7}{2}$,$\frac{9}{2}$ \\
$\frac{11}{2}$,$\frac{13}{2}$ %
\end{tabular} &
\begin{tabular}{l}
$\frac{1}{2}$,$\frac{3}{2}$,$\frac{5}{2}$,$\frac{7}{2}$ \\
$\frac{5}{2}$,$\frac{7}{2}$,$\frac{9}{2}$,$\frac{11}{2}$ \\
$\frac{9}{2}$,$\frac{11}{2}$,$\frac{13}{2}$,$\frac{15}{2}$%
\end{tabular}
& - \\

\hline 3099 &
\begin{tabular}{l}
(21-1,2-1-1, \\
-21-1,-2-11)%
\end{tabular}
& 0 & 0 & 0 &
\begin{tabular}{l}
$\Delta(3099)$ \\
N(3099)%
\end{tabular} &
\begin{tabular}{l}
$\frac{7}{2}$,$\frac{9}{2}$ \\
$\frac{11}{2}$,$\frac{13}{2}$ %
\end{tabular} &
\begin{tabular}{l}
$\frac{5}{2}$,$\frac{7}{2}$,$\frac{9}{2}$,$\frac{11}{2}$ \\
$\frac{9}{2}$,$\frac{11}{2}$,$\frac{13}{2}$,$\frac{15}{2}$%
\end{tabular}
& - \\

\hline 3099 &
\begin{tabular}{l}
(11-2,1-1-2, \\
-11-2,-1-1-2)%
\end{tabular}
& 0 & 0 & 0 &
\begin{tabular}{l}
$\Delta(3099)$ \\
N(3099)%
\end{tabular} &
\begin{tabular}{l}
$\frac{5}{2}$,$\frac{7}{2}$ \\
$\frac{9}{2}$,$\frac{11}{2}$ \\
$\frac{13}{2}$,$\frac{15}{2}$ %
\end{tabular} &
\begin{tabular}{l}
$\frac{3}{2}$,$\frac{5}{2}$,$\frac{7}{2}$,$\frac{9}{2}$ \\
$\frac{7}{2}$,$\frac{9}{2}$,$\frac{11}{2}$,$\frac{13}{2}$\\
$\frac{11}{2}$,$\frac{13}{2}$,$\frac{15}{2}$,$\frac{17}{2}$%
\end{tabular}
& + \\

\hline
\end{tabular}%
\end{tabular}%
$\newline \\

\subsection{The baryons on the $\Lambda$-axis}

$\qquad$The $\Lambda$-axis is a three-fold rotary axis (see Fig. 1),
R = 3, from (\ref{Strange Number}) and Postulate II, any energy band
has
\begin{equation}
\mathcal{B} = 1 \,\,\,and\,\,\,S_{\Lambda} = -1. \label{B+S of
Lamb-axis}
\end{equation}
Thus all energy bands of the $\Lambda$-axis have $\mathcal{B}$ = 1
and $S_{\Lambda}$ = -1.

From Fig. 3, we see that there is a single energy band with
$\vec{n}$ = (0, 0, 0); it is the baryon N(939) from (\ref{N(939)}).
All other bands are three-fold or six-fold energy bands. From
(\ref{Subdeg}), the six-fold energy bands will divide into two
three-fold energy bands. Thus we only need to discuss the three-fold
energy bands.

For a three-fold energy band, I = 1 from (\ref{IsoSpin}); we have
already found the strange number of all energy bands S = -1 and the
baryon number $\mathcal{B}$ = +1 in (\ref{B+S of Lamb-axis}). Thus
from (\ref{GMN}), we have Q = 1, 0, -1. Comparing the quantum
numbers of the three-fold energy bands with the quantum numbers of
the baryons in Table 2 and the experimental baryon spectrum in
\cite{Baryon}, we find there is only the $\Sigma$-baryon having the
same quantum numbers as the band. According to Postulate III, the
$\Sigma$-baryons will excite from the vacuum into the three-fold
energy bands to form the observable $\Sigma(M)$-baryons
($\Sigma^{+}$, $\Sigma ^{0}$, $\Sigma ^{-}$). Since de = 0 from
Table 7, the mass M = $M_{Fig}$ (from Fig. 3) as appearing in Table
9.

At the same time, from (\ref{I-2}), there is another baryon with I =
1 - 1 = 0 for each three-fold energy band. The baryon has
$\mathcal{B}$ = 1, S = -1, I = 0 and Q = 0 from (\ref{GMN}).
Comparing the quantum numbers of the baryon with the quantum of the
baryons in Table 2 and the experimental baryon spectrum in
\cite{Baryon}, we find that the baryon is $\Lambda(M)$-baryon. The
mass M = $M_{Fig}$ as appearing in Table 9. Thus there is a
$\Sigma$-baryon and a $\Lambda$-baryon in each three-fold energy
band as appearing in Table 9.

On the $\Lambda$-axis, there is a $\Sigma(1209)$ and a
$\Lambda(1209)$ in the lowest three-fold band with $\vec{n}$ =
(110,101,011) and $M_f$ = 1209. The $\Lambda(1209)$ corresponds to
the band with (1,1,0). From (\ref{Lambda(1119)}), the band with
(1,1,0) forms a $\Lambda(1119)$-baryon as appearing in Table 9 also.

Table A2 gives the possible $J^{P}$ values of the deduced
$\Sigma$-baryons and $\Lambda$-baryons; and Table 6B limits out the
correct $J^{P}$ values as appearing in Table 9:

At $M_{fig}$ = 1659, a six fold band with $\vec{n}$ =
(1-10,-110,01-1 0-11,10-1,-101) divided into two three-fold bands.
The first three-fold band with $\vec{n}$ = (1-10,-110,01-1) has I =
1, S = -1 and Q = 1, 0, -1 from (\ref{GMN}). This is a
$\Sigma$(1659). At same time, there is a $\Lambda(1659)$ also. From
Table B2, corresponding to this band the representation is a three
fold $\Gamma'_{25}$ (L = 2). Thus $J^{P}$ = $\frac{3}{2}^{-}$,
$\frac{5}{2}^{-}$; $\frac{1}{2}^{-}$, $\frac{3}{2}^{-}$,
$\frac{5}{2}^{-}$. The second three-fold sub-band with $\vec{n}$ =
(0-11,10-1,-101), in Table B2, spin s = $\frac{1}{2}$ and
$\frac{3}{2}$, representation = $\Gamma_{12}$ (L = 2) and
$\Gamma_{1}$, (L = 0). Thus for three-fold $\Sigma$(1659),  $J^P$
will take the common values $\frac{1}{2}^{-}$, $\frac{3}{2}^{-}$.
For single $\Lambda(1659)$, $J^{P}$ will take $\frac{1}{2}^{-}$,
$\frac{3}{2}^{-}$ for L = 0 and $\frac{3}{2}^{-}$,
$\frac{5}{2}^{-}$; $\frac{1}{2}^{-}$, $\frac{3}{2}^{-}$,
$\frac{5}{2}^{-}$ for L = 2. The $\Sigma$(1659), $J^P$ values and
$\Lambda(1659)$, $J^{P}$ values will appear in Table 9.
$\newline$\\

\begin{tabular}{l}
$\qquad$$\qquad$$\qquad$$\qquad$$\qquad$\ \ \ Table 9.\ The
baryons on the $\Lambda$-axis \\
\begin{tabular}{|l|l|l|l|l|l|l|l|l|l|l|}
\hline $M_{Fig}$ & ($n_1,n_2,n_3$) & I & S & $\Sigma(M)$ or
$\Lambda$(M) & J(1/2) &  J(3/2) & P \\
\hline 1209 & (110,101,011) & 1 & -1 & $\Sigma(1209)$ &
$\frac{1}{ 2}$, \,$\frac{3}{2}$ & & + \\
\hline 1209 & (1, 1, 0) & 0 & -1 &
$\Lambda$(1119) & $\frac{1}{ 2}$ &  & + \\
\hline 1659 &
\begin{tabular}{l}
(1-10,-110,01-1  \\
0-11,10-1,-101) %
\end{tabular} &   & -1 &  &  &  &  \\
\hline 1659 & (1-10,-110,01-1) & 1 or 0 & -1 & $\Sigma(1659)$or
$\Lambda$(1659) & $\frac{ 3}{2}$,$\frac{5}{2}$ &
$\frac{1}{2}$,$\frac{3}{2}$,$\frac{5}{2}$ & - \\
\hline 1659 & (0-11,10-1,-101) & 1 & -1 & $\Sigma(1659)$ &
$\frac{1}{2}$ & $\frac{3}{2}$ & - \\
\hline 1659 & (0-11,10-1,-101) & 0 & -1 & $\Lambda$(1659) &
\begin{tabular}{l}
$\frac{1}{2}$  \\
$\frac{3}{2}$,$\frac{5}{2}$ %
\end{tabular} &
\begin{tabular}{l}
$\frac{3}{2}$ \\
$\frac{1}{2}$,$\frac{3}{2}$,$\frac{5}{2}$ %
\end{tabular}
& - \\
\hline 1659 & (-10-1,0-1-1,-1-10) & 1 or 0 & -1 & $\Sigma(1659)$ or
$\Lambda$(1659) & $ \frac{1}{2}$,$\frac{3}{2}$ &
$\frac{1}{2}$,$\frac{3}{2}$,$\frac{5}{2}$ & +  \\
\hline 1929 & (020,002,200) & 1 or 0 & -1 & $\Sigma(1929)$ or
$\Lambda$(1929) &
\begin{tabular}{l}
$\frac{1}{2}$,$\frac{3}{2}$ \\
$\frac{5}{2}$,$\frac{7}{2}$ \\
$\frac{3}{2}$,$\frac{5}{2}$  %
\end{tabular}
&
\begin{tabular}{l}
$\frac{1}{2}$,$\frac{3}{2}$,$\frac{5}{2}$ \\
$\frac{3}{2}$,$\frac{5}{2}$,$\frac{7}{2}$ \\
$\frac{1}{2}$,$\frac{3}{2}$,$\frac{5}{2}$,$\frac{7}{2}$%
\end{tabular}
&
\begin{tabular}{l}
+ \\
- %
\end{tabular} \\

\hline 1929 & (020,002,200) & 1 or 0 & -1 & $\Sigma(1929)$ or
$\Lambda(1929)$ &
\begin{tabular}{l}
$\frac{1}{2}$,$\frac{3}{2}$;\\
$\frac{5}{2}$,$\frac{7}{2}$ \\
$\frac{3}{2}$,$\frac{5}{2}$%
\end{tabular}
&
\begin{tabular}{l}
$\frac{1}{2}$,$\frac{3}{2}$,$\frac{5}{2}$ \\
$\frac{3}{2}$,$\frac{5}{2}$,$\frac{7}{2}$ \\
$\frac{1}{2}$,$\frac{3}{2}$,$\frac{5}{2}$,$\frac{7}{2}$%
\end{tabular}
&
\begin{tabular}{l}
+ \\
-%
\end{tabular} \\

\hline 2379 & (00-2,0-20,-200) & 1 or 0 & -1 & $\Sigma(2379)$ or
$\Lambda$(2379) &
\begin{tabular}{l}
$\frac{1}{2}$,$\frac{3}{2}$ \\
$\frac{5}{2}$,$\frac{7}{2}$ %
\end{tabular}
&
\begin{tabular}{l}
$\frac{1}{2}$,$\frac{3}{2}$,$\frac{5}{2}$; \\
$\frac{3}{2}$,$\frac{5}{2}$,$\frac{7}{2}$,$\frac{9}{2}$ %
\end{tabular} &
\begin{tabular}{l}
+ \\
+  %
\end{tabular} \\

\hline 2649 &
\begin{tabular}{l}
(12-1,1-12,21-1  \\
2-11,-121,-112)   %
\end{tabular} &  & -1 &   &   & & \\

\hline 2649 & (12-1,1-12,21-1) & 1 or 0 & -1 & $\Sigma(2649)$ or
$\Lambda(2649)$ &
\begin{tabular}{l}
$\frac{1}{2}$,$\frac{3}{2}$, \\
$\frac{5}{2}$,$\frac{7}{2}$;  \\
$\frac{3}{2}$,$\frac{5}{2}$, \\
$\frac{7}{2}$,$\frac{9}{2}$;  %
\end{tabular}&
\begin{tabular}{l}
$\frac{1}{2}$,$\frac{3}{2}$,$\frac{5}{2}$, \\
$\frac{3}{2}$,$\frac{5}{2}$,$\frac{7}{2}$,$\frac{9}{2}$; \\
$\frac{1}{2}$,$\frac{3}{2}$,$\frac{5}{2}$,$\frac{7}{2}$, \\
$\frac{3}{2}$,$\frac{5}{2}$,$\frac{7}{2}$,$\frac{9}{2}$. %
\end{tabular} &
\begin{tabular}{l}
+  \\
+  \\
-  \\
-  %
\end{tabular}\\

\hline 2649 & (2-11,-121,-112) & 1 or 0 & -1 & $\Sigma(2649)$ or
$\Lambda(2649)$ &
\begin{tabular}{l}
$\frac{1}{2}$,$\frac{3}{2}$, \\
$\frac{5}{2}$,$\frac{7}{2}$;  \\
$\frac{3}{2}$,$\frac{5}{2}$, \\
$\frac{7}{2}$,$\frac{9}{2}$;  %
\end{tabular}&
\begin{tabular}{l}
$\frac{1}{2}$,$\frac{3}{2}$,$\frac{5}{2}$, \\
$\frac{3}{2}$,$\frac{5}{2}$,$\frac{7}{2}$,$\frac{9}{2}$; \\
$\frac{1}{2}$,$\frac{3}{2}$,$\frac{5}{2}$,$\frac{7}{2}$, \\
$\frac{3}{2}$,$\frac{5}{2}$,$\frac{7}{2}$,$\frac{9}{2}$. %
\end{tabular} &
\begin{tabular}{l}
+  \\
+  \\
-  \\
-  %
\end{tabular}\\

\hline 2649 & (202,220,022) & 1 or 0 & -1 & $\Sigma(2649)$ or
$\Lambda(2649)$ &
\begin{tabular}{l}
$\frac{5}{2}$,$\frac{7}{2}$;  \\
$\frac{9}{2}$,$\frac{11}{2}$, \\
$\frac{7}{2}$,$\frac{9}{2}$;  %
\end{tabular}&
\begin{tabular}{l}
$\frac{3}{2}$,$\frac{5}{2}$,$\frac{7}{2}$,$\frac{9}{2}$, \\
$\frac{7}{2}$,$\frac{9}{2}$,$\frac{11}{2}$,$\frac{13}{2}$; \\
$\frac{5}{2}$,$\frac{7}{2}$,$\frac{9}{2}$,$\frac{11}{2}$. %
\end{tabular} &
\begin{tabular}{l}
+  \\
+  \\
-  %
\end{tabular} \\

\hline
\end{tabular}%
\end{tabular}%
\newline \\

\subsection{The baryons on the $\Sigma$-axis}

$\qquad$The $\Sigma$-axis is a two-fold rotary axis, R = 2,
$S_{\Sigma}$ = -2 from (\ref{Strange Number}). Thus all energy bands
of the $\Sigma$-axis have $S_{\Sigma}$ = -2 and $\mathcal{B}$ = 1
from Postulate II.
\begin{equation}
S_{\Sigma} = -2\,\,\, and\,\,\, \mathcal{B} = 1. \label{S=-2 and
B=1}
\end{equation}

For low energy levels, there are two-fold degenerate energy bands,
four-fold degenerate energy bands and single energy bands on the
axis (see Fig. 4). From (\ref{Subdeg}), each four degenerate energy
band on the symmetry $\Sigma$-axis will divide into two two-fold
degenerate bands. Thus we only need to discuss the two-fold
degenerate bands for four-fold and two-fold degenerate bands.

For the two-fold degenerate energy bands, I = $\frac{1}{2}$ from
(\ref{IsoSpin}); S = -2 from (\ref{Strange Number}); Q = 0, -1 from
(\ref{GMN}). Comparing the quantum numbers of the energy bands with
the quantum numbers of the baryons in Table 2 and the experimental
baryon spectrum in \cite{Baryon}, we find there are only the
$\Xi$-baryons having the same quantum numbers as the energy band.
According to Postulate III, the $\Xi$-baryons will excite from the
vacuum into the two-fold energy bands to form the observable
$\Xi(M)$-baryons. The masses M =$M_{Fig}$ +de = $M_{Fig}$ since de =
0 from Table 7 ($\Delta$S = 0).

Table B3 gives the possible J and parity P for each energy band and
Table 6B limits off extra J values for each deduced $\Xi$-baryon.
Fig. 4 gives the masses of the $\Xi(M)$. Thus we deduce the
$\Xi(M)$-baryons with correct $J^{P}$ values as appearing in Table
10: $\newline$\\

\begin{tabular}{l}

$\qquad$$\qquad$$\qquad$  Table 10.\ The baryons$\Xi(M)$ on the
$\Sigma-$axis \\
\begin{tabular}{|l|l|l|l|l|l|l|l|l|l|l|}
\hline $M_{Fig}$ & ($n_1,n_2,n_3$) & I & S & Baryon(M) &
J(1/2) & J(3/2) & P  \\
\hline 1479 & (101,10-1,011,01-1) &  & -2 &  &  & & \\
\hline 1479 & (1,0,1;0,1,1) & $\frac{1}{2}$ & -2 & $\Xi(1479)$ &
$\frac{1}{2}$ & $\frac{3}{2}$& -  \\
\hline 1479 & (1,0,-1;0,1,-1) & $\frac{1}{2}$ & -2 & $\Xi(1479)$ &
$\frac{1}{2}$$ \frac{3}{2}$ &
$\frac{1}{2}$,$\frac{3}{2}$,$\frac{5}{2}$ &  + \\
\hline 1659 & (1,-1,0),(-1,1,0) & $\frac{1}{2}$ & -2 & $\Xi(1659)$ &
$\frac{1}{2}$,$ \frac{3}{2}$ &
$\frac{1}{2}$,$ \frac{3}{2}$,$\frac{5}{2}$ & + \\
\hline 1659 & (-101,-10-1,0-11,0-1-1) &  & -2 &  &  &  & \\
\hline 1659 & (-101,0-11) & $\frac{1}{2}$ & -2 & $\Xi(1659)$ &
$\frac{3}{2}$,$ \frac{5}{2}$ &
$\frac{1}{2}$,$\frac{3}{2}$,$\frac{5}{2}$ & - \\
\hline 1659 & (-10-1,0-1-1) & $\frac{1}{2}$ & -2 & $\Xi(1659)$ &
$\frac{3}{2}$,$ \frac{5}{2}$ &
$\frac{1}{2}$,$\frac{3}{2}$,$\frac{5}{2}$ & - \\
\hline 1839 & (2, 0, 0);(0,2,0) & $\frac{1}{2}$ & -2 & $\Xi(1839)$ &
$\frac{3}{2}$,$\frac{5}{2}$ &
$\frac{1}{2}$,$\frac{3}{2}$,$\frac{5}{2}$$\frac{7}{2}$ & - \\
\hline 2199 & (121,12-1,211,21-1) & & -2 &  & & & \\
\hline 2199 & (1,2,1;2,1,1) & $\frac{1}{2}$ & -2 & $\Xi(2199)$ &
\begin{tabular}{l}
$\frac{3}{2}$,$\frac{5}{2}$\\
$\frac{7}{2}$,$\frac{9}{2}$ %
\end{tabular} &
\begin{tabular}{l}
$\frac{1}{2}$,$\frac{3}{2}$,$\frac{5}{2}$,$\frac{7}{2}$ \\
$\frac{5}{2}$,$\frac{7}{2}$,$\frac{9}{2}$ %
\end{tabular} & -  \\

\hline 2199 & (1,2,-1;2,1,-1) & $\frac{1}{2}$ & -2 & $\Xi(2199)$ &
\begin{tabular}{l}
$\frac{1}{2}$,$\frac{3}{2}$\\
$\frac{5}{2}$,$\frac{7}{2}$ %
\end{tabular} &
\begin{tabular}{l}
$\frac{1}{2}$,$\frac{3}{2}$,$\frac{5}{2}$ \\
$\frac{3}{2}$,$\frac{5}{2}$,$\frac{7}{2}$,$\frac{9}{2}$ %
\end{tabular} & +  \\

\hline 2379 & (0,0,2;0,0,-2) & $\frac{1}{2}$ & -2 & $\Xi(2379)$ &
\begin{tabular}{l}
$\frac{1}{2}$,$\frac{3}{2}$\\
$\frac{5}{2}$,$\frac{7}{2}$ %
\end{tabular} &
\begin{tabular}{l}
$\frac{1}{2}$,$\frac{3}{2}$,$\frac{5}{2}$ \\
$\frac{3}{2}$,$\frac{5}{2}$,$\frac{7}{2}$,$\frac{9}{2}$ %
\end{tabular}& +  \\

\hline 2379 & (-2,0,0;0,-2,0) & $\frac{1}{2}$ & -2 & $\Xi(2379)$ &
\begin{tabular}{l}
$\frac{3}{2}$,$\frac{5}{2}$ \\
$\frac{7}{2}$,$\frac{9}{2}$ %
\end{tabular} &
\begin{tabular}{l}
$\frac{1}{2}$,$\frac{3}{2}$,$\frac{5}{2}$,$\frac{7}{2}$ \\
$\frac{5}{2}$,$\frac{7}{2}$,$\frac{9}{2}$,$\frac{11}{2}$
\end{tabular} & -  \\

\hline 2559 & (112;11-2) & $\frac{1}{2}$ & -2 & $\Xi(2559)$ &
\begin{tabular}{l}
$\frac{3}{2}$,$\frac{5}{2}$; \\
$\frac{7}{2}$,$\frac{9}{2}$. \\
\end{tabular}
&
\begin{tabular}{l}
$\frac{1}{2}$,$\frac{3}{2}$,$\frac{5}{2}$,$\frac{7}{2}$. \\
$\frac{5}{2}$,$\frac{7}{2}$,$\frac{9}{2}$,$\frac{11}{2}$.%
\end{tabular}
& - \\

\hline
\end{tabular}%
\end{tabular} \newline \\

\subsection{The baryons of the single energy bands on the
$\Sigma $-axis}

$\qquad$The single energy bands of the $\Sigma$-axis appear in Fig.
8. For the single energy bands, d = 1, I = 0 from (\ref{IsoSpin});
for the $\Sigma$-axis, the strange number $S_{\Sigma}$ = -2 from
(\ref{Strange Number}) (see Fig.1) and $L_{axis}$ = 0 from
(\ref{Laxis}) and Fig. 8. All single energy bands of the
$\Sigma$-axis have
\begin{equation}
\mathcal{B}\,\, = \,\,1,\,\,I\,\,=\,\,0,\,\,S_{\Sigma} = -2,\,\,and
\,\,L\,\,= \,\,0.  \label{Single bands}
\end{equation}

In Fig. 8, at $M_{Fig}$ = 1119 Mev , the energy band with $\vec{n}$
= (1, 1, 0) is the $\Lambda(1119)$ from (\ref{Lambda(1119)})
\begin{equation}
\Lambda(1119)\,\,\,with\,\, I = 0,\,\, S = -1,\,\, Q = 0\,\, and\,\,
M = 1119 .
\end{equation}
It is very important to pay attention to \textbf{the strange baryon
$\Lambda(1119)$\,$ \frac{1}{2}^{+}$ born on the single energy band
of the $\Sigma $-axis from $ \Delta $S = +1.}

At $M_{Fig}$ = 1659 Mev in Fig. 8, the band with $\vec{n}$ = (-1,
-1, 0) has $\Delta $S = -1 from (\ref{DS}). Thus S = $S_{\Sigma}$
+$\Delta $S = -3, I = 0, Q = -1 from (\ref{GMN}). Comparing the
quantum numbers of the energy band with the quantum numbers of the
baryons in Table 3 and in \cite{Baryon}, we find there is only the
$\Omega$-baryon having the same quantum numbers as the energy band.
According to Postulate III, the $\Omega$-baryon will excite from the
vacuum into the energy band to form a $\Omega$(1659)-baryon. From
Table 3, its spin s = $\frac{3}{2}$; $L_{axis}$ = 0 from Fig. 8 and
(\ref{Laxis}). Thus $J^{P}$ = $\frac{3}{2}^{+}$ from (\ref{Ground
P}) and (\ref{J}).

At $M_{Fig}$ = 2559 Mev, the band with $\vec{n}$ = (2, 2, 0) has
$\Delta S = +1$ from (\ref{DS}), S = -2 + 1 = -1 and Q = 0.
Similarly to $M_{Fig}$ = 1119 Mev, there is a
$\Lambda(2559)$-baryon. $J^{P}$ = $\frac{1}{2}^{+}$, from Table 2
(spin s = 1/2),(\ref{Laxis}) ($L_{axis}$ = 0), (\ref{J}) and
(\ref{Ground P}).

Similarly to $M_{Fig}$ = 1659 Mev, at $M_{Fig}$ = 3819, the band
with ($\vec{n}$ = (-2, -2, 0)) has $ \mathcal{B}$ = 1, S = -3, I = 0
and Q = -1. It is a $\Omega(3819)$-baryon with $J^{P}$ =
$\frac{3}{2} ^{+}$.

At $M_{Fig}$ = 5439 (see Fig.8), the band with $\vec{n}$ = (3, 3, 0)
has $\mathcal{B}$ = 1, I = 0, S = -2 + 1 = -1 from (\ref{S+DS}) and
(\ref{DS}) and Q = 0 from (\ref{GMN}). From (\ref{Bottom}), we know
that the energy band of the b-particle has a bottom number B = -1.
Comparing the quantum numbers of the band with the quantum numbers
of the baryons in \cite{Baryon}, we find there is only the
$\Lambda_b$-baryon having the same quantum numbers as the energy
band. Thus the $\Lambda_b$-baryon will be excited from the vacuum
into the energy band to form a $\Lambda_b(5554)$-baryon since M =
$M_{Fig}$ + de = 5439 + 115 = 5554 Mev from Table 7. Its $J^{P}$ =
$\frac{1 }{2}^{+}$, from (\ref{s = 1/2}), (\ref{Laxis})($L_{axis}$ =
0), (\ref{J}) and (\ref{Ground P}). Using Fig. 8, we have the baryon
as appearing in Table 11:$\qquad $\newline

\begin{tabular}{l}
\ \ \ \ \ \  Table 11.\ The baryons of the single bands on
the $\Sigma$-axis \\
\begin{tabular}{|l|l|l|l|l|l|l|l|l|l|l|l|l|}
\hline $M_{fig}$ & $\vec{n}$ & $S_{axis}$ & $\Delta$S & S & B &
Baryon(M) & L & J & P\\
\hline 1119 & (1, 1, 0) & -2 & +1 & -1 & 0 &  $\Lambda$(1119) &
0 & $\frac{1}{2}$ & +  \\
\hline 1659 & (-1, -1, 0) & -2 & -1 & -3 & 0 &  $\Omega$(1659) &
0 & $\frac{3}{2} $ & +  \\
\hline 2559 & (2, 2, 0) & -2 & +1 & -1 &
0 &  $\Lambda$(2559) & 0 & $\frac{1}{2}$ & + \\
\hline 3819 & (-2,-2, 0) & -2 & -1 & -3 & 0 & $\Omega$(3819) & 0 &
$\frac{3}{2}$ & + \\
\hline 5439 & (3, 3, 0) & -2 & +1 & 0 &
-1 & $\Lambda_b$(5554) & 0 & $\frac{1 }{2}$ & + \\
\hline
\end{tabular}%
\end{tabular}%
\newline $\qquad $ \\
It is very important to pay attention to \textbf{the bottom baryon
$\Lambda_b(5554)$ born on the single energy band with $\vec{n}$ =
(3, 3, 0) and $M_{\Sigma}$ = 5439 Mev from the $\Delta S\,=\,+1$}
and de = 115.

\subsection{Three axes on the surfaces of the first Brillouin zone}
$\qquad$For the three symmetry axes that are on the surface of the
first Brillouin zone (the D-axis (P-N), the F-axis (P-H) and the
G-axis (M-N), see Fig.1), the energy bands with the same energy
might have asymmetric $\vec{n}$ values (see Fig. 5, Fig. 6 and Fig.
7). For symmetrical $\vec{n}$ values we give a definition: for two
$\vec{n}$ values [such as $\vec{n}$ = ($n_1$, $n_2$, $n_3$) and
$\vec{n^{\prime}}$ = ($n^{\prime}_1$, $n^{\prime}_2$,
$n^{\prime}_3$)], if using the permutation of the three components
and changing the signs of the components (one, two, or three), they
can be interchanged each other, then we call they are symmetrical
$\vec{n}$ values. For example, the following eight $\vec{n}$ values
are all symmetric: (1,1,0), (1,-1,0), (-1,1,0), (-1,-1,0),(1,0,-1),
(-1,0,-1), (0,1,-1), (0,-1,-1). While (1,-1,1) and (1,1,0) are
asymmetric. (0,0,2) and (1,1,2) are asymmetric also.

For an energy band with asymmetric $\vec{n}$ values, for example
(1-10,-110,020,200), we cannot use (\ref{IsoSpin}) to find I =
$\frac{3}{2}$. We have to divide it into two symmetric values
(1-10,-110) and (020,200) first. Then, using (\ref{IsoSpin}), we can
get two sub-bands with I = $\frac{1}{2}$.

For this division ($\gamma = \frac{d}{R}$), if the R-fold
``sub-degeneracy" bands have R-fold symmetry $\vec{n}$ values, the
strange number of the sub-energy band will not change ($\Delta$S =0,
$\to$ $S_{sub}$ = $S_{ax}$) since the sub-band has the same R-fold
symmetries as the axis has.
\begin{equation}
If\,\, a\,\,sub-band\,\, has\,\,R-fold\,\, symmetry\,\,
\vec{n}\,\,values\,\,\,\, ,\Delta S = 0. \label{DS=0}
\end{equation}

If the R-fold ``sub-degeneracy" bands have asymmetric $\vec{n}$
values, however, the strange number of the sub-energy band will
change ($S_{sub} \neq$ $S_{ax}$) since the sub-band has not the
R-fold symmetries.  $\Delta$S can be found using the below formula
(\ref{DSf}).

Sometimes, after dividing, although the $\vec{n}$ $\to$
Sign($n_{1}+n_{2}+n_{3}$) = 0 $\to$ $\Delta$S = 0 from (\ref{DS}),
this is not correct. For example, on the F-axis ($S_F$ = -1), the
energy band $M_{Fig}$ = 1929 with $\vec{n}$ = (0,0,2;-1,0,1;0,-1,1).
This three-fold band with three asymmetric $\vec{n}$ values divide
into two bands with (2,0,0) and (-1,0,1;0,-1,1). The energy band
with two symmetric $\vec{n}$ values (-1,0,1;0,-1,1) $\to$
Sign($n_{1}+n_{2}+n_{3}$) = 0 $\to$ $\Delta$S = 0 from (\ref{DS}).
This result is not correct. The F-axis $S_F$= -1, the two-fold
degeneracy band I = 1/2 from (\ref{IsoSpin}), Q = $\frac{1}{2}$,
-$\frac{1}{2}$ from (\ref{GMN}). Since this division has broken the
three-fold symmetry of the F-axis, the strange number $S_F$= -1 has
to change. We have to change Sign($n_{1}+n_{2}+n_{3}$) into
Sign($n_{1}+n_{2}+n_{3}$+f). When the sub-band with d $\neq$ R and
R-d $\neq$ 2 as well as Sign($n_{1}+n_{2}+n_{3}$) = 0, the formula
(\ref{DS}) is changed into
\begin{equation}
\Delta S = [1 - 2\delta({S_{ax}})] Sing(n_{1}+n_{2}+n_{3}+f)\time 1,
\label{DSf}
\end{equation}
where f = 1 for the surface axes (the D-axis, the F-axis and the
G-axis), and f = 0 for inside three axes (the $\Delta$-axis, the
$\Lambda$-axis and the $\Sigma$ -axis) as appearing in Table 7.

Generally, if an energy band has d asymmetric $\vec{n}$ values, the
energy band will divide into subgroups of energy bands with $d_1$,
$d_2$ $\ldots$ symmetric (or single) $\vec{n}$ values. For each
subgroup of energy bands, using (\ref{S+DS}) and (\ref{DS}) as well
as (\ref{DSf}), we can find the strange numbers of the sub-bands.
Finally we recognize the baryons using the quantum numbers (I, S and
Q). Generally, the two (or three or four) sub-bands have different S
(or C or B) $\to$ different baryons. If the two (or three or four)
sub-energy bands with $d_1$ and $d_2$ (or $d_3$ or $d_4$) get the
same S (or C or B), they might combine back to the energy band with
d = $d_1$ + $d_2$:
\begin{equation}
d = d_1 + d_2 + \ldots. \label{Combine sub-bands}
\end{equation}

\subsection{The baryons on the D-axis(P-N)}

$\qquad$The energy bands of the D-axis appear in Fig. 5. The strange
number of the energy bands on the D-axis has $S_{D}$ = 0 from
(\ref{S-D}). All energy bands have
\begin{equation}
\mathcal{B}\,\, = \,\,1,\,\,\,and\,\,\,S_{D} = 0. \label{B=1 and
S=0}
\end{equation}

For low energy levels, there are four-fold energy bands and two-fold
energy bands on the axis (see Fig. 5). In Fig. 5, the lowest energy
band has two fold asymmetric $\vec{n}$ = (000, 110). The two-fold
band will divide into two single bands. The band with $\vec{n}$ =
(0, 0, 0) is N(939)$\frac{1}{2}^+$ from (\ref{N(939)}). The band
with $\vec{n}$ = (1,1,0) is the $\Lambda(1119)$$\frac{1}{2}^+$  from
(\ref{1119}). In Fig. 5, the second lowest band with two fold
$\vec{n}$ (011,101) and $M_{Fig}$ = 1209, but uncertain L, J and P
from Table B4. It cannot form a two-fold N(1209)-baryon. They might
be regarded as $\Delta(1232)^{+}$ and $\Delta(1232)^{0}$.

From Fig. 5, we can see that each four-fold energy band has 4
asymmetry $\vec{n}$ values. They all can be divided into two groups.
Each of them has two symmetric $\vec{n}$ values. At the same time,
there are many two-fold energy bands with symmetry $\vec{n}$ values.
For these two-fold bands with symmetry $\vec{n}$ values, I =
$\frac{1}{2}$ from (\ref{IsoSpin}), I$_{Z}$ = 1/2, -1/2; Q = 1, 0
from (\ref{GMN}) and $S_D$ = 0. Comparing the quantum numbers (I, S
and Q) of the two-fold energy bands with the quantum numbers of the
baryons in Table 2 and in \cite{Baryon}, we find only the N-baryons
with completely the same quantum numbers as the two-fold energy
bands. According to Postulate III, the N-baryons will excite from
the vacuum into the two-fold energy bands to form the
N($M_{Fig}$)-baryon. Fig. 5 gives the mass M = M$_{Fig}$ of the
N-baryons as appearing in Table 12.

There is a four-fold energy band with $M_{Fig}$ = 1839 and $\vec{n}$
= (1-10,-110,020,200) as well as the lowest energy at the point N.
Since the four-fold $\vec(n)$ values are asymmetrical, it divide
into two two-fold sub-bands with symmetrical $\vec{n}$ = (1-10,-110)
and (020,200). Since the two sub-bands have d = R and two
symmetrical $\vec{n}$ values, the $\Delta$S =0 from the formula
(\ref{DS=0}). The two sub-bands have the same masses M = 1839 and
same S = 0, from (\ref{Combine sub-bands}), they should combine back
to form a four-fold $\Delta$-baryon. There, however, is not any
four-fold single and double representation \cite{Double} in the
N-group Table A3. Thus the two sub-bands only form two N(1839) as
appearing in Table 12.

There is another four-fold energy band with $M_{Fig}$ = 1929 and
$\vec{n}$ = (-101,0-11,211,121) as well as the lowest energy at the
point P (a special case). The band with four asymmetry $\vec{n}$
should divide into two bands with two-fold symmetry n-values, the
two two-fold bands still have the same strange number S = 0 since
the two sub-bands have d = R $\Delta$S = 0 from (\ref{DS=0}). From
(\ref{Combine sub-bands}), the two two-fold bands combine back to
form a four-fold band with S = 0, M =1929, I = $\frac{3}{2}$ and Q =
+2, +1, 0, -1 since there are four-fold double representations in
the P-group \cite{Double}. This is a $\Delta$(1929)-baryon.  At the
same time, there is a N(1929) also from (\ref{I-2}). The $J^{P}$ =
$\frac{1}{2}^-$,2$\frac{3}{2}^-$, 2$\frac{5}{2}^-$,$\frac{7}{2}^-$
and 2$\frac{1}{2}^+$, 3$\frac{3}{2}^+$, 3$\frac{5}{2}^+$ and
2$\frac{7}{2}^+$ as appearing in Table 12.

At $E_P$ =1929, there is a two-fold band with asymmetry (002,112).
It will divide into two single bands with (0,0,2) and (1,1,2). From
(\ref{DSf}), the band with (0,0,2) and $M_{fig}$ = 1929 has S
=$S_{D}$ + $\Delta$S = 0 + (-1) = -1, I = 0 from (\ref{IsoSpin}) and
Q = 0 from (\ref{GMN}). Similarly to the band with (1,1,2) and
$M_{fig}$ = 1929, the band with (2,2,0) and $M_{fig}$ = 2559 and the
band with (1,1,-2) and $M_{fig}$ = 2559, they all have S = -1, I = 0
and Q = 0. Comparing the quantum numbers of these energy bands with
the quantum numbers of the baryons in Table 2 and in \cite{Baryon},
we find there are only the $\Lambda$-baryons having completely the
same quantum numbers as these energy bands. According to Postulate
III, the $\Lambda$-baryons will excite into these energy bands to
form observable $\Lambda$-baryons. There is a adjust energy de = -
115 S$\times$ $\Delta$S = -115 (Mev) for each $\Lambda$-baryon from
Table 7.  The masses of the baryons are M = $M_{Fig}$ + de =
$M_{Fig}$ - 115. So that the baryons are $\Lambda(1814)$--the band
with (002), $\Lambda(1814)$--the band with (112),
$\Lambda(2444)$--the band with (220), $\Lambda(2444)$--the band with
(11-2) as appearing in Table 12.

At $E_N$ = 2559, there are two two-folds bands with two asymmetric
(220,-1-10) and (11-2,00-2). The band with (2,2,0,-1,-1,0) divides
to two bands with (2,2,0) and (-1,-1,0). The band with (2,2,0) has I
= 0 from (\ref{IsoSpin}), $\Delta$S = -1 from (\ref{DS}), Q = 0 from
(\ref{GMN}) and de = -115 from Table 7. The band $\to$
$\Lambda(2444)$ as appears above. The band with (-1,-1,0), $\Delta$S
= +1 from (\ref{DS}). Thus the band has I = 0 from (\ref{IsoSpin}),
C = + 1 from (\ref{Charmed}) and Q = +1 from (\ref{GMN}). Comparing
the quantum numbers of the band with the quantum numbers of the
baryons in Table 2 and in \cite{Baryon}, we find there are the
$\Lambda_c$-baryons having completely the same quantum numbers as
the band. According to Postulate III, the $\Lambda_c$-baryons will
excite from the vacuum into the band to form three $\Lambda(M)_c$s.
The mass M = M$_{Fig}$ + de = 2559+0 (2559); 2559+115 (2674); 2559+
230 (2789) from Table 7. Similarly, another two-fold band gives out:
the band with (1,1,-2) $\to$ $\Lambda(2444)$ from(\ref{DSf}) and
Table 7 as appearing in above and the band with (0,0,-2) $\to$
$\Delta$S = +1 from (\ref{DS}) and $\Lambda_c(M)$ from (\ref{DS}).
From Table 7 de = 345; 460; 575 since $n_c$ = 3, 4, 5 for the second
$\Lambda_c$-baryon as appearing in Table 12.
\newline

\begin{tabular}{l}
$\qquad$$\qquad$$\qquad$Table 12.\ The Baryons of
the Energy Bands on the D-axis (P-N) \\
\begin{tabular}{|l|l|l|l|l|l|l|l|l|l|}
\hline $M_{Fig}$ & ($n_1,n_2,n_3$) & I & $\Delta$S & S & C & de &
Baryon(M) & J & P \\
\hline 1479 & (10-1,01-1) & $\frac{1}{2}$ & 0 & 0 & 0 & 0 & N(1479)
& $\frac{3}{2}$,$\frac{5}{2}$; $\frac{2}{2}$,$\frac{3}{2}$,
$\frac{5}{2}$ & - \\
\hline 1839 &
\begin{tabular}{l}
(1-10,-110, \\
020,200)%
\end{tabular}
&  & 0 &  &  &  &  & &  \\
\hline 1839 & (1-10,-110) & $\frac{1}{2}$ & 0 & 0 & 0 & 0 & N(1839)
& $\frac{3}{2}$,$\frac{5}{2}$;$\frac{1}{2}$,$\frac{3}{2}$,
$\frac{5}{2}$,$\frac{7}{2}$ & - \\
\hline 1839 & (020,200) & $\frac{1}{2}$ & 0 & 0 & 0 & 0 & N(1839) &
$\frac{1}{2}$,$\frac{3}{2}$;
$\frac{1}{2}$,$\frac{3}{2}$,$\frac{5}{2}$ & + \\
\hline 1929 &
\begin{tabular}{l}
(-101,0-11, \\
211,121)%
\end{tabular}
&$\frac{3}{2}^*$ & 0 & 0  & 0 & 0 &
\begin{tabular}{l}
$\Delta{1929)}$ \\
N(1929)         %
\end{tabular}&
\begin{tabular}{l}
$\frac{1}{2}$,$\frac{3}{2}$;
$\frac{1}{2}$,$\frac{3}{2}$,$\frac{5}{2}$ \\
$\frac{5}{2}$,$\frac{7}{2}$;
$\frac{3}{2}$,$\frac{5}{2}$,$\frac{7}{2}$ \\
$\frac{3}{2}$,$\frac{5}{2}$;
$\frac{1}{2}$,$\frac{3}{2}$,$\frac{5}{2}$,$\frac{7}{2}$ %
\end{tabular} &
\begin{tabular}{l}
+  \\
+  \\
-  %
\end{tabular} \\
\hline 1929 & (002,112) & &  &  &  &  &  & &\\
\hline 1929 & (0,0,2) & 0 & -1 & -1 & 0 & -115 & $\Lambda$(1814) &
$\frac{3}{2}$,$\frac{5}{2}$;
$\frac{1}{2}$,$\frac{3}{2}$,$\frac{5}{2}$& - \\

\hline 1929 & (1,1,2) & 0 & -1 & -1 & 0 & -115 & $\Lambda$(1814) &
\begin{tabular}{l}
$\frac{1}{2}$,$\frac{3}{2}$;
$\frac{1}{2}$,$\frac{3}{2}$,$\frac{5}{2}$ \\
$\frac{3}{2}$,$\frac{5}{2}$;
$\frac{1}{2}$,$\frac{3}{2}$,$\frac{5}{2}$ %
\end{tabular} &
\begin{tabular}{l}
+  \\
-   %
\end{tabular} \\

\hline 2199 & (12-1,21-1) & $\frac{1}{2}$ & 0 & 0 & 0 & 0 & %
N(2199)&
\begin{tabular}{l}
$\frac{3}{2}$,$\frac{5}{2}$;
$\frac{1}{2}$,$\frac{3}{2}$,$\frac{5}{2}$,$\frac{7}{2}$.\\
$\frac{7}{2}$,$\frac{9}{2}$;
$\frac{5}{2}$,$\frac{7}{2}$,$\frac{9}{2}$.%
\end{tabular} & -  \\

\hline 2199 & (-10-1,0-1-1) & $\frac{1}{2}$ & 0 & 0 & 0 & 0 &
N(2199) & \begin{tabular}{l} $\frac{1}{2}$,$\frac{3}{2}$;
$\frac{1}{2}$,$\frac{3}{2}$,$\frac{5}{2}$.\\
$\frac{5}{2}$,$\frac{7}{2}$;
$\frac{3}{2}$,$\frac{5}{2}$,$\frac{7}{2}$,$\frac{9}{2}$.%
\end{tabular} & + \\

\hline 2559 & (220,-1-10) &  &  &  &  &  & & & \\
\hline 2559 & (2,2,0) & 0 & -1 & -1 & 0 & -115 & $\Lambda(2444)$ &
\begin{tabular}{l}
$\frac{3}{2}$,$\frac{5}{2}$;
$\frac{1}{2}$,$\frac{3}{2}$,$\frac{5}{2}$. \\
$\frac{7}{2}$,$\frac{9}{2}$;
$\frac{5}{2}$,$\frac{7}{2}$,$\frac{9}{2}$,$\frac{11}{2}$. %
\end{tabular} & - \\

\hline 2559 & (-1,-1,0) & 0 & +1 & 0 & +1 & 0 & $\Lambda_c(2559)$ &
$\frac{1}{2}$ & -  \\
\hline 2559 & (-1,-1,0) & 0 & +1 & 0 & +1 & 115 & $\Lambda_c(2674)$
& $\frac{3}{2}$ & -  \\
\hline 2559 & (-1,-1,0) & 0 & +1 & 0 & +1 & 230 & $\Lambda_c(2789)$
& $\frac{5}{2}$,$\frac{7}{2}$ & -  \\

\hline 2559 & (11-2,00-2) &  &  &  &  &  &  & & \\
\hline 2559 & (1,1,-2) & 0 & -1 & -1 & 0 & -115 & $\Lambda(2444)$ &
\begin{tabular}{l}
$\frac{1}{2}$,$\frac{3}{2}$;
$\frac{1}{2}$,$\frac{3}{2}$,$\frac{5}{2}$.\\
$\frac{5}{2}$,$\frac{7}{2}$;
$\frac{3}{2}$,$\frac{5}{2}$,$\frac{7}{2}$,$\frac{9}{2}$.%
\end{tabular} & +  \\
\hline 2559 & (0,0,-2) & 0 & +1 & 0 & +1 & 345 & $\Lambda_c(2904)$ &
2$\frac{1}{2}$ & +  \\
\hline 2559 & (0,0,-2) & 0 & +1 & 0 & +1 & 460 & $\Lambda_c(3019)$ &
2$\frac{3}{2}$ & +\\
\hline 2559 & (0,0,-2) & 0 & +1 & 0 & +1 & 575 & $\Lambda_c(3134)$ &
$\frac{5}{2}$ & +  \\

\hline 2649 & (-121,2-11) & $\frac{1}{2}$ & 0 & 0 & 0 & 0 & N(2649)&
\begin{tabular}{l}
$\frac{1}{2}$,$\frac{3}{2}$;
$\frac{1}{2}$,$\frac{3}{2}$,$\frac{5}{2}$.\\
$\frac{5}{2}$,$\frac{7}{2}$;
$\frac{3}{2}$,$\frac{5}{2}$,$\frac{7}{2}$,$\frac{9}{2}$.\\
$\frac{3}{2}$,$\frac{5}{2}$;
$\frac{1}{2}$,$\frac{3}{2}$,$\frac{5}{2}$,$\frac{7}{2}$.\\
\end{tabular}
&
\begin{tabular}{l}
+ \\
+ \\
- %
\end{tabular}\\

\hline 2649 &
\begin{tabular}{l}
(-112,1-12, \\
202,022)
\end{tabular}&
\begin{tabular}{l}
$\frac{3}{2}$  \\
$\frac{1}{2}$   %
\end{tabular} & 0 & 0 & 0 &  0 &
\begin{tabular}{l}
$\Delta$(2649) \\
N(2649)         %
\end{tabular} &
\begin{tabular}{l}
$\frac{5}{2}$,$\frac{7}{2}$;
$\frac{3}{2}$,$\frac{5}{2}$,$\frac{7}{2}$,$\frac{9}{2}$.\\
$\frac{9}{2}$,$\frac{11}{2}$;
$\frac{7}{2}$,$\frac{9}{2}$,$\frac{11}{2}$,$\frac{13}{2}$.\\
$\frac{7}{2}$,$\frac{9}{2}$;
$\frac{5}{2}$,$\frac{7}{2}$,$\frac{9}{2}$,$\frac{11}{2}$.\
\end{tabular} &
\begin{tabular}{l}
+ \\
+  \\
-  %
\end{tabular} \\
\hline 2919 & (2-1-1,-12-1) & $\frac{1}{2}$ & 0 & 0 & 0 & 0 &
N(2919) &
\begin{tabular}{l}
$\frac{3}{2}$,$\frac{5}{2}$;
$\frac{1}{2}$,$\frac{3}{2}$,$\frac{5}{2}$,$\frac{7}{2}$.\\
$\frac{7}{2}$,$\frac{9}{2}$;
$\frac{5}{2}$,$\frac{7}{2}$,$\frac{9}{2}$,$\frac{11}{2}$.\\
$\frac{11}{2}$,$\frac{13}{2}$;
$\frac{9}{2}$,$\frac{11}{2}$,$\frac{13}{2}$,$\frac{15}{2}$.%
\end{tabular} & - \\
\hline
\end{tabular}%
\end{tabular}
\newline
$\qquad $ $\qquad $ $\qquad $ $\qquad $ $\qquad $ $\qquad $ $\qquad
$ $\qquad $
\newline\\

At $M_f$ = 2649, a band with $\vec{n}$ = (-121,2-11) has I =
$\frac{1}{2}$ from (\ref{IsoSpin}), S = 0 from (\ref{B=1 and S=0}),
Q = 1, 0 from (\ref{GMN}). Similar to above, it is a N(2649).

At $M_f$ = 2649, a band with $\vec{n}$ = (-112,1-12,202,022) is
similar to the four-fold band with $M_{Fig}$ = 1929 and $\vec{n}$ =
(-101,0-11,211,121). The four-fold band forms a $\Delta(2649)$ and
N(2649) from (\ref{I-2}) as appearing in Table 12.

Table B4 gives the possible $J^P$ values of each energy band, Table
6B limits extra $J^P$ values off. The deduced baryons with correct
$J^P$ values appear in Table 12.

\subsection{The baryons on the F-Axis}

$\qquad$The F-axis is a three-fold symmetric axis, from (\ref{S-F}),
$S_{F}$ = -1. All energy bands of the F-axis have
\begin{equation}
\mathcal{B}\,\, = \,\,1,\,\,\,and\,\,\,S_{F} = - 1. \label{B=1 and
S=-1}
\end{equation}
There are six-fold energy bands, three-fold energy bands and single
energy bands on the axis (see Fig. 6).

For the single bands, $S_{F}$ = -1, the isospin I = 0 from
(\ref{IsoSpin}) and Q = 0 from (\ref{GMN}). Comparing the quantum
numbers of the energy bands with the quantum numbers of the baryons
in Table 2 and in \cite{Baryon}, we find there is only the
$\Lambda$-baryon having the same quantum numbers as the single
bands. According to Postulate III, the $\Lambda$-baryons will excite
from the vacuum into the single bands to form the $\Lambda$-baryons.
The single band with $\vec{n}$ = (1,1,0) forms the strange
$\Lambda(1119)$-baryon from (\ref{Lambda(1119)}). The single band
with $\vec{n}$ = (-1,-1,2) forms a strange baryon $\Lambda(2019)$.

In Fig. 6, the three-fold energy bands all have the three asymmetry
$\vec{n}$ values. Each of the three-fold asymmetry $\vec{n}$ values
composes of a 2-fold symmetry $\vec{n}$ values and a single
$\vec{n}$ value. The single energy bands, similarly to the above
paragraph, are the $\Lambda(M)$-baryons with S = -1, I = 0 and Q = 0
as well as M = $M_{Fig}$ as appearing in Table 13. Using (\ref{DS})
or (\ref{DSf}), if the two-fold symmetry bands have $\Delta$S = +1,
the two-fold energy bands form N-baryons; otherwise, if the two-fold
symmetry bands have $\Delta$S = -1, the two-fold energy bands form
$\Xi$-baryons as appearing in Table 13.

The six-fold bands all have six asymmetry $\vec{n}$-values. Each of
them composes of three two-fold symmetry $\vec{n}$-values. There are
always two ($n_1 + n_2 + n_3$) values larger than 0 and two ($n_1 +
n_2 + n_3$) values smaller than 0. It will divide into two three
fold bands since the rotary fold R = 3 of the F-axis and
(\ref{Subdeg}). The one contains the two ($n_1 + n_2 + n_3$) larger
than 0, while the another contains the two ($n_1 + n_2 + n_3$)
smaller or equal 0.

For example, at $E_P$ = 2649, a six fold-band with $\vec{n}$
=(202,022,-121,2-11,0-1-1,-10-1) will divided into two three-fold
sub-bands from (\ref{Subdeg}). The first sub-band with $\vec{n}$ =
(202,022,-121) has two n-values larger than zero; the second
sub-band with $\vec{n}$ = (2-11,0-1-1,-10-1) has two n-values
smaller than zero.

For first division, using (\ref{DS}), the first sub-band with an
asymmetry $\vec{n}$ = (202,022,-121) will get a strange number
$\Delta$S = +1 for each of the three values (202,022,-121). Then
they become a three-fold band with S = $S_{F}$ + $\Delta$S= -1 + 1 =
0. Since the three $\vec{n}$ values are asymmetric, the band will
divide again. For the second time division, using (\ref{DS}), we get
$\Delta$S = +1 = C from (\ref{Charmed}) for each band of the three
bands. Although the three bands with asymmetry $\vec{n}$ values have
been divided into a single band and a two-fold band, the single band
and the two-fold band have the same quantum numbers S and C as well
as the same mass. From (\ref{Combine sub-bands}), the single band
and the two-fold bands will combine back to form a three-fold band
with S = 0, C = +1 and I = 1 from (\ref{IsoSpin}) as well as Q = +2,
+1, 0 from (\ref{GMN}). Comparing the quantum numbers of the
three-fold band with the quantum numbers of the baryons in Table 2
and in \cite{Baryon}, we find there is only the $\Sigma_c$-baryon
with the same quantum numbers as the three-fold band. According to
Postulate III the $\Sigma_c$-baryons will excite from the vacuum
into the band to form $\Sigma_c(M)$-baryons. The M = $M_{fig}$ + de
= 2649 -115 ($n_c$ = 0, 2534); 2649 +0 ($n_c$ = 1, 2649); 2649 + 115
($n_c$ = 2, 2764) from Table 7 appear in Table 13.

For the second sub-bands with asymmetry $\vec{n}$ =
(2-11,0-1-1,-10-1), using (\ref{DS}), it will divide into a single
band and a two-fold band with different $\Delta$S from (\ref{DSf}).
Thus the single band and the two-fold bands cannot combine back into
a three-fold band with the same S, C and mass as the first
sub-bands. The second sub-bands will divide into two bands: a
two-fold band with symmetry $\vec{n}$ values (0,-1,-1;-1,0,-1) and a
single band with (2,-1,1). Using (\ref{DS}) for the two-fold band,
we get $\Delta$S = -1. Since the $S_{F}$ = -1, the strange number of
the band S = $S_{F}$ + $\Delta$S = -1 + (-1) = -2. The band has S =
-2, I = $\frac{1}{2}$ from (\ref{IsoSpin}), Q = 0 and -1 from
(\ref{GMN}). Comparing the quantum numbers of the band with the
quantum numbers of the baryons in Table 2 and in \cite{Baryon}, we
find there is only the $\Xi$-baryon having the same quantum numbers
as the band. According to Postulate III, the $\Xi$ baryon will
excite from the vacuum into the energy band to form a $\Xi(M)$
baryon. The mass of the baryon M = $M_{Fig}$ + de = $M_{Fig}$ since
de = 0 from Table 7. The single band with (2, -1, 0) and d =1, R - d
= 3-1 = 2. From (\ref{R-d=2}), the strange number S = $S_{F}$ = -1
and Q = 0 from (\ref{GMN}) as well as I = 0 from (\ref{IsoSpin}).
The band with I = 0, S = -1 and Q = 0 leads to a
$\Lambda(2649)$-baryon as appearing in Table 13.

At $M_{Fig}$ = 2649, the three-fold band with $\vec{n}$ =
(21-1,12-1,220) will divide into a two-fold symmetry band with
$\vec{n}$ = (21-1,12-1) and a single band with $\vec{n}$ = (2, 2,
0). The single band, similarly to the above single bands, is a
$\Lambda(2649)$. The two-fold band has I = $\frac{1}{2}$ from
(\ref{IsoSpin}), S = $S_{F}$ = -1, $\Delta$S = +1 = C from
(\ref{DS}) and (\ref{Charmed}) and Q = +1, 0 from (\ref{GMN}).
Comparing the quantum numbers of the band with the quantum numbers
of the baryons in Table 2 and in \cite{Baryon}, we find there is
only a $\Xi_c$-baryon with the same quantum numbers as the band.
According to Postulate III, the $\Xi_c$-baryon will excite from the
vacuum into the band to form $\Xi_c(M)$-baryons. The M = $M_{Fig}$
+de = 2649 - 115 ($n_c$ = 0, 2534); 2649 + 0 ($n_c$ = 1, 2649); 2649
+ 115 ($n_c$ = 2, 2764) from (\ref{Total Mass}) and Table 7 as
appearing in Table 13. Since there are $\Xi_c$ and $\Xi'_c$ in Table
2, there are two $\Xi_c(2534)$.

There is second six-fold energy bands with $M_{Fig}$ = 2739 and
$\vec{n}$ = (1-21,013,103;0-20,-200,-211). From (\ref{Subdeg}), the
six fold band divides into two three sub-bands with $\vec{n}$ =
(1-21,013,103) and (0-20,-200,-211) respectively. From (\ref{DS}),
the first sub-band with (1-21,013,103) gives a sub-sub-bands with
$\vec{n}$ = (013,103) and $\Delta$S = +1 = C from (\ref{Charmed})
and another sub-sub-band with (1,-2,1) and $\Delta$S = 0.  The
sub-sub-band with (013,103) has I = $\frac{1}{2}$ from
(\ref{IsoSpin}), S = $S_{F}$ = -1 , C =+1 from (\ref{Charmed}) and Q
= +1, 0. These are $\Xi_c$(M) baryons with M = $M_f$ + de = 2739-115
(2624 $n_c$ = 0); 2739 + 0 (2739 $n_c$ = 1); 2739+115 (2875 $n_c$ =
2) from Table 7. The second sub-band with $\vec{n}$ =
(0-20,-200,-211) divide to a sub-sub-band with $\vec{n}$ =
(0-20,-200) and a sub-sub-band with $\vec{n}$ = (-2, 1, 1). The
sub-sub-band with $\vec{n}$ = (-2, 1, 1) has I = 0 from
(\ref{IsoSpin}), S = $S_F$ = -1 from (\ref{R-d=2}) and Q = 0 from
(\ref{GMN}). It is a $\Lambda(2739)$. The sub-sub-band with
$\vec{n}$ = (0-20,-200) has I = $\frac{1}{2}$ from (\ref{IsoSpin}),
S = -2 from (\ref{DS}) and Q = 0, -1 from (\ref{GMN}). This is a
$\Xi(2739)$.

\begin{tabular}{l}
$\qquad$$\qquad$$\qquad$$\qquad$$\qquad$ Table 13a .\ The
baryons on the F-axis (P-H) \\
\begin{tabular}{|l|l|l|l|l|l|l|l|l|l|l|l|l|l|}
\hline $M_{Fig}$ & ($n_1, n_2, n_3)$ & I & $\Delta$S & S & C &
de & baryon & J & P \\
\hline 1209 & (0,1,1;1,0,1) & $\frac{1}{2}$ & +1 & 0 & 0 &  &
N(1209) & $\frac{1}{2}$,$\frac{3}{2}$ & +   \\
\hline 1299 & (002,-101,0-11) &  &  &  &  &  &  & & \\
\hline 1299 & (0, 0, 2) & 0 & 0 & -1 & 0 & 0 & $\Lambda$(1299) &
$\frac{1}{2}$ & - \\
\hline 1299 & (-1,0,1;0,-1,1) & $\frac{1}{2}$ & +1 & 0 & 0 & 0 &
N(1299) & $\frac{3}{2}$ & - \\
\hline 1929 & (112,-110,1-10) &  &  &  &  &  &  & &  \\
\hline 1929 & (1,1,2) & 0 & 0 & -1 & 0 & 0 & $\Lambda$(1929) &
$\frac{1}{2}$; $\frac{ 3}{2}$ & - \\
\hline 1929 & (-1,1,0;1,-1,0) & $\frac{1}{2}$ & +1 & 0 & 0 & 0 &
N(1929) & $\frac{3}{2}$,$\frac{5}{2}$;
$\frac{1}{2}$$\frac{3}{2}$$\frac{5}{2}$$\frac{7}{2}$ & - \\
\hline 1929 &
\begin{tabular}{l}
(121,211,200 \\
020,01-1,10-1)%
\end{tabular} &  &  &  &  &  &  &  &  \\
\hline 1929 & (020,01-1,10-1) &  &  &  & &  &  &  &    \\
\hline 1929 & (0, 2, 0) & 0 & 0 & -1 & 0 &  & $\Lambda$(1929) &
\begin{tabular} {l}
$\frac{1}{2}$,$\frac{3}{2}$;
$\frac{1}{2}$,$\frac{3}{2}$,$\frac{5}{2}$  \\
$\frac{3}{2}$,$\frac{5}{2}$;
$\frac{1}{2}$,$\frac{3}{2}$,$\frac{5}{2}$,$\frac{7}{2}$   %
\end{tabular} &
\begin{tabular} {l}
+  \\
-   %
\end{tabular} \\

\hline 1929 & (01-1,10-1) & $\frac{1}{2}$ & +1 & 0 & 0 & 0 & N(1929)
&
\begin{tabular} {l}
$\frac{1}{2}$,$\frac{3}{2}$;
$\frac{1}{2}$,$\frac{3}{2}$,$\frac{5}{2}$  \\
$\frac{3}{2}$,$\frac{5}{2}$;
$\frac{1}{2}$,$\frac{3}{2}$,$\frac{5}{2}$,$\frac{7}{2}$   %
\end{tabular} &
\begin{tabular} {l}
+  \\
-   %
\end{tabular} \\

\hline 1929 & (121,211,200) &  &  &  &  &  &  &  &  \\
\hline 1929 & (2, 0, 0) & 0 & 0 & -1 & 0 & 0 & $\Lambda$(1929) &
$\frac{5}{2}$,$\frac{7}{2}$;
$\frac{3}{2}$$\frac{5}{2}$$\frac{7}{2}$ & + \\
\hline 1929 & (1,2,1;2,1,1) & $\frac{1}{2}$ & +1 & 0 & 0 & 0 &
N(1929) & $\frac{5}{2}$$\frac{7}{2}$;
$\frac{3}{2}$,$\frac{5}{2}$$\frac{7}{2}$ & + \\
\hline 2019 & (-1-10,-112,1-12) &  &  &  &  &  &  &  & \\
\hline 2019 & (-1,-1,0) & 0 & 0 & -1 & 0 & 0 & $\Lambda$(2019) &
\begin{tabular} {l}
$\frac{3}{2}$,$\frac{5}{2}$;
$\frac{1}{2}$,$\frac{3}{2}$,$\frac{5}{2}$,$\frac{7}{2}$  \\
$\frac{7}{2}$  ;
$\frac{5}{2}$,$\frac{7}{2}$  %
\end{tabular} & - \\
\hline 2019 & (-1,1,2;1,-1,2) & $\frac{1}{2}$ & +1 & 0 & 0 & 0 &
N(2019) &
\begin{tabular} {l}
$\frac{3}{2}$,$\frac{5}{2}$;
$\frac{1}{2}$,$\frac{3}{2}$,$\frac{5}{2}$,$\frac{7}{2}$  \\
$\frac{7}{2}$,$\frac{9}{2}$;
$\frac{5}{2}$,$\frac{7}{2}$,$\frac{9}{2}$ %
\end{tabular} & - \\
\hline 2019 & (-1,-1,2) & 0 & 0 & -1 & 0 & 0 & $\Lambda$(2019) &
$\frac{5}{2}$,$\frac{7}{2}$;
$\frac{3}{2}$,$\frac{5}{2}$,$\frac{7}{2}$ & +\\

\hline 2649 &
\begin{tabular}{l}
(202,022,-121, \\
2-11,0-1-1,-10-1)%
\end{tabular} &  &  &  &  &  &  &  &  \\
\hline 2649 & (202,022,-121) & 1 & +1 & 0 & +1 & -115 &
$\Sigma_c(2534)$ & $\frac{1}{2}$ & +  \\
\hline 2649 & (202,022,-121) & 1 & +1 & 0 & +1 & 0&
$\Sigma_c(2649)$ & 2$\frac{3}{2}$ & +  \\
\hline 2649 & (202,022,-121) & 1 & +1 & 0 & +1 &+115 &
$\Sigma_c(2764)$ & $\frac{5}{2}$ & +  \\

\hline 2649 & (2-11,0-1-1,-10-1) & & & & & & & & \\
\hline 2649 & (0-1-1,-10-1) & $\frac{1}{2}$ & -1 & -2 & 0 & 0 &
$\Xi$(2649) & $\frac{5}{2}$,$\frac{7}{2}$;
$\frac{3}{2}$,$\frac{5}{2}$,$\frac{7}{2}$,$\frac{9}{2}$ & + \\
\hline 2649 & (2,-1,-1) & 0 & 0 & -1 & 0 & 0 & $\Lambda(2649)$ &
$\frac{5}{2}$,$\frac{7}{2}$;
$\frac{3}{2}$,$\frac{5}{2}$,$\frac{7}{2}$,$\frac{9}{2}$ & + \\

\hline 2649 & (21-1,12-1,220) &  &  &  &  &  &  & & \\
\hline 2649 & (21-1,12-1) & $\frac{1}{2}$ & +1 & -1 & 1 & -115 &
$\Xi_c(2534)$ & $\frac{1}{2}$ & + \\
\hline 2649 & (21-1,12-1) & $\frac{1}{2}$ & +1 & -1 & 1 & -115 &
$\Xi'_c(2534)$ & $\frac{1}{2}$ & + \\
\hline 2649 & (21-1,12-1) & $\frac{1}{2}$ & +1 & -1 & 1 & 0 &
$\Xi_c(2649)$ & 2$\frac{3}{2}$ & + \\
\hline 2649 & (21-1,12-1) & $\frac{1}{2}$ & +1 & -1 & 1 & +115 &
$\Xi_c(2764)$ & $\frac{5}{2}$ & + \\
\hline 2649 & (220) & 0 & 0 & -1 & 0 & 0 & $\Lambda(2649)$ &
\begin{tabular}{l}
$\frac{1}{2}$,$\frac{3}{2}$;
$\frac{1}{2}$,$\frac{3}{2}$,$\frac{5}{2}$; \\
$\frac{5}{2}$,$\frac{7}{2}$;
$\frac{3}{2}$,$\frac{5}{2}$,$\frac{7}{2}$,$\frac{9}{2}$; \\
$\frac{3}{2}$,$\frac{5}{2}$;
$\frac{1}{2}$,$\frac{3}{2}$,$\frac{5}{2}$,$\frac{7}{2}$ %
\end{tabular}
&
\begin{tabular}{l}
+ \\
+ \\
-%
\end{tabular} \\

\hline
\end{tabular}%
\end{tabular}

\begin{tabular}{l}
$\qquad$$\qquad$$\qquad$$\qquad$$\qquad$ Table 13b .\ The
baryons on the F-axis (P-H) \\
\begin{tabular}{|l|l|l|l|l|l|l|l|l|l|l|l|l|l|}
\hline $M_{Fig}$ & ($n_1, n_2, n_3)$ & I & $\Delta$S & S &
C & de & baryon & J & P \\
\hline 2739 &
\begin{tabular}{l}
(013,103,1-21, \\
-211,0-20,-200)%
\end{tabular} &  &  &  &  &  &  &  &   \\

\hline 2739 & (013,103,1-21) &  &  &  &  &   &   &  &   \\
\hline 2739 & (013,103) & $\frac{1}{2}$ & +1 & -1 & +1 & 0 &
$\Xi_c$(2739) &  $\frac{1}{2}$ & - \\
\hline 2739 & (013,103) & $\frac{1}{2}$ & +1 & -1 & +1 & 115 &
$\Xi_c$(2854) &  2$\frac{3}{2}$ & - \\
\hline 2739 & (013,103) & $\frac{1}{2}$ & +1 & -1 & +1 & 230 &
$\Xi_c$(2969) &  2$\frac{5}{2}$, $\frac{7}{2}$ & - \\
\hline 2739 & (-2,1,1) & 0 & 0 & -1 & 0 & 0 & $\Lambda(2739)$ &
\begin{tabular}{l}
$\frac{3}{2}$,$\frac{5}{2}$;
$\frac{1}{2}$,$\frac{3}{2}$,$\frac{5}{2}$,$\frac{7}{2}$ \\
$\frac{7}{2}$,$\frac{9}{2}$;
$\frac{5}{2}$,$\frac{7}{2}$,$\frac{9}{2}$,$\frac{11}{2}$ %
\end{tabular} & - \\
\hline 2739 & (-211,0-20,-200) & & & & & & & & \\
\hline 2739 & (-2,1,1) & 0 & 0 & 0 & 0 & 0 & $\Lambda(2739)$ &
\begin{tabular}{l}
$\frac{1}{2}$,$\frac{3}{2}$;
$\frac{1}{2}$$\frac{3}{2}$$\frac{5}{2}$ \\
$\frac{5}{2}$,$\frac{7}{2}$;
$\frac{3}{2}$,$\frac{5}{2}$,$\frac{7}{2}$,$\frac{9}{2}$%
\end{tabular} & + \\
\hline 2739 & (0-20,-200) & $\frac{1}{2}$ & -1 & -2 & 0 & 0 &
$\Xi$(2739) &
\begin{tabular}{l}
$\frac{1}{2}$,$\frac{3}{2}$;
$\frac{1}{2}$,$\frac{3}{2}$,$\frac{5}{2}$ \\
$\frac{5}{2}$,$\frac{7}{2}$;
$\frac{3}{2}$,$\frac{5}{2}$,$\frac{7}{2}$,$\frac{9}{2}$ %
\end{tabular} & + \\
\hline 2739 &
\begin{tabular}{l}
(0-13,-103,0-22, \\
-202,-2-11,-1-21)%
\end{tabular}
&  &  &  &  &  &  &  &   \\
\hline 2739 & (0-13,-103,0-22) &  &  &  &  &  &  &  & + \\
\hline 2739 & (0,-2,2) & 0 & 0 & -1 & 0 & 0 &
$\Lambda(2739)$ &
\begin{tabular}{l}
$\frac{5}{2}$,$\frac{7}{2}$;
$\frac{3}{2}$,$\frac{5}{2}$,$\frac{7}{2}$,$\frac{9}{2}$ \\
$\frac{9}{2}$,$\frac{11}{2}$;
$\frac{7}{2}$,$\frac{9}{2}$,$\frac{11}{2}$ %
\end{tabular}& + \\
\hline 2739 & (0-13,-103) & $\frac{1}{2}$ & +1 & -1 & +1 & +345 &
$\Xi_c(3084)$ & $\frac{3}{2}$ & + \\
\hline 2739 & (0-13,-103) & $\frac{1}{2}$ & +1 & -1 & +1 & +460 &
$\Xi_c(3199)$ & 2$\frac{5}{2}$ & + \\
\hline 2739 & (0-13,-103) & $\frac{1}{2}$ & +1 & -1 & +1 & +575 &
$\Xi_c(3319)$ & 2$\frac{7}{2}$ & + \\
\hline 2739 & (0-13,-103) & $\frac{1}{2}$ & +1 & -1 & +1 & +690 &
$\Xi_c(3429)$ & $\frac{9}{2}$ & + \\

\hline 2739 & (-202,-2-11,-1-21) &  &  &  &  &  & & &   \\
\hline 2739 & (-2-11,-1-21) & $\frac{1}{2}$ & -1 & -2 & 0 & 0 &
$\Xi$(2739) &
\begin{tabular}{l}
$\frac{7}{2}$,$\frac{9}{2}$;
$\frac{5}{2}$,$\frac{7}{2}$,$\frac{9}{2}$,$\frac{11}{2}$ \\

\end{tabular}
& - \\
\hline 2739 & (-2,0,2), & 0 & 0 & -1 & 0 & 0 & $\Lambda(2739)$ &
\begin{tabular}{l}
$\frac{7}{2}$,$\frac{9}{2}$;
$\frac{5}{2}$,$\frac{7}{2}$,$\frac{9}{2}$,$\frac{11}{2}$ \\
\end{tabular}
& - \\
\hline
\end{tabular}%
\end{tabular}
$\newline$ \\

Similarly to above paragraph, the third six-fold band with $M-{Fig}$
= 2739 and $\vec{n}$ = (-202,0-13,-103;-2-10,-1-21,0-22) leads to a
$\Xi_c$(M) M = $M_{Fig}$ + de = 2739 + 230 (2969, $n_c$ = 3); 2739 +
345 (3314, $n_c$ = 4); 2739 + 460 (3199, $n_c$ = 5) from Table 7.
There are $\Xi(2739)$ and two $\Lambda(2739)$. They appear in Table
13b.

Table B5 gives the J and P of the energy bands (the baryons) as
appearing in Table 13. Using Table 6B, we can find the $J_{Max}$ for
each baryon. Thus we omit the J larger than the $J_{Max}$ from Table
B5 as appearing in Table 13.

\subsection{The baryons on the G-axis (M-N)}

$\qquad$The G-axis is a two-fold symmetric axis, $S_G$ = - 2 from
(\ref{S-G}) and $\mathcal{B}$ = 1 from Postulate II. All energy
bands have a baryon number $\mathcal{B}=1$ and $S_G$ = -2:
\begin{equation}
\mathcal{B}\,\, = \,\,1\,\, and \,\,\,S_G = -2. \label{B=1 and S=-2}
\end{equation}
There are two-fold, four-fold and six-fold energy bands on the axis
(see Fig. 7). At $ M_{Fig}$ = 1119, there is a two-fold energy band
with asymmetry $\vec{n}$ = (0, 0, 0; 1,1,0). The band with
$\vec{n}$= (0, 0, 0) is the N(939)-baryon from(\ref{N(939)}), and
the band with $\vec{n}$ = (1, 1, 0) is the baryon $\Lambda $(1119)
from (\ref{Lambda(1119)}).

The four-fold bands with asymmetric $\vec{n}$ values divide into two
two-fold bands with symmetric $\vec{n}$ values from (\ref{Subdeg}).
Thus we only need to discuss the two-fold bands.

The two-fold bands with symmetric $\vec{n}$ values have S = -2 from
(\ref {S-G}), I = $\frac{1}{2}$ from (\ref{IsoSpin}), Q = 0 and -1
from (\ref{GMN}). Comparing the I, S and Q of the bands with the I,
S and Q of the baryons in Table 2 and in \cite{Baryon}, we find
there are only the $\Xi$-baryons with the same quantum numbers as
the band. According to Postulate III, the $\Xi$-baryons will excite
from the vacuum into the two-fold energy bands to form
$\Xi(M_{Fig})$-baryons as appearing in Table 14.

There are two two-fold bands with asymmetric $\vec{n}$ values: at
$E_M$ = 1299, a band with $\vec{n }$ = (200,1-10); at $E_N$ = 1839,
a band with (020,-110). The two two-fold bands will divide into four
single bands. Each single band has S = $S_{F}$ + $\Delta$S = -2 + 1
= -1 from (\ref{DSf}) and (\ref{S+DS}), I = 0 from (\ref{IsoSpin})
and Q = 0 from (\ref{GMN}). Comparing the quantum numbers of the
single bands with the quantum numbers of the baryons in Table 2 and
in \cite{Baryon}, we find there is only the $\Lambda$ -baryons with
the same quantum numbers as the single bands. According to Postulate
III, the four $\Lambda$-baryons will excite from the vacuum into the
four single bands to form four $\Lambda(M)$-baryons. The mass M =
$M_{Fig}$ + de = 1299 + 115 = 1414 for the two bands with (200) and
(1-10); the M = 1839 + 115 = 1954 for the two bands with (020) and
(-110) as appearing in Table 14.

At $M_{Fig}$ = 2559, there is a six-fold band with asymmetry
$\vec{n}$ = (112,11-2,002,00-2,220,-1-10). From (\ref{Subdeg}), the
band is divided into two two-fold bands with symmetric $\vec{n}$
values (112,11-2) and (002,00-2)) as well as one two-fold band with
asymmetric $\vec{n}$ = (220,-1-10). The two two-fold bands with
symmetric $\vec{n}$ values have I = 1/2, S = - 2, Q = 1, 0. They are
two $\Xi(2559)$-baryons as mentioned above. The third two-fold
sub-band with asymmetric $\vec{n}$ = (220,-1-10) divide into a
single sub-sub-band with $\vec{n}$ = (2,2,0) and another single
sub-sub-band with $\vec{n}$ = (-1,-1, 0). The single sub-band with
(2, 2, 0) has I = 0, S = -2, $\Delta$ S = +1 = C from
(\ref{Charmed}) and Q = 0 from (\ref{GMN}). Comparing the quantum
numbers of the sub-sub-band with the quantum numbers of the baryons
in Table 2 and in \cite{Baryon}, we find there is only the
$\Omega_c$-baryons with the same quantum numbers as the
sub-sub-band. According to Postulate III, the $\Omega_c$-baryons
will excite from the vacuum into the sub-band to form
$\Omega_c(M)$-baryons. The M = $M_{Fig}$ + de = 2559 + 230 ($n_c$ =
0,2789);  2559 + 3450 ($n_c$ = 1, 2904);  2559 + 230 ($n_c$ =
0,30199)from Fig. 7 and Table 7 (de = 115($n_c$-S)$\Delta$S). The
$\Omega_c$-baryons are shown in Table 14. The sub-sub-band with
(-1,-1, 0) has I = 0, S =$S_{G}$ + $\Delta$S = -2 + (-1) = -3 from
(\ref{DSf}) and Q = -1 from (\ref{GMN}). Comparing the quantum
numbers of the band with the quantum numbers of the baryons, we find
there is only an $\Omega$-baryon with the same quantum numbers as
the band with. According to Postulate III, The $\Omega$-baryon will
be excited from the vacuum into the sub-sub-band to form an
$\Omega(M)$-baryon. The M = $M_{Fig}$ + de = 2559 + (-345) = 2214
from Figure 7 and Table 7. The $\Omega(2214)$ ($\$n_c$ = 0) appear
in Table 14.

At $M_{f}$ = 2739 there is another six-fold energy band with
(310,0-20,202,20-2,1-12,1-1-2). Similarly to the six-fold band at
$M_{f}$ = 2559 as mentioned above, we can deduce the baryons of the
six-fold band. The six-fold band divide into two two-fold symmetry
sub-bands with $\vec{n}$ = (202,20-2) and (1-12,1-1-2) as well as an
asymmetry two-fold sub-band with (310,0-20). The two two-fold
sub-bands lead to two $\Xi(2739)$-baryons. The asymmetry two-fold
sub-band (310,0-20) divides into two single sub-sub-bands with
$\vec{n}$ = (3,1,0) and (0,-2,0). The single sub-sub-band with
$\vec{n}$ = (3,1,0) leads to $\Omega_{c}$(M)-baryons with M =
$M_{Fig}$ + de = 2739 + 230 ($\$n_c$ = 0, 2969); 2739 + 345 ($\$n_c$
= 1, 3084); 2739 + 460 ($\$n_c$ = 2, 2199)  from de = 115($n_c$
-S)$\Delta$S in Table 7. The other single sub-sub-band with
$\vec{n}$ = (0,-2,0) leads to an $\Omega(2394)$-baryon (de = -345).
These baryons appear in Table 14.

\begin{tabular}{l}
$\qquad$$\qquad$$\qquad$$\qquad$$\qquad$\,\,\,\,\,Table 14a.\ The
Baryons on the G-axis \\
\begin{tabular}{|l|l|l|l|l|l|l|l|l|l|l|}
\hline $M_{Fig}$ & ($n_1, n_2, n_3$) & I & $\Delta$S & S & C & de &
baryon & J & P \\
\hline 1299 & (1,0,1;0,1,-1) & $\frac{1}{2}$ & 0 & -2 & 0 & 0 &
$\Xi(1299)$ & $ \frac{1}{2}$ $\frac{3}{2}$ & + \\
\hline 1299 & (2,0,0;1,-1,0) &  &  &  &  &  &  &  & \\
\hline 1299 & (2,0,0) & 0 & +1 & -1 & 0 & 115 & $\Lambda$(1414) &
$\frac{1}{2}$,$\frac{3}{2}$ & + \\
\hline 1299 & (1,-1,0) & 0 & +1 & -1 & 0 & 115 & $\Lambda$(1414) &
$\frac{1}{2}$ $\frac{3}{2}$ & - \\
\hline 1479 & (0,1,1;0,1,-1) & $\frac{1}{2}$ & 0 & -2 & 0 & 0 &
$\Xi$(1479) & $\frac{1}{2}$,$\frac{3}{2}$;
$\frac{1}{2}$,$\frac{3}{2}$,$\frac{5}{2}$ & + \\
\hline 1839 & (0,2,0;-1,1,0) & & &  &  &  &  &  &  \\
\hline 1839 & (0,2,0,) & 0 & +1 & -1 & 0 & 115 & $\Lambda$(1954) &
$\frac{3}{2}$,$\frac{5}{2}$;
$\frac{1}{2}$,$\frac{3}{2}$,$\frac{5}{2}$,$\frac{7}{2}$ & - \\
\hline 1839 & (-1,1,0) & 0 & +1 & -1 & 0 & 115 & $\Lambda$(1954) &
\begin{tabular} {l}
$\frac{1}{2}$;$\frac{3}{2}$ \\
$\frac{3}{2}$,$\frac{5}{2}$;
$\frac{1}{2}$,$\frac{3}{2}$$\frac{5}{2}$,$\frac{7}{2}$ %
\end{tabular}& - \\
\hline 2019 & (0-11,0-1-1;21-1,211) &  &  &  &  &  &  &  & \\
\hline 2019 & (0,-1,1;0,-1,-1) & $\frac{1}{2}$ & 0 & -2 & 0 & 0 &
 $\Xi(2019)$ & $\frac{3}{2}$; $\frac{1}{2}$,$\frac{3}{2}$,
$\frac{5}{2}$ & ? \\
\hline 2019 & (2,1,-1;2,1,1) & $\frac{1}{2}$ & 0 & -2 & 0 &  &
$\Xi(2019)$ &
\begin{tabular} {l}
$\frac{3}{2},\frac{5}{2}$;
$\frac{1}{2},\frac{3}{2},\frac{5}{2},\frac{7}{2}$  \\
$\frac{7}{2}$; $\frac{5}{2}$,$\frac{7}{2}$ %
\end{tabular}& - \\
\hline 2019 & (2,-1,1;2,-1,-1) & $\frac{1}{2}$ & 0 & -2 & 0 & 0 &
$\Xi(2019)$ &
\begin{tabular} {l} $\frac{1}{2},\frac{3}{2}$;
$\frac{1}{2},\frac{3}{2},\frac{5}{2}$\\
$\frac{5}{2},\frac{7}{2}$;
$\frac{3}{2},\frac{5}{2},\frac{7}{2}$ %
\end{tabular}& + \\

\hline 2199 & (-101,-10-1,121,12-1) &  & & & &  &  &  & \\

\hline 2199 & (-101,-10-1) & $\frac{1}{2}$ & 0 & -2 & 0 & 0 &
$\Xi(2199)$ &
\begin{tabular} {l}
$\frac{1}{2},\frac{3}{2}$;
$\frac{1}{2},\frac{3}{2},\frac{5}{2}$\\
$\frac{5}{2},\frac{7}{2}$;
$\frac{3}{2},\frac{5}{2},\frac{7}{2}$ %
\end{tabular}& + \\

\hline 2199 & (121,12-1) & $\frac{1}{2}$ & 0 & -2 & 0 & 0 &
$\Xi(2199)$ &
\begin{tabular} {l}
$\frac{3}{2},\frac{5}{2}$;
$\frac{1}{2},\frac{3}{2},\frac{5}{2},\frac{7}{2}$  \\
$\frac{7}{2},\frac{9}{2}$; $\frac{5}{2},\frac{7}{2},\frac{9}{2}$ %
\end{tabular}& - \\

\hline 2559 &
\begin{tabular}{l}
(112,11-2,002 \\
00-2, 220,-1-10)%
\end{tabular}
&  &  &  &  &  &  &  & \\

\hline 2559 & (112,11-2) & $\frac{1}{2}$ & 0 & -2 & 0 & 0 &
$\Xi(2559)$ &
\begin{tabular} {l}
$\frac{1}{2},\frac{3}{2}$;
$\frac{1}{2},\frac{3}{2},\frac{5}{2}$\\
$\frac{5}{2},\frac{7}{2}$;
$\frac{3}{2},\frac{5}{2},\frac{7}{2}$ %
\end{tabular}& + \\

\hline 2559 & (002,00-2) & $\frac{1}{2}$ & 0 & -2 & 0 & 0 &
$\Xi(2559)$ &
\begin{tabular} {l}
$\frac{3}{2},\frac{5}{2}$;
$\frac{1}{2},\frac{3}{2},\frac{5}{2},\frac{7}{2}$  \\
$\frac{7}{2},\frac{9}{2}$;
$\frac{5}{2},\frac{7}{2},\frac{9}{2},\frac{11}{2}$
\end{tabular}& - \\

\hline 2559 & (220,-1-10) &  &  &  &  &  &  &  &  \\
\hline 2559 & (-1, -1, 0) & 0 & -1 & -3 & 0 & -345 & $\Omega(2214)$
& $\frac{1}{2},\frac{3}{2},\frac{5}{2}$,$\frac{7}{2}$ & - \\

\hline 2559 & (2, 2, 0) & 0 & +1 & -2 & +1 & +230 & $\Omega_c(2789)$
& $\frac{1}{2}$ & + \\
\hline 2559 & (2, 2, 0) & 0 & +1 & -2 & +1 & +345 & $\Omega_c(2904)$
& $\frac{3}{2}$& + \\
\hline 2559 & (2, 2, 0) & 0 & +1 & -2 & +1 & +460 & $\Omega_c(3019)$
& $\frac{5}{2}$& + \\
\hline 2739 &
\begin{tabular}{l}
(310,0-20,202 \\
20-2, 1-12,1-1-2)%
\end{tabular}
&  &  &  &  &  &  &  & \\

\hline 2739 & (3,1,0;0,-2,0)
&  &  &  &  &  &  &  & \\

\hline 2739 & (3, 1, 0) & 0 & +1 & -2 & +1 & +230 & $\Omega_c(2969)$
& 2$\frac{1}{2}$ & + \\
\hline 2739 & (3, 1, 0) & 0 & +1 & -2 & +1 & +345 & $\Omega_c(3084)$
& 2$\frac{3}{2}$ & + \\
\hline 2739 & (3, 1, 0) & 0 & +1 & -2 & +1 & +460 & $\Omega_c(3199)$
& $\frac{1}{2}$,$\frac{3}{2}$;
$\frac{1}{2}$,$\frac{3}{2}$,$\frac{5}{2}$ & + \\

\hline 2739 & (0, -2, 0) & 0 & -1 & -3 & 0 & -345 & $\Omega(2394)$ &
$\frac{3}{2}$,$\frac{5}{2}$,$\frac{7}{2}$,$\frac{9}{2}$& + \\

\hline 2739 & (202,20-2) & $\frac{1}{2}$ & 0 & -2 & 0 & 0 &
$\Xi(2739)$ &
\begin{tabular} {l}
$\frac{1}{2}$,$\frac{3}{2}$;
$\frac{1}{2}$,$\frac{3}{2}$,$\frac{5}{2}$  \\
$\frac{5}{2}$,$\frac{7}{2}$;
$\frac{3}{2}$,$\frac{5}{2}$,$\frac{7}{2}$,$\frac{9}{2}$ %
\end{tabular} & + \\

\hline 2739 & (1-12,1-1-2) & $\frac{1}{2}$ & 0 & -2 & 0 & 0 &
$\Xi(2739)$ &
\begin{tabular} {l}
$\frac{5}{2},\frac{7}{2};
\frac{3}{2},\frac{5}{2}$,$\frac{7}{2},\frac{9}{2}$  \\
$\frac{9}{2},\frac{11}{2}$;
$\frac{7}{2}$,$\frac{9}{2}$,$\frac{11}{2}$ %
\end{tabular}& +\\
\hline
\end{tabular}%
\end{tabular}
\newline \\

Table B6 gives the J and P values for these baryons of the G-axis.
The formula (\ref{J}) and Table 6B give the maximum $J_{Max}$ for
each baryon. Using it we can take off the extra J values (larger
than $J_{Max}$) as appearing in Table 14.

\section{Comparison of the Deduced Baryons with
the Experimental Baryons}

$\qquad$We have deduced the baryons appearing in (\ref{N(939)}),
(\ref{1119}) and Table 8--Table 14 in Section 5. Using Table
15--Table 22, we will compare the deduced baryons with the
experimental results \cite{Baryon}. In the comparisons, we use the
baryon name to represent the intrinsic quantum numbers (I, S, C, B
and Q) of baryons as appearing in Table 2 and Table 3 as well as in
\cite{Baryon}. If the name is the same between theory and
experiment, this means that the intrinsic quantum numbers (I, S, C,
B and Q) of the deduced and experimental baryons are completely the
same. In the comparison, the same kind of baryons (with same I, S,
C, B and Q) will be divided into groups with the same J and P as
well as similar masses (about inside the experimental width). For
each group, we deduce the average mass and use it to compare with
the experimental mass of the corresponding baryon.

\subsection{The most important baryons}

$\qquad$First we compare the most important baryons. From
(\ref{N(939)}), (\ref{1119}) and Table 8--Table 14, we get the most
important deduced baryons N(939), $\Lambda(1119)$, $\Sigma(1209)$,
$\Xi(1318)$), $\Omega(1659)$, $\Lambda_c(2264)$ and
$\Lambda_b(5554)$ as appearing in Table 15. The corresponding
experimental results are from \cite{Baryon}.
\newline\\

\begin{tabular}{l}
\ \ \ \ Table 15.\ The comparison of the most important baryons \\
\begin{tabular}{|l|l|l|l|l|l|l|}
\hline Experiment & N(939)$\frac{1}{2}^{+}$ &
$\Lambda(1116)\frac{1}{2}^{+}$ & $ \Sigma(1193)\frac{1}{2}^{+}$ &
$\Xi(1318)\frac{1}{2}^{+}$ \\
\hline Theory & N(939)$\frac{1}{2}^{+}$ &
$\Lambda(1119)\frac{1}{2}^{+}$ & $\Sigma(1209)\frac{1}{2}^{+}$ &
$\Xi(1299)\frac{1}{2}^{+}$ \\
\hline Where & ($\ref{N(939)}$)
& Table 11 & Table 9 & Table 14 \\
\hline $\Delta$M/$M_{Exp.}$
& 0.0 & 0.3 & 1.3 & 1.4 \\
\hline Experiment & $\Omega(1672)\frac{3}{2}^{+}$&
$\Lambda_c(2286)\frac{1}{2}^{+}$ &
$\Lambda_b(5624)\frac{1}{2}^{+}$&\\
\hline Theory & $\Omega(1659)\frac{3}{2}^{+}$ &
$\Lambda_c(2264)\frac{1}{2}^{+}$ & $\Lambda_b(5554)\frac{1}{2}^{+}$
& \\
\hline Where & Table 11 & Table 8 & Table 11 & \\
\hline $\Delta$M/$M_{Exp.}$ & 0.8 & 1.0 & 1.2 & \\

\hline
\end{tabular}%
\end{tabular}
$\newline$ \\
Table 15 shows that the deduced quantum numbers (I, S,
C, B, Q) are exactly the same as the experimental results of the
seven most important baryons. Their masses are consistent with
experimental results more than 98.5 percent using only three
constant parameters (M$_0$ =939 Mev, $\alpha$ = 360 Mev and $\beta$
= 115 Mev ).

\subsection{The comparison between the deduced and experimental
$\Delta$-baryons} $\qquad$The deduced $\Delta$-baryons are from
Table 8 and Table 12 ($\Delta(1929)$ and $\Delta(2649)$). The
experimental $\Delta$-baryons are from \cite{Baryon}. We compare the
deduced $\Delta$-baryons with the experimental $\Delta$-baryons in
Table 16:
\newline
\begin{tabular}{l}
\\
$\qquad$$\qquad$$\qquad$$\qquad$$\qquad$Table 16.\ The Comparison of
the $\Delta$-Baryons \\
\begin{tabular}{|l|l|l|l|l|l|l|l|l|l|l|}
\hline Order,Compose & Theory & Experiment, $\Gamma$ &
$\frac{\Delta M}{M}$\\
\hline 1. $\Delta(1299)$,$(\frac{3}{2})^{+}$ &
$\Delta(1299)$,$(\frac{3}{2})^{+}$ &
$\Delta(1232)$,$(\frac{3}{2})^{+}$, 118**** & 5.4 \\
\hline 2. $\Delta(1659)$,($\frac{1}{2})^{-}$ &
$\Delta(1659)$,($\frac{1}{2})^{-}$ &
$\Delta(1630)$,$(\frac{1}{2})^{-}$, 145**** & 1.8 \\
\hline 3. 2$\Delta(1659)$,($\frac{3}{2})^{-}$ &
$\Delta(1659)$,($\frac{3}{2})^{-}$ &
$\Delta(1700)$,$(\frac{3}{2})^{-}$, 300**** & 2.4 \\
\hline 4. $\frac{1}{2}$($\Delta(1659)$,+$\Delta(1929)$)\,
$(\frac{1}{2})^{+}$ & $\Delta(1794)$,\,$(\frac{1}{2})^{+}$ &
$\Delta(1750)$,\,$(\frac{1}{2})^{+}$, 300 * & 2.5 \\
\hline 5. $\Delta(1659)$,\,$(\frac{3}{2})^{+}$ &
$\Delta(1659)$,\,$(\frac{3}{2})^{+}$ &
$\Delta(1600)$,\,$(\frac{3}{2})^{+}$,350*** & 3.7 \\
\hline 6. $\frac{1}{4}$[$\Delta(1659)$+$\Delta(1929)$+
2$\Delta(2019)$] $\frac{1}{2}^{+}$ &
$\overline{\Delta(1907)}$,$(\frac{1}{2})^{+}$ &
$\Delta(1910)$,$(\frac{1}{2})^{+}$, 200**** & 0.2 \\
\hline 7. $\frac{1}{7}$[$\Delta(1659)$+3$\Delta(1929)$+
3$\Delta(2019)$] $\frac{3}{2}^{+}$&
$\overline{\Delta(1929)}$,$(\frac{3}{2})^{+}$ &
$\Delta(1920),(\frac{3}{2})^{+} $, 200*** & 0.5 \\
\hline 8. $\frac{1}{5}$[$\Delta(1659)$+3$\Delta(1929)$+
$\Delta(2019)$]$\frac{5}{2}^{+}$ &
$\overline{\Delta(1893)}$,$(\frac{5}{2})^{+}$ &
$\Delta(1890),(\frac{5}{2})^{+} $, 330**** & 0.2 \\
\hline 9. $\frac{1}{4}$ (2$\Delta(1929)$+2$\Delta(2019))$,
($\frac{7}{2})^{+}$ & $\overline{\Delta(1974)}$, ($\frac{7}{2})^{+}$
&
$\Delta(1930),(\frac{7}{2})^{+}$, 285**** & 2.3 \\
\hline 10. $\Delta(1929)$, $(\frac{1}{2})^{-}$ &
$\Delta(1929)$,$(\frac{1}{2})^{-}$ &
$\Delta(1900),(\frac{1}{2})^{-} $, 200** & 1.5\\
\hline 11. 2$\Delta(1929)$,$\frac{3}{2}^{-}$ &
$\Delta(1929),(\frac{3}{2})^{-}$ &
$\Delta(1940),(\frac{3}{2})^{-} $, 300* & 0.6\\
\hline 12. $\frac{1}{6}$[2$\Delta(1659)$+2$\Delta(1929)$+
2$\Delta(2379)] \frac{5}{2}^{-}$ &
$\overline{\Delta(1989)}$,$(\frac{5}{2})^{-}$ &
$\Delta(1960,(\frac{5}{2})^{-}$, 360*** & 1.5 \\

\hline 13. $\frac{1}{4}$[$\Delta(1929)$ + 3$\Delta(2379)$]
$\frac{7}{2}^{-} $ & $\overline{\Delta(2267)}$,($\frac{7}{2})^{-}$ &
$\Delta(2200)$,($\frac{7}{2})^{-}$, 400* & 3.0 \\
\hline 14. 2$\Delta(2019),(\frac{5}{2})^{+}$ &
$\Delta(2019),(\frac{5}{2})^{+}$ &
$\Delta(2000),(\frac{5}{2})^{+}$, 200** & 1.0 \\
\hline 15. $\Delta(2379)$,$\frac{1}{2}^{-}$ &
$\Delta(2379),(\frac{1}{2})^{-}$ &
$\Delta(2150),(\frac{1}{2})^{-} $, 200* & 10.6\\
\hline 16.  2$\Delta(2379),(\frac{7}{2})^{+}$ &
$\Delta(2379),(\frac{7}{2})^{+}$ &
$\Delta(2390),(\frac{7}{2})^{+}$, 300* & 0.5 \\
\hline 17. $\frac{1}{4}$[$\Delta(2019)$ + 3$\Delta(2379)$],
$\frac{9}{2}^{+}$ & $\overline{\Delta(2289)}$,$(\frac{9}{2})^{+}$ &
$\Delta(2300),(\frac{9}{2})^{+}$, 400** & 0.5 \\
\hline 18. $\frac{1}{4}$[2$\Delta(2379)$+2$\Delta(2649)$]
$(\frac{11}{2})^{+}$ &
$\overline{\Delta(2514)}$,($\frac{11}{2})^{+}$ &
$\Delta(2420),(\frac{11}{2})^{+}$, 400**** & 3.9 \\
\hline 19. $\Delta(2379),(\frac{5}{2})^{-}$ &
$\Delta(2379),(\frac{5}{2})^{-}$ &
$\Delta(2350),(\frac{5}{2})^{-}$, 300* & 1.3 \\
\hline 20. $\frac{1}{4}$[2$\Delta(2379)$+2$\Delta(2649)$]
$\overline{\Delta(2514)}$, $(\frac{9}{2})^{-}$ &
$\Delta(2379)$,($\frac{9}{2})^{-}$ &
$\Delta(2400),(\frac{9}{2})^{-}$, 400** & 4.8 \\

\hline 21. 4$\Delta(2739),(\frac{13}{2})^{-}$ &
$\Delta(2739),(\frac{13}{2})^{-}$ &
$\Delta(2750),(\frac{13}{2})^{-}$, 400** & 0.4  \\
\hline 22. 4$\Delta(3099),(\frac{15}{2})^{+}$ &
$\Delta(3099),(\frac{15}{2})^{+}$ &
$\Delta(2950),(\frac{15}{2})^{}$, 500** & 5.1  \\

\hline 23. 3$\Delta(2739),(\frac{11}{2})^{-}$ &
$\Delta(2739),(\frac{11}{2})^{-}$ &
?,    ?,  ? & ?  \\
\hline 24. 4$\Delta(3099),(\frac{17}{2})^{+}$ &
$\Delta(3099),(\frac{17}{2})^{+}$ &
?, ? & ?  \\

\hline
\end{tabular}%
\end{tabular}%
\newline

The quantum numbers (I, S, C, B, Q, J and P) of all deduced
$\Delta$-baryons are completely the same as the experimental
results. The masses of the ten established (four or three stars)
$\Delta$-baryons agree more than average 97.8 percent with the
experimental results. The largest error of the ten established
$\Delta$-baryons is 5.4 percent. All 22 deduced $\Delta$-baryons
mass agree average more then 97.5 percent with the 22 observed
$\Delta$-baryons.
\subsection{The comparison between the deduced and experimental
$\Sigma$-baryons}

$\qquad$The deduced $\Sigma$-baryons are from Table 9. The
experimental $\Sigma$-baryons are from \cite{Baryon}. The comparison
appears in Table 17:
\newline

\begin{tabular}{l}
$\qquad$$\qquad$$\qquad$$\qquad$ Table 17.\ Comparing theory
$\Sigma$s with the experimental
results \\
\begin{tabular}{|l|l|l|l|l|l|l|l|l|l|l|}
\hline Order,Compose & Theory $J^P$ & Experiment $J^P$,$\Gamma$ &
$\frac{\Delta M}{M}$ \\
\hline 1. $\Sigma(1209)$, $(\frac{1}{2})^{+}$ &
$\Sigma(1209)$,$(\frac{1}{2})^{+}$ &
$\Sigma(1193)$, $(\frac{1}{2})^{+}$,**** & 1.4 \\
\hline 2. $\frac{1}{2}$[$\Sigma(1209)$ +$\Sigma(1659)$],
($\frac{3}{2})^{+}$ &$\overline{\Sigma(1434)}$,$(\frac{3}{2})^{+}$ &
$\Sigma(1385), (\frac{3}{2})^{+}$,36**** & 3.5\\
\hline 3. $\Sigma(1659)$, $(\frac{1}{2})^{-}$ &
$\Sigma(1659)$,$(\frac{1}{2})^{-}$ &
$\Sigma(1620)$, $(\frac{1}{2})^{-}$,43** & 2.4 \\
\hline 4. 2$\Sigma(1659)$, ($\frac{3}{2})^{-}$
&$\Sigma(1659)$,($\frac{3}{2})^{-}$ &
$\Sigma(1580)$, $(\frac{3}{2})^{-}$,15* & 5.0 \\
\hline 5. $\Sigma(1659)$, $(\frac{1}{2})^{+}$ &
$\Sigma(1659)$,$(\frac{1}{2})^{+}$ &
$\Sigma(1660)$, $(\frac{1}{2})^{+}$,$100***$ & .06  \\
\hline 6. 2$\Sigma(1659)$, $(\frac{3}{2})^{-}$
&$\Sigma(1659)$,$(\frac{3}{2})^{-}$ &
$\Sigma(1670)$, $(\frac{3}{2})^{-}$,60**** & 0.7 \\
\hline 7. $\Sigma(1659)$, ($\frac{5}{2})^{+}$ &
$\Sigma(1659)$,($\frac{5}{2})^{+}$ &
$\Sigma(1690)$, $(?)^{?}$,130** & 1.8\\
\hline 8. $\frac{1}{2}$[$\Sigma(1659)$ +$\Sigma(1929)$],
($\frac{1}{2})^{-}$ & $\overline{\Sigma(1794)}$,$(\frac{1}{2})^{-}$
& $\Sigma(1750)$, $(\frac{1}{2})^{-}$,90*** & 2.5  \\
\hline 9. $\frac{1}{8}$[4$\Sigma(1659)$+4$\Sigma(1929)$],
($\frac{5}{2})^{-}$ & $\overline{\Sigma(1794)}$,$(\frac{5}{2})^{-}$
&
$\Sigma(1775)$, $(\frac{5}{2})^{-}$,120**** & 1.1 \\
\hline 10. 4$\Sigma(1929)$, $(\frac{3}{2})^{-}$
&$\Sigma(1929)$,$(\frac{3}{2})^{-}$ &
$\Sigma(1940)$, $(\frac{3}{2})^{-}$ ,220*** & 0.6 \\
\hline 11. $\frac{1}{2}$[$\Sigma(1659)$+$\Sigma(1929)$],
$(\frac{1}{2})^{+}$ &$\overline{\Sigma(1794)}$,$(\frac{1}{2})^{+}$ &
$\Sigma(1770)$, $(\frac{1}{2})^{+}$ ,100* & 1.4 \\
\hline 12. $\frac{1}{3}$[$\Sigma(1659)$+2$\Sigma(1929)$],
($\frac{3}{2})^{+}$ &$\overline{\Sigma(1839)}$,$(\frac{3}{2})^{+}$ &
$\Sigma(1840)$, $(\frac{3}{2})^{+}$,100* & .05 \\
\hline 13. $\Sigma(1929)$$(\frac{5}{2})^{+}$,
&$\Sigma(1929)$,$(\frac{5}{2})^{+}$ & $\Sigma(1915)$,
$(\frac{5}{2})^{+}$ ,120**** & 0.7 \\
\hline 14. 2$\Sigma(1929)$, $(\frac{7}{2})^{-}$
&$\Sigma(1929)$,$(\frac{7}{2})^{-}$ &
$\Sigma(2100)$, $(\frac{7}{2})^{-}$,100* & 8.1 \\
\hline 15. 2$\Sigma(1929)$, $(\frac{1}{2})^{+}$
&$\Sigma(1929)$,$(\frac{1}{2})^{+}$ & $\Sigma(1880)$,
$(\frac{1}{2})^{+}$ ,100** & 2.6 \\
\hline 16. $\Sigma(1929)$, ($\frac{1}{2})^{-}$ &
$\Sigma(1929)$,$(\frac{1}{2})^{-}$
& $\Sigma(2000)$, $(\frac{1}{2})^{-}$,200* & 3.6  \\
\hline 17.  $\frac{1}{6}$[4$\Sigma(1929)$ +2$\Sigma(2379)$],
$(\frac{3}{2})^{+}$ & $\overline{\Sigma(2079)}$, $(\frac{3}{2})^{+}$
& $\Sigma(2080)$, $(\frac{3}{2})^{+}$, 200** & .04 \\
\hline 18. $\frac{1}{6}$[4$\Sigma(1929)$ +2$\Sigma(2379)$],
($\frac{5}{2})^{+}$ &$\overline{\Sigma(2079)}$,$(\frac{5}{2})^{+}$ &
$\Sigma(2070)$, $(\frac{5}{2 })^{+}$,300* & 0.4 \\
\hline 19. $\frac{1}{5}$[4$\Sigma(1929)$ + $\Sigma(2379)$],
$(\frac{7}{2})^{+}$ &$\overline{\Sigma(2019)}$,$(\frac{7}{2})^{+}$ &
$\Sigma(2030)$, $(\frac{7}{2})^{+}$ ,180**** & 0.5 \\
\hline 20. $\frac{1}{4}$[$\Sigma(1929)$+3$\Sigma(2379)$],
$\frac{3}{2}^{+}$$\frac{5}{2}^{+}$$\frac{7}{2}^{+}$&
$\overline{\Sigma(2267)}$,$(?)^{?}$ &
$\Sigma(2250)$, $(?)^{?}$,100** & 0.8 \\

\hline 21. $\frac{1}{2}$[$\Sigma(2379)$+$\Sigma(2649)$],
($\frac{9}{2})^{+}$ &$\overline{\Sigma(2514)}$,($\frac{9}{2})^{+}$ &
$\Sigma(2455)$ Bumps,$(\frac{3}{2})^{-}$,120** & 2.4 \\

\hline 22. $\Sigma(2649)$, ($\frac{11}{2})^{+}$
&$\Sigma(2649)$,($\frac{11}{2})^{+}$ &
$\Sigma(2620)$ Bumps $(?)^{?}$,200** & 1.1 \\

\hline 23. 2$\Sigma(2649)$, $\frac{3}{2}$,$\frac{5}{2}$,
$\frac{7}{2}$,$\frac{9}{2}$ &
$\Sigma(2649)$,$(?)^{?}$&$\Sigma(?)$,$(?)^{?}$,
$?^{?}$ & ? \\

\hline
\end{tabular}%
\end{tabular}
$\qquad $ $\qquad $ $\qquad$ \newline

The quantum numbers (I, S, C, B, Q, J and P) of all 22 deduced
$\Sigma$-baryons are completely the same as the experimental
results. The masses of the all 22 observed $\Sigma$-baryons agree
more than 98.2 percent with the experimental results. The largest
error of the nine established $\Sigma$-baryons is only 3.5 percent.

\subsection{The comparison of the N-baryons}
$\qquad $ The deduced N-baryons are from Table 8, Table 12 and Table
13; and the experimental N-baryons are from \cite{Baryon}. The
comparison between the deduced N-baryons with the experimental
N-baryons appears in Table 18.

\begin{tabular}{l}
$\qquad$$\qquad$$\qquad$Table 18.\ Comparing theory N-baryons with
the experimental results \\
\begin{tabular}{|l|l|l|l|l|l|l|l|l|l|l|l|l|}
\hline Order,J$^{P}$,Compose & N-baryon & expr & $\Gamma $ & * &
$\frac{\Delta M}{M}$ \\
\hline 0. $\frac{1}{2}^{+}$, N(939)& N(939) &
N(939)& &**** & 0.0 \\
\hline 1. $\frac{1}{2}^{+}$,$\frac{1}{3}$[N(1209)+N(1299)+N(1659)]
& $\overline{N(1389)}$ & N(1440) & 300 & **** & 3.5 \\
\hline 2. $\frac{3}{2}^{+}$,$\frac{1}{2}$[N(1209)+N(1299)] &
$\overline{N(1254)}$ &$\Delta (1232)$ & 118 & **** & 1.8 \\
\hline 3. $\frac{1}{2}^{-}$, $\frac{1}{2}$[N(1479)+ N(1659)] &
$\overline{N(1569)}$ & N(1535)&150 &**** &2.2\\
\hline 4. $\frac{3}{2}^{-}$,  2N(1479)& N(1479)&
N(1520) & 115 &****&2.7\\
\hline 5. $\frac{1}{2}^{-}$, N(1659) &N(1659)&
N(1655) & 165 & **** &0.5\\
\hline 6. $\frac{3}{2}^{-}$,$\frac{1}{7}$[N(1299)+4N(1659)+2N(1839)]
& $\overline{N(1659)}$ & N(1700) &100 & *** & 2.4\\
\hline 7.
$\frac{5}{2}^{-}$,$\frac{1}{8}$[2N(1479)+4N(1659)+2N(1839)]
& $\overline{N(1659)}$ & N(1675) &150 & **** & 1.0\\
\hline 8. $\frac{1}{2}^{+}$,$\frac{1}{2}$[N(1659)+N(1839)] &
$\overline{N(1749)}$ & N(1710) &100 & *** & 2.3\\
\hline 9. $\frac{3}{2}^{+}$,$\frac{1}{2}$[N(1659)+N(1839)] &
$\overline{N(1749)}$ & N(1720) &200 & **** &1.7\\
\hline 10. $\frac{5}{2}^{+}$,N(1659) & N(1659) &
N(1680) &130 & **** &1.3\\
\hline 11. $\frac{1}{2}^{+}$,
\begin{tabular} {l}
$\frac{1}{11}$[N(1839)+4N(1929)+2N(2019) \\
\,\,\,\,\, + 2N(2199)+ 2N(2379)] %
\end{tabular} & $\overline{N(2068)}$ & N(2100) &
394 & * & 1.5 \\

\hline 12. $\frac{3}{2}^{+}$, $\frac{1}{10}$[N(1839)+
6N(1929)+3N(2019)]& $\overline{N(1947)}$ &  N(1900) &
500 & ** & 2.5 \\
\hline 13.
$\frac{5}{2}^{+}$,$\frac{1}{10}$[N(1839)+6N(1929)+3N(2019)] &
$\overline{N(1947)}$ &  N(2000) &
200 & ** & 2.7 \\
\hline 14.
$\frac{7}{2}^{+}$,$\frac{1}{8}$[4N(1929)+2N(2019)+2N(2199)] &
$\overline{N(2019)}$ &  N(1990) &
350 & ** & 1.5 \\
\hline 15. $\frac{9}{2}^{+}$,$\frac{1}{5}$[N(2019)+ N(2199)+
3N(2379)] & $\overline{N(2271)}$ &  N(2220) &
400 &****  & 2.3 \\
\hline 16.
$\frac{11}{2}^{+}$,$\frac{1}{10}$[2N(2379)+2N(2649)+6N(2739)] &
$\overline{N(2649)}$ &  ? & ? & ? &? \\
\hline 17. $\frac{13}{2}^{+}$,$\frac{1}{4}$[N(2649)+ 3N(2739)] &
$\overline{N(2717)}$ & N(2700) &
625 & ** & 0.6 \\
\hline 18. $\frac{1}{2}^{-}$,
\begin{tabular} {l}
$\frac{1}{7}$[N(1839)+3N(1929)+N(2019) \\
\,\,\,\,\,+ N(2199)+ N(2379)] %
\end{tabular} & $\overline{N(2032)}$ & N(2090) &
300 & * & 2.8 \\
\hline 19. $\frac{3}{2}^{-}$,
\begin{tabular} {l}
$\frac{1}{12}$[6N(1929)+2N(2019)+ \\
\,\,\,\,2N(2199)+ 2N(2379)] %
\end{tabular} & $\overline{N(2064)}$ & N(2080) &
400 & ** & 0.8 \\
\hline 20. $\frac{5}{2}^{-}$,\begin{tabular} {l}
$\frac{1}{18}$[6N(1929)+3N(2019)+ \\
\,\,\,\,\,3N(2199)+ 3N(2379)+3N(2649)]  %
\end{tabular} & $\overline{N(2184)}$ & N(2200) &
364 & ** & 0.7 \\
\hline 21. $\frac{7}{2}^{-}$,
\begin{tabular} {l}
$\frac{1}{13}$[N(1839)+3N(1929)+3N(2019+ \\
\,\,\,\,\,3N(2199)+3N(2379)] %
\end{tabular} & $\overline{N(2109)}$ & N(2190) &
500 & **** & 3.7 \\

\hline 22. $\frac{9}{2}^{-}$,
\begin{tabular} {l}
$\frac{1}{8}$[2N(2019)+2N(2199)+  \\
\,\,\,\,\,2N(2379) + 2N(2649)]
\end{tabular}
& $\overline{N(2312)}$ &
N(2275) &
500 & **** & 1.6 \\
\hline 23. $\frac{11}{2}^{-}$, $\frac{1}{6}$ [ N(2379) + N(2649)+
4N(2739)]
 & $\overline{N(2664)}$ & N(2600) &
650 & *** & 1.1 \\
\hline 24. $\frac{13}{2}^{+}$, $\frac{1}{4}$ [ N(2649)+ 3N(2739)]
 & $\overline{N(2717)}$ & N(2700) &
625 & ** & 0.6 \\
\hline 25. $\frac{15}{2}^{+}$, N(3099) &
N(3099) & ? &  ?  &  ? & ? \\

\hline
\end{tabular}
\end{tabular}\\
\newline

The all quantum numbers (I, S, C, B, Q, J and P) of all deduced 23
N-baryons are completely the same as the experimental results. The
masses of 23 observed N-baryons agree more than average 98 percent
with the experimental results. The largest error of 15 established
N-baryons is only 3.7 percent. $\qquad $ $\qquad $ $\qquad $ $\qquad
$ $\qquad $\newline

\subsection{The comparison of the $\Lambda$-baryons}
$\qquad $ The deduced $\Lambda$-baryons are from Table 9, Table 12,
Table 13 and Table 14. The experimental $\Lambda$-baryons are from
\cite{Baryon}. The comparison between the deduced and the
experimental $\Lambda$-baryons appears in Table 19.$\newline$\\

\begin{tabular}{l}
$\qquad$$\qquad$$\qquad$$\qquad$$\qquad$ \ Table 19. \ The
comparison of the $\Lambda$-baryons\\
\begin{tabular}{|l|l|l|l|l|l|}
\hline Order, $J^{P}$, Compose & Theory & Exper.$\Lambda(M),
\Gamma$ & $\frac{\Delta M}{M}$ \\
\hline 1. $\frac{1}{2}^{+}$, $\Lambda(1119)$ & $\Lambda(1119)$ &
$\Lambda(1116)$ **** & 0.3  \\
\hline 2. $\frac{1}{2}^{-}$, $\frac{1}{2}$[$\Lambda(1299)$+
2$\Lambda(1414)$] &$\overline{\Lambda(1376)}$&
$\Lambda(1405)$,50**** & 2.1 \\
\hline 3. $\frac{3}{2}^{-}$, $\frac{1}{2}$[$\Lambda(1414)$+
$\Lambda(1659)$]&
$\overline{\Lambda(1537)}$ & $\Lambda(1520)$,16**** & 1.1 \\
\hline 4. $\frac{1}{2}^{-}$, $\Lambda(1659)$ & $\Lambda(1659)$ &
$\Lambda(1670)$,35**** & 0.7 \\
\hline 5. $\frac{3}{2}^{-}$,4$\Lambda(1659)$ &
$\overline{\Lambda(1659)}$&$\Lambda(1690)$,60**** & 1.8 \\
\hline 6. $\frac{1}{2}^{+}$, $\frac{1}{2}$[$\Lambda(1414)$+
2$\Lambda(1659)$]& $\overline{\Lambda(1577)}$ &
$\Lambda(1600)$,150*** & 1.4 \\
\hline 7. $\frac{1}{2}^{+}$, 2$\Lambda(1814)$ &
$\overline{\Lambda(1814)}$ &
$\Lambda(1810)$,150*** & 0.2 \\
\hline 8. $\frac{3}{2}^{+}$,
\begin{tabular} {l}
$\frac{1}{15}$[$\Lambda(1414)$ + 2$\Lambda(1659)$ +
2$\Lambda(1814)$\\
+ 9$\Lambda(1929)$ + $\Lambda(2019)$] %
\end{tabular} &
$\overline{\Lambda(1849)}$ & $\Lambda(1890)$,100**** & 2.2 \\
\hline 9. $\frac{5}{2}^{+}$, $\frac{1}{4}$[$\Lambda(1659)$ +
$\Lambda(1814)$ + 2$\Lambda(1929)$] & $\overline{\Lambda(1833)}$ &
$\Lambda(1820)$,80**** & 0.7 \\
\hline 10. $\frac{1}{2}^{-}$,  $\frac{1}{7}$[2$\Lambda(1659)$ +
2$\Lambda(1814)$+3$\Lambda(1929)$] & $\overline{\Lambda(1819)}$ &
$\Lambda(1800)$,300*** & 1.1 \\
\hline 11. $\frac{3}{2}^{-}$,
\begin{tabular} {l}
$\frac{1}{18}$[4$\Lambda(1814)$ + 7$\Lambda(1929)$ + \\
5$\Lambda(1954)$+2$\Lambda(2019)$]   %
\end{tabular} & $ \overline{\Lambda(1920)}$&
$\Lambda(2000)(?^?)$,125* & 4.0 \\
\hline 12. $\frac{5}{2}^{-}$,
\begin{tabular} {l}
$\frac{1}{18}$[4$\Lambda(1659)$+4$\Lambda(1814)$+  \\
6$\Lambda(1929)$+ 4$\Lambda(1954)$]   %
\end{tabular} & $\overline{\Lambda(1849)}$ &
$\Lambda(1830)$,95**** & 1.0 \\
\hline 13. $\frac{7}{2}^{-}$,
\begin{tabular} {l}
$\frac{1}{11}$[3$\Lambda(1929)$ + 2$\Lambda(1954)$ +\\
\,\,\,\,3$\Lambda(2019)$+3$\Lambda(2444)$]   %
\end{tabular} & $\overline{\Lambda(2099)}$ &
$\Lambda(2100)$,200**** & 0.0 \\
\hline 14. $\frac{3}{2}^{+}$, \,3$\Lambda(2379)$ &
$\Lambda(2379)$  & ? ? & ? \\
\hline 15. $\frac{5}{2}^{+}$,
\begin{tabular} {l}
$\frac{1}{12}$[7$\Lambda(1929)$ + 2$\Lambda(2019)$ \\
+3$\Lambda(2379)$]  %
\end{tabular} & $ \overline{\Lambda(2056)}$
& $\Lambda(2110)$,200*** & 2.6 \\
\hline 16. $\frac{7}{2}^{+}$, $\frac{1}{10}$[6$\Lambda(1929)$ +
2$\Lambda(2019)$ + 2$\Lambda(2379)$] & $\overline{\Lambda(2037)}$ &
$\Lambda(2020)$,140* & 0.8 \\

\hline 17. $\frac{9}{2}^{+}$,
$\frac{1}{2}$[$\Lambda(2379)$+$\Lambda(2444)$] &
$\overline{\Lambda(2412)}$ & $\Lambda(2350)$,150*** & 2.6 \\

\hline 18. $\frac{3}{2}^{-}$, 2$\Lambda(2444)$ &
$\overline{\Lambda(2444)}$ &
$\Lambda(2325)$,169* & 5.1 \\

\hline 19. $\frac{7}{2}^{+}$, $\frac{1}{6}$[2$\Lambda(2444)$ +
4$\Lambda(2649)$] & $\overline{\Lambda(2581)}$ &
$\Lambda(2585)$,225** & 0.2 \\
\hline 20. $\frac{9}{2}^{-}$, $\frac{1}{4}$[2$\Lambda(2444)$ +
2$\Lambda(2649)$] & $\overline{\Lambda(2547)}$ &
?, ?, ? & ? \\

\hline \hline
\end{tabular}%
\end{tabular}
\newline

The quantum numbers (I, S, C, B, Q, J and P) of all deduced
$\Lambda$-baryons are completely the same as the experimental
results. The masses of the 14 established (four or three stars)
$\Lambda$-baryons agree more than 97.8 percent with the experimental
results.

\subsection{The comparison of the $\Xi$-baryons}
$\qquad$From Table 10, Table 13 and Table 14, we can find the
deduced $\Xi(M)$-baryons and from Table B3, Table B5 and Table B6,
we can find $J^{p}$ value for each $\Xi(M)$. Using \cite{Baryon}, we
can find the experimental $\Xi(M)$$J^{p}$ -baryons. Then we
comparing the deduced $\Xi$ $J^{p}$-baryons with the experimental
$\Xi(M)$ $J^{p}$-baryons \cite {Baryon} as appearing in Table 20.

For many experimental $\Xi(M)$-baryons, the $J^{p}$ are uncertainly
and the widthes are very small. There are three possibility: The
first possible, there is a main $\Xi(m)$ with the largest number of
the same mass $\Xi(M)$-baryons; for example,
$\frac{1}{7}$[$\Xi(1299)$+4$\Xi(1479)$+2$\Xi(1659)$]$\frac{3}{2}^+$,
the 4$\Xi(1479)$ $\frac{3}{2}^+$ is the main $\Xi(m)$ with the
largest number 4 of the same mass $\Xi(M)$-baryons. The second
possible, there are two nearest different mass kinds of
$\Xi(M)$-baryons with many different $J^{P}$ values; for examples,
$\frac{1}{18}$[8$\Xi(1479)$+10$\Xi(1659)$],
$\frac{1}{5}$[3$\Xi(1659)$+2$\Xi(1839)$]
$\frac{1}{5}$[3$\Xi(2019)$+2$\Xi(1839)$] and
$\frac{1}{14}$[6$\Xi(2199)$+8$\Xi(2019)$]. Third possible, there are
many energy bands with the same mass but many different $J^{P}$
values; for example $\frac{1}{34}$[34$\Xi(2199)$] and
$\frac{1}{23}$[23$\Xi(2379)$] and $\frac{1}{23}$[23$\Xi(2559)$].

\begin{tabular}{l}
$\qquad$$\qquad$$\qquad$$\qquad$$\qquad$\ \ \ \ Table 20.\
The comparison of the $\Xi(M)$-baryons \\
\begin{tabular}{|l|l|l|l|l|l|l|l|l|l|l|}
\hline Order,Compose & Deduced $\Xi(M)$ &
Experiment$\Xi(M)$,$J^{P}$,$\Gamma$
& $\frac{\Delta M}{M}$ \\
\hline 1. $\Xi(1299)$ $\frac{1}{2}^{+}$ &
$\Xi(1299)$,$\frac{1}{2}^{+}$ & $\Xi(1318)$,$\frac{1}{2}^{+}$
**** & 1.5  \\
\hline 2. $\frac{1}{7}$[$\Xi(1299)$+4$\Xi(1479)$+2$\Xi(1659)$],
$\frac{3}{2}^{+}$ & $\overline{\Xi(1505)}$$\frac{3}{2}^{+}$ &
$\Xi(1532)$, $\frac{3}{2}^{+}$ 9.1 **** & 1.8 \\
\hline 3. $\frac{1}{18}$[8$\Xi(1479)$+10$\Xi(1659)$] &
$\overline{\Xi(1579)}$ &
$\Xi(1620)$,${?}^{?},27*$ & 2.5 \\
\hline 4. $\frac{1}{5}$[3$\Xi(1659)$+2$\Xi(1839)$]
& $\overline{\Xi(1731)}$ & $\Xi(1690)$,(?), 25*** & 2.4 \\
\hline 5. 2$\Xi(1839)$,$\frac{3}{2}^{-}$ &
$\Xi(1839)$)$\frac{3}{2}^{-}$ &
$\Xi(1820)$,$\frac{3}{2}^{-}$ ,24*** & 1.0  \\
\hline 6. $\frac{1}{5}$[3$\Xi(2019)$+2$\Xi(1839)$] &
$\overline{\Xi(1947)}$
& $\Xi(1950)$,(?),60** & 0.2  \\
\hline 7. $\frac{1}{12}$[12$\Xi(2019)$,$\ge\frac{5}{2}^{?}$] &
$\overline{\Xi(2019)}$  &
$\Xi(2030)$,$\ge\frac{5}{2}^{?}$,20*** & 0.5 \\
\hline 8. $\frac{1}{14}$[6$\Xi(2199)$+8$\Xi(2019)$] &
$\overline{\Xi(2096)}$
& $\Xi(2120)$,(?),20* & 0.8  \\
\hline 9. $\frac{1}{34}$[34$\Xi(2199)$] & $\overline{\Xi(2199)}$
& $\Xi(2250)$ ,(?),46** & 2.3  \\
\hline 10. $\frac{1}{23}$[23$\Xi(2379)$] & $\overline{\Xi(2379)}$
& $\Xi(2370)$ ,(?),80** & 0.4 \\
\hline 11. $\frac{1}{23}$[23$\Xi(2559)$] & $\overline{\Xi(2559)}$
& $\Xi(2500)$ ,(?), 105*  & 2.3 \\

\hline 12. $\frac{1}{6}$[6$\Xi(2649)$] & $\overline{\Xi(2649)}$
& $\Xi(?)$ ,(?), ?  & ? \\

\hline
\end{tabular}%
\end{tabular}\\

The all quantum numbers (I, S, C, B and Q ) of all deduced
$\Xi$-baryons are completely the same as the experimental results.
The all deduced masses agree more than average 98.5 percent with the
experimental results. The largest error of the established is only
2.4 percent.

\subsection{The comparison of $\Omega$-baryons}
$\qquad$Table 11 and Table 14b have given the deduced
$\Omega$-baryons, and \cite{Baryon} has already shown the
experimental $\Omega$-baryons. We compare the deduced and
experimental $\Omega$- baryons in Table 21: $\newline$

\begin{tabular}{l}

\ \ \ \ \ \ Table 21.\ The comparison of $\Omega$-baryons \\
\begin{tabular}{|l|l|l|l|l|l|l|l|}
\hline Theory & Experiment & $\frac{\Delta M}{M}$  \\
\hline $\Omega(1659)$,$\frac{3}{2}^{+}$ &
$\Omega(1672)$, $\frac{3}{2}^{+}$ ****& 0.8 \\
\hline $\Omega$(2214),$\frac{1}{2}^{-}$,$\frac{3}{2}^{-}$,
$\frac{5}{2}^{-}$,$\frac{7}{2}^{-}$& $\Omega$(2250), ? 55 ***& 1.6 \\
\hline $\Omega$(2394),$\frac{3}{2}^{+}$,$\frac{5}{2}^{+}$,
 & $\Omega(2380)$, ? ? ** & 1.6 \\
\hline $\Omega(2394)$,$\frac{7}{2}^{+}$,$\frac{9}{2}^{+}$, &
$\Omega(2470)$, ? 26 ** & 0.6 \\

\hline $\Omega(3819)$,$\frac{3}{2}^{+}$ &
?  ?  & ? \\

\hline
\end{tabular}%
\end{tabular}%
\newline

The deduced quantum numbers (I, S, C, B and Q) of the four
$\Omega$-baryons are completely the same with the corresponding
experimental results. The deduced masses of the four
$\Omega$-baryons agree with average more then 98.4 percenter
experimental masses.
\subsection{The comparison of charmed-baryons}
$\qquad$The $\Lambda_c(2264)\frac{1}{2}$ is from Table 8. In Table
12, the band with $M_{Fig}$ = 2559 and $\vec{n}$ = (-1,-1,0) has a
$\Lambda_c$ with $\frac{1}{2}^{-}$,$\frac{3}{2}^{-}$,
$\frac{5}{2}^{-}$ and $\frac{7}{2}^{-}$. The band with $M_{Fig}$ =
2559 and $\vec{n}$ = (0,0,-2) has  a $\Lambda_c$ with
2$\frac{1}{2}^{+}$,2$\frac{3}{2}^{+}$, $\frac{5}{2}^{+}$. From Table
7, they form $\Lambda_c$(2559),$\frac{1}{2}^{-}$; $\Lambda_c$(2674),
$\frac{3}{2}^{-}$; $\Lambda_c$(2789), $\frac{5}{2}^{-}$,
$\frac{7}{2}^{-}$ and $\Lambda_c$(2904),2$\frac{1}{2}^{+}$,
2$\frac{3}{2}^{+}$,$\frac{5}{2}^{+}$ as appearing in Table 22.

In Table 13, the band with $M_{Fig}$ = 2649 and $\vec{n}$ =
(21-1,12-1) has $\Xi_c$ with 2$\frac{1}{2}^{+}$,2$\frac{3}{2}^{+}$,
$\frac{5}{2}^{+}$. Using adjusted energy de in Table 7, we get
$\Xi_c$(2534),$\frac{1}{2}^{+}$; $\Xi'_c$(2534), $\frac{1}{2}^{+}$;
$\Xi_c$(2649),2$\frac{3}{2}^{+}$; $\Xi_c$(2649), $\frac{5}{2}^{+}$.
The band with $M_{Fig}$ = 2739 and $\vec{n}$ = (013,103) has
$\Xi_c(2739)$, $\frac{1}{2}^{-}$; $\Xi_c(2854)$, $\frac{3}{2}^{-}$;
$\Xi_c(2969)$, 2$\frac{5}{2}^{-}$, $\frac{7}{2}^{-}$ from de of the
F-axis in Table 7. The band with $M_{Fig}$ = 2739 and $\vec{n}$ =
(0-13,-103) has $\Xi_c$ with $\frac{3}{2}^{+}$,2$\frac{5}{2}^{+}$,
2$\frac{7}{2}^{+}$,$\frac{9}{2}^{+}$. From de in Table 7, we get
$\Xi(3084)$,$\frac{3}{2}^{+}$,2$\frac{5}{2}^{+}$,
2$\frac{7}{2}^{+}$,$\frac{9}{2}^{+}$. The deduced $\Xi_c$-baryons
appears in Table 22.

In Table 13, at $m_{Fig}$ = 2649, the sub-band with $\vec{n}$ =
(202,022,-121) has $\Xi_c$ with
$\frac{1}{2}^{+}$,2$\frac{3}{2}^{+}$, 2$\frac{5}{2}^{+}$. From de in
Table 7, we get $\Xi_c(2534)$,$\frac{1}{2}^{+}$;
$\Xi_c(2649)$,$\frac{3}{2}^{+}$; $\Xi_c(2764)$,$\frac{5}{2}^{+}$ as
appearing in Table 22.

In Table 14, at $M_{Fig}$ = 2559 the band with $\vec{n}$ = (2,2,0)
has $\Omega_c(2789)$, 2$\frac{1}{2}^{+}$, $\Omega_c(2904)$,
2$\frac{3}{2}^{+}$, $\Omega_c(3019)$,$\frac{5}{2}^{+}$ from de =
115($n_c$-S)$\Delta$S in Table 7. The $\Omega_c$-baryons appear in
Table 22.\newline\\

\begin{tabular}{l}
\ \ \ \ Table 22.\ The Comparison of charmed baryons\\
\begin{tabular}{|l|l|l|l|l|l|l|l|}
\hline $M_{Fig}$, $n_c$ & Theory & Experiment & $\frac{\Delta M}{M}$  \\
\hline 2379, 0 & $\Lambda_c$(2264), $\frac{1}{2}^{+}$ &
$\Lambda_c$(2286)$, \frac{1}{2}^{+}$****& 1.0 \\
\hline 2559, 0 & $\Lambda_c$(2559), $\frac{1}{2}^{-}$ &
$\Lambda_c$(2593), $\frac{1}{2}^{-}, 3.6,$*** & 1.3\\
\hline 2559, 1 & $\Lambda_c$(2674), $\frac{3}{2}^{-}$ &
$\Lambda_c$(2625), $\frac{3}{2}^{-}$, 1.9,*** & 1.8 \\
\hline 2559, 2 & $\Lambda_c$(2789),
 $\frac{5}{2}^{-}$,$\frac{1}{2}^{+}$ &
$\Lambda_c$(2765), ?, 50,* & 0.9 \\
\hline 2559, 3 & $\Lambda_c$(2904), $\frac{7}{2}^{-}$,
$\frac{3}{2}^{+}$, & $\Lambda_c$(2880), ? ** &
0.8 \\
\hline 2559, 4 & $\Lambda_c$(3019), $\frac{5}{2}^{+}$, &
$?^{?}$, $?^{?}$  & ? \\
\hline

\hline 2649, 0 & $\Xi_c$(2534), $\frac{1}{2}^{+}$ &
$\Xi_c$(2470), $\frac{1}{2}^{+}$***& 2.6   \\
\hline 2649, 0 & $\Xi^{\prime}_c$(2534), $\frac{1}{2}^{+}$&
$\Xi^{\prime}_c$(2577), $\frac{1}{2}^{+}$,*** & 1.7  \\
\hline 2649, 1 & $\Xi_c$(2649), $\frac{3}{2}^{+}$ &
$\Xi_c$(2645), $\frac{3}{2}^{+}$, $\le$ 4.6 ***& 0.2  \\
\hline 2649, 2 & $\Xi_c$(2764), $\frac{5}{2}^{+}$ &
$?^{?}$,   $?^{?}$ & ?  \\

\hline 2739, 1 & $\Xi_c$(2739), $\frac{1}{2}^{-}$ &
$\Xi_c$(2790), $\frac{1}{2}^{-}$, $\le$ 14, *** & 1.8  \\
\hline 2739, 2 & $\Xi_c$(2854), $\frac{3}{2}^{-}$ &
$\Xi_c$(2815)$, \frac{3}{2}^{-}$, $\le$ 5, *** & 1.4 \\
\hline 2739, 3 & $\Xi_c$(2969), $\frac{5}{2}^{-}$,
$\frac{7}{2}^{-}$ & ?,    $?^{?}$  & ? \\
\hline 2739, 4 & $\Xi_c$(3084), $\frac{3}{2}^{+}$,$\frac{5}{2}^{+}$,
$\frac{7}{2}^{+}$,$\frac{9}{2}^{+}$ & $?^{?}$, $?^{?}$  & ? \\
\hline

\hline 2649, 0 & $\Sigma_c$(2534), $\frac{1}{2}^{+}$ &
$\Sigma_c$(2455), $\frac{1}{2}^{+}$, $\sim$2.5, **** & 3.2 \\
\hline 2649, 1 & $\Sigma_c$(2649), $\frac{3}{2}^{+}$ &
$\Sigma_c$(2520), $\frac{3}{2}^{+}$, $\sim$ 16 ***& 5.1\\
\hline 2649, 2 & $\Sigma_c$(2764), $\frac{5}{2}^{+}$ &
$\Sigma_c$(2800), ? , $\sim$ 66, *** & 1.3 \\
\hline

\hline $\Omega_c$ & $\Omega_c$(2789), $\frac{1}{2}^{+}$ &
$\Omega_c$(2698), $\frac{1}{2}^{+}$ ***& 3.4 \\
\hline $\Omega_c$ & $\Omega_c$(2904), $\frac{3}{2}^{+}$ &
$?^{?}$,  $?^{?}$ & ? \\
\hline $\Omega_c$ & $\Omega_c$(3019), $\frac{5}{2}^{+}$ &
$?^{?}$,  $?^{?}$, & ? \\

\hline
\end{tabular}%
\end{tabular}%
\newline \\

The $\Sigma_c$-baryons are from Table 13 also. The
$\Omega_c$(2789)-baryons are from Table 14. The experimental
$\Lambda_c$-baryons, $\Xi_c$-baryons, $\Sigma_c$-baryons,
$\Omega_c$-baryons and $\Lambda_b$-baryon are all found from
\cite{Baryon}. We show the comparisons of these deduced baryons with
the experimental baryons in Table 22.

The $J^{P}$ values of the charmed-baryons have not been measured by
experiment, we will show the $J^{P}$ values of the quark model in
Table 22.\newline \\

Table 22 shows that the all quantum numbers (I, S, C, Q) of all 14
deduced baryons (5 $\Lambda_c$, 5 $\Xi_c$, 3 $\Sigma_c$ and one
$\Omega_c$) are completely the same with the corresponding
experimental results. The deduced masses of the $\Lambda_c$-baryons
agree average more then 98.8 percent. The deduced masses of the
$\Xi_c$-baryons consistent with the corresponding experimental
results more then average 98.4 percent. The deduced masses of the
$\Sigma_c$-baryons agree average more then 96.8 percenter the
experimental masses. The deduced mass of the $\Omega_c$-baryon
consists with the experimental result 96,6 percent.

\subsection{In Summary}

$\qquad$Table 15 shows that the all deduced quantum numbers (I, S,
C, B, Q) of the seven most important baryons (N(939),
$\Lambda(1119)$, $\Sigma(1209)$, $\Xi(1299)$, $\Omega$(1659),
$\Lambda_c(2264)$ and $\Lambda_b(5554)$) are exactly the same as the
experimental results. Their masses are consistent with experimental
results more than 98.5 percent using only three constant parameters
(M$_0$ =939 Mev, $\alpha$ = 360 Mev and $\beta$ = 115 Mev ).

The all deduced quantum numbers (I, S, C, B, Q, J and P) of all
observed 116 baryons are completely the same as corresponding
experimental result. The masses of the 68 established baryons (10
$\Delta$-baryons, nine $\Sigma$-baryons, 15 N-baryons, 14
$\Lambda$-baryons, five $\Xi$-baryons, two $\Omega$-baryons, three
$\Lambda_c$-baryons, three $\Sigma_c$-baryons, five $\Xi_c$-baryons,
one $\Omega_C$-baryon and one $\Lambda_b$-baryon) agree with more
than the average 98.3 percenter with the experimental results as
appearing in Table 23. The masses of all observed 116 baryons agree
with more than the average 98 percenter with the experimental
results as appearing in Table 23.
\\$\newline$

\begin{tabular}{l}
Table 23.\ The Comparison of the Deduced Masses with the
Experimental Masses of Baryons \\
\begin{tabular}{|l|l|l|l|l|l|l|l|}
\hline Baryon & Established &  Average of $\Delta M$ &
Max. $\Delta$M & Total number & Total Average \\
\hline $\Delta(M)$ & 10 & 2.19 & 5.4 & 22 & 2.44 \\
\hline $\Sigma(M)$ & 9 & 1.23 & 3.5 & 23 & 1.77 \\
\hline N(M)        & 15 & 1.87 & 3.7 & 23 & 1.81 \\
\hline $\Lambda(M)$ & 14 & 1.27 & 2.6 & 18 & 1.55 \\
\hline $\Xi(M)$ & 5 & 1.44 & 2.4 & 11 & 1.43 \\
\hline $\Omega(M)$ & 2 & 1.20 & 1.6 & 4 & 1.15  \\
\hline $\Lambda_c(M)$ & 3 & 1.37 & 1.8 & 5 & 1.16  \\
\hline $\Xi_c(M)$ & 5 & 1.54 & 2.6 & 5 & 1.54   \\
\hline $\Sigma_c(M)$ & 3 & 3.2 & 5.1 & 3 & 3.2  \\
\hline $\Omega_c(M)$ & 1 & 3.4 & 3.4 & 1 & 3.4  \\
\hline $\Lambda_b(M)$ & 1 & 1.21 & 1.21 & 1 &1.21  \\
\hline All Baryons & 68 & 1.68 & 5.4 & 116 & 1.82  \\
\hline
\end{tabular}%
\end{tabular}

Inside Table 23, the term ``Established" means the numbers of the
experimental established baryons. The term``Max. $\Delta$M" is the
largest $\Delta$M of all established baryons belong to one kind. The
term ``Total Number" is the numbers of the all observed baryons. The
term ``Total average" is the average of $\frac{\Delta\,M}{M}$ of all
observed baryons. $\newline$

All 116 observed baryons(22 $\Delta$-baryons, 23 $\Sigma$-baryons,
23 N-baryons, 18 $\Lambda$-baryons, 11 $\Xi$-baryons, four
$\Omega$-baryons, five $\Lambda_c$-baryons, three
$\Sigma_c$-baryons, five $\Xi_c$-baryons, one $\Omega_C$-baryon and
one $\Lambda_b$-baryon) have been deduced. All deduced baryons with
M $\le$ 2019 Mev have been found in experiments (see Table 15-22)
except the $\Lambda$-baryons with J =$\frac{1}{2}$. All deduced 116
baryons have completely the same quantum numbers (I, S, C, B, Q, J
and P) as the corresponding experimental baryons. The deduced masses
of the 116 baryons agree average more than 98 percent with
experimental result. These results are the results of the expanded
Schrodinger equation and BCCP condition using only four constant
parameters ($\alpha$ = 360 Mev, $M_0$ = 939 Mev, $\beta$ = 115 Mev
and f = 0 for inside axis or f = 1 for the axes on the surface of
the first Brillouin zone Fig. 1). There is not any established
baryon outside the deduced baryons. Considering the large widthes of
the baryons, we have to say the deduced baryon masses consistent
well with the experimental results.

\section{Deducing the Masses of the Quarks from the Masses of
the Most Important Deduced Baryons}

$\qquad$In the first postulate, we assumed that the five quarks have
the same unknown large bare mass $m_q$. When a baryon excite from
the vacuum state into an energy band, it becomes an observable
baryon with the same quantum numbers as in the vacuum state as
appearing in Table 2 or 3, but different physical masses from its
bare mass. The three quarks inside the baryons are excited from
vacuum state into the observable physical state with the same
quantum numbers as in the vacuum (Table 14.1 of \cite{Quarks}), but
different observable masses from the bare mass $m_q$. The mass of a
quark inside a observable hadron relating to the laboratory
reference system is the ``quark mass." The ``quark masses" are
different values for different quarks since they are inside
different baryons with different energies.

Estimating the ``quark masses" is a difficult job since all quarks
are confined inside hadrons. We can estimate the quark masses from
the masses of the baryons composed of the quarks. In order to
distinguish the masses of u-quark with d-quark, we use the measured
masses of proton and newtron. For the five most important baryons
(proton, neutron, $\Lambda(1119)$, $\Lambda_c(2264)$ and
$\Lambda_b(5554)$), there are five quark mass equations: $\newline$

\begin{equation}
m_u + m_d +m_u + E_{bin}(uud) = 938,  \label{m939}
\end{equation}
\begin{equation}
m_u +m_d +m_d + E_{bin}(udd) = 940,  \label{m939'}
\end{equation}
\begin{equation}
m_u +m_d +m_s + E_{bin}(uds) = 1119,  \label{m1119}
\end{equation}
\begin{equation}
m_u +m_d +m_c + E_{bin}(udc) = 2264,  \label{m2379}
\end{equation}
\begin{equation}
m_u +m_d +m_b + E_{bin}(udb) = 5554.  \label{m5539}
\end{equation}
Where $E_{bin}(uud)$ is the binding energy of three quarks inside
the proton, $E_{bin}(udd)$ is the binding energy of three quarks
inside the neutron, $E_{bin}(uds)$ is the binding energy of three
quarks inside $\Lambda(1119)$, $ \ldots$ . Using the perfect
$SU(5)_f$ symmetry, we can have

\begin{equation}
E_{bin}(uud)\,=\,E_{bin}(udd)\,=\,E_{bin}(uds)\,=\,
E_{bin}(udc)\,=\,E_{bin}(udb).  \label{BindE}
\end{equation}

From the above six equations (\ref{m939})--(\ref{BindE}), we can get

\begin{equation}
m_d - m_u = 2 , m_s-m_u = 181, m_c-m_u = 1326, m_b-m_u = 4616.
\label{Dm}
\end{equation}

Assuming $E_{bin}$(uud)\,=\,$E_{bin}$(udd)\,=\,$E_{bin}$(uds)\,=
$\ldots$ =0 (it is impossible!), we can obtain the masses of the
quarks:

\begin{equation}
m_u = 312,\,\,\, m_d\,= 314 ,\,\,\, m_s = 493 \,\,\, m_c = 1638,
\,\,\,m_b = 4928.  \label{differnces of m}
\end{equation}

Since the baryons have never been divided into free quarks, the
binding energies $E_{bin}(uud)$, $E_{bin}(udd)$, $E_{bin}(uds)$,
$E_{bin}(udc)$ and $E_{bin}(udb)$ shall be very big. If making
strong stable baryons, the masses of quarks will be much larger than
the values as appearing in Table 24: $\newline$

\begin{tabular}{l}
\ \ Table 24.\ The quark masses for various possible binding
energies \\
\begin{tabular}{|l|l|l|l|l|l|l|l|}
\hline $E_{bind}(Mev)$ & $m_u$ & $m_d$ & $m_s$ & $m_c$ &$m_b$ \\
\hline 0 & 312 & 314 & 493 & 1638 & 4928 \\
\hline 30000 & 10312 & 10314 & 10493 & 11638 & 14928 \\
\hline 300000 & 100312 & 100314 & 100493 & 101638 & 104928 \\
\hline 3000000 & 1000312 & 1000314 & 1000493 & 1001638 & 1004928 \\
\hline 30000000 & 10000312 & 10000314 & 10000493 & 10001638 &
10004928  \\
\hline
\end{tabular}%
\end{tabular}%
$\newline$

Historically, the first determinations of quark masses were
performed using quark models. The resulting masses only make sense
in the limited context of a particular quark model, and cannot be
related to the quark mass parameters of the standard Model.
Similarly, these masses of the quarks are deduced using the quark
model also. They are in the limited context of the quark model. For
the quark model with binding energy $E_{bind}$ = 300000 Mev, $m_d$ =
100312, $m_u$ = 100314, $m_s$ = 100493, $m_c$ = 101653 and $m_b$ =
104913. The deduced quark masses are really different as formula
(\ref{Dm}). The mass differences of the five quarks are independent
from the binding energies. They are very small constants. Comparing
with the huge quark masses, they can neglect. The masses of the
quarks are almost the same for the five quarks (see Table 24) since
the binding energies very big to confine the quarks. Thus the SU(3),
SU(4) and SU(5) are really perfect symmetry groups.

Using the deduced masses in Table 23 (laboratory system) of the
quarks, we can deduced the quark mass parameters of the standard
model\cite{Mass parameter}. The mass parameters describe the
scatting and the decays of the quarks. Since they are measured in
mass center inside hadrons, if subtracting the binding energies
$E_{bind}$ and the energy differences between the mass center system
and the laboratory system from the quark masses, we will get the
mass parameters of the quarks:
\begin{equation}
m(q)_{par.}\,= \,m_q\,-\,E_{bind} \,-\,DE\, \label{m-para}
\end{equation}
where DE phenomenologically = $\frac{1}{3}M_p$ -
2.7\,+\,100$|S+C+B|$ = 310 + 100$|S+C+B|$. Using (\ref{m-para}), we
find

\begin{equation}
m(u)_{par.}\,= \,2,\,\,\,\,m(d)_{par.}\,=
\,4,\,\,\,\,m(c)_{par.}\,=\, 83,\,\,\,\, m(c)_{par.}\,=
\,1233,\,\,\,\, m(b)_{par.}\,=\,4503.\label{m(q)para}
\end{equation}

Comparing the deduced mass parameters in (\ref{m(q)para}) with
experimental results in \cite{Mass parameter}, we find the deduced
mass parameters of the standard model agree well with experimental results:\\

\begin{tabular}{l}
\ \ Table 25.\ The comparison of the quark mass parameters \\
\begin{tabular}{|l|l|l|l|l|l|l|l|}
\hline Mass Parameter & $m(u)_P$ & $m(d)_P$ & $m(s)_P$ &
$m(c)_P$ &$m(b)_P$ \\
\hline Deduced M-Par. & 2 & 4 & 83 & 1233 & 4503 \\
\hline Measured M-Par. & 1.5 to 3.0 & 3 to 7 & 95$\pm$25 &
1250 $\pm$ 90 &
\begin{tabular}{l}
4200 $\pm$ 70 \\
4700 $\pm$ 70 %
\end{tabular} \\

\hline
\end{tabular}%
\end{tabular}%
$\newline$

Table 25 and formula (\ref{m(q)para}) not only use new method to
deduce the mass parameters, but also give the relationship between
the masses and the mass parameters of the quarks. Using the
relationship, we can deduce the masses of quarks from the
corresponding mass parameters and we can deduce the quark mass
parameters from the corresponding quark masses.

\section{Discussion}
$\qquad$1. Although the bare mass $M_b$ of the b-particle is
unknown, we have the constant $\alpha$ = $\frac{h^2}{2M_{b}a^{2}}$ =
360 Mev from (\ref{alpha}) and (\ref{360 Mev}). Thus we can get
\begin{equation}
\\
M_b = \frac{h^2}{2\alpha a^2}.
\end{equation}%
\newline
Here the periodic constant a of the body-centred cubic periodic
symmetry condition, a $\sim$ $10^{-23}$ m. The bare mass of the
b-particle will be
\begin{equation}
\\
M_b\,\,\, \sim \,\,\,\frac{h^2}{2\alpha a^2} \sim 2.14 \times
10^{19} (Mev)\gg M_p = 938 (Mev).
\end{equation}%
\newline
The estimated bare masses $M_b$ $\sim$ 2.14 $\times 10^{19}$
(Mev)$\gg M_p$. Thus we can use the Schrodinger equation
(\ref{b-particle}) instead of the Dirac equation to deduce the
baryon spectrum. In fact the results of this paper in Table 15-Table
22 also show that this is a very good approximation.

2. The experimental facts of the quark confinement and the stability
of the proton (proton p with $\tau$ about $10^{30}$ years)
inevitably result in huge binding energies of the three quarks in
baryons. According to the E = M$C^{2}$, the huge binding energies
and the small masses of baryons inevitably lead to huge quark
masses. The huge masses of quarks and small mass differences of
baryons inevitably bring about that differences of the quark masses
can be neglected as appearing in Table 24. Thus, the new
experimental facts have already shown clearly that the $SU(N)_f$ (N
= 3, 4 and 5) symmetry groups are perfect since the five quark
masses are essentially the same.

3. In the deduction of quantum numbers, this paper does not make any
approximation. The degeneracy d of the energy bands, the rotary fold
number R of the symmetry axis and the index number $\vec{n}$ of the
energy bands are exactly the results of the Schrodinger equation
with BCCP boundary condition. The isospins of the baryons are from
the energy band degeneracy d (\ref {IsoSpin}); the strange numbers
of the baryons are from the R of the symmetry axes (\ref{Strange
Number}); the charmed number and bottom number are from the fold
numbers R and the energy band index number $\vec{n}$ (\ref{Charmed})
and (\ref{Bottom}). Thus the strange number, the charmed number and
the bottom number of the energy bands are the products of the BCCP
symmetry and the expanded Schrodinger equation. In fact, the quantum
numbers originate from the body-centred cubic periodic symmetries.

4. Although there are infinite high energy bands in theory, this
case means there will be infinite high baryons in theory. In fact,
however, since the density of the energy bands is higher and higher
(the distances between energy bands is smaller and smaller) as the
energy increases and the experimental width of the baryons are
larger and larger as the energy grows, so that many energy bands
will mix together to form the physical back ground in the higher
energy case. For example, for N-baryons, at M = 2200 Mev, N(2190)
$\Gamma$ = 500 Mev, N(2220) $\Gamma$ = 400 Mev and N(2250) $\Gamma$
= 500 Mev; but the distances between the energy bands are about
180--360 Mev in the theory. Thus there are the upper limits of
various observable baryons that are dependent on the experimental
technique and equipments. As improvements of the experimental
technique and equipments occur, many new baryons might be discovered
in the future.

5. This paper can help us to understand the instability of the
baryons since all energy bands (baryons) are not the eigenstates of
angular momentum and all baryons are the excited states of the same
b-particle. The higher excited states decay into lower energy states
and the energy excited states decay into the ground state is a
natural law in physics. Thus the baryons do so.

6. How about the nature at distance scale larger than $10^{-18}m$?
Since the distance scale is larger than the periodic constant ``a"
($\sim$ $10^{-23}$) by more than $10^5$ times, physicists using the
nature at distance scale larger than $10^{-18}m$ will not see the
body-centred cubic symmetries of the local spaces around baryons.
Thus there is not the BCCP symmetry in their physical theory--the
standard model.

\section{Conclusions}

$\qquad$1. \textbf{There is a new symmetry beyond the standard
model-the body-centred cubic periodic symmetry (BCCP)} with a
\textbf{periodic constant ``a" $\sim$ $10^{-23}$m} in the local
space around baryons. The masses and the space quantum numbers (L, J
and P) of baryons result from the BCCP symmetries.

2. The $SU(N)_{f}$ (N = 3, 4 and 5) symmetry groups not only really
exist, but also \textbf{they are perfect}. The $SU(N)_{f}$ groups
give the intrinsic quantum numbers (I, S, C, B, Q and spin s) of the
baryons.

3. \textbf{The free b-particle expanded Schrodinger equation with
large bare mass $M_{b}$ is a new motion equation beyond the standard
model. It is available to the distance scale $\sim$ $10^{-23}m$ and
BCCP condition. Using it and the BCCP symmetry condition, this paper
deduces the full (including M, I, S, C, B, Q, J and P) baryon
spectrum (116 observed baryons, 68 established baryons) with only
four constant parameters ($M_0$ = 939 Mev, $\alpha$ = 360 Mev,
$\beta$ = 115 Mev and f = 1 or 0). The deduced quantum numbers are
consistent with experimental results completely; the deduced masses
agree with the experimental results average more than average 98
percent the experimental results.}

4. This paper has deduced the masses of the five quarks (u, d, s, c
and b), from the masses of the most important deduced baryons, in
terms of the qqq baryon model of the quark model. The deduced quark
masses are huge. The huge masses not only provide a necessary
condition for the quark confinement, but also provide a solid
physical foundation of perfect $SU_{f}(N)$ symmetry. The huge mass
quarks stabilize the proton with the lifetime $10^{30}$ years. This
paper also deduces the mass parameters (of the standard model) which
agree with the measured results as appearing in Table 25.

5. The experimental large width of baryons (special N-baryons)
result from the several same kind of baryons with the same I, S, C,
B, Q, J and P as well as similar masses (see Table 18).\\

\section{Prediction--``Zeeman Effect'' of Baryons}
$\qquad$There are many baryons with large widthes ($\Gamma$) as 300
Mev, 400 Mev, 500 Mev and 650 Mev. Table 18 shows that the big
widthes of the N-baryons result from the composition of several
nearby energy bands with the same I, S, C, B, Q, J and P. There are
several baryons that are the composition baryons such as
$\Delta$-baryons in Table 16, $\Sigma$-baryons in Table 17,
N-baryons in Table 18 and $\Lambda$-baryons in Table 19. This
phenomena we call the``Particle Zeeman Effect.'' For example, the
baryon N(2190) $\frac{7}{2}^-$ with width $\Gamma$ = 500(Mev)**** is
composed of [N(1839) + 4N(1929) +3N(2019) +3N(2199) + 3N(2379) +
2N(2649)]$\frac{7}{2}^-$ $\to$ $\overline{N(2165)}$$\frac{7}{2}^-$
with $\Gamma$ $\ge$ [(2379-1929)] = 450 Mev. There are 16 energy
bands with fluctuation in a region about 500 Mev. They are
difficultly separated by experiments. For other example, N(1720)
$\frac{3}{2}^{+}$ with $\Gamma$ = 200 Mev and ****, however,
composed of only two energy bans: $\frac{1}{2}$[N(1659)+N(1839)]
$\frac{3}{2}^{+}$ = $\overline{N(1749)}$,$\frac{3}{2}^{+}$ with
$\Gamma$ = 1839 - 1659 = 180 (Mev). Thus we suggest that using
experiments to separate the N(1720)-baryon $\frac{3}{2}^{+}$, we
will observe one N(1659) $\frac{3}{2}^{+}$ and one
N(1839)$\frac{3}{2}^{+}$. If we actually get the two N-baryons with
$J^{P}$ = $\frac{3}{2}^{+}$, this will show that the BCCP symmetry
really exists.

\begin{center}
\bigskip \textbf{Acknowledgments}
\end{center}

I sincerely thank Professor Robert L. Anderson for his valuable
advice. I acknowledge\textbf{\ }my indebtedness to Professor D. P.
Landau for his help also. I would like to express my heartfelt
gratitude to Dr. Xin Yu. I sincerely thank Professor Kang-Jie Shi
for his important advice. I wholeheartedly thank Professor Han-yin
Guo, Professor Zhan Xu, Professor Xing-chang Song for their useful
advice.\ \ \ \newline \ \ \

\appendix

\section{The Irreducible Representations and Their Compatibility Relations}

\begin{tabular}{l}
\\
Table A1.\ The irreducible representations and basis
functions of $O_h$ \\
\begin{tabular}{|l|l|l|l|l|l|l|l|l}
\hline $O_h$ & basis function & Deg. & L \\
\hline $\Gamma_1,H_1,M_1$ & 1 & 1 & 0  \\
\hline $\Gamma_2,H_2,M_2$ & $x^4(y^2-z^2)+y^4(z^2-x^2)+
z^4(x^2-y^2)$ & 1 & 6  \\
\hline $\Gamma_{12},H_{12},M_{12}$ & $z^2-\frac{1}{2}(x^2+y^2)$,
$(x^2-y^2)$ & 2 & 2 \\
\hline $\Gamma_{15},H_{15},M_{15}$ & (x, y, z) & 3 & 1 \\
\hline $\Gamma_{25},H_{25},M_{25}$ & $x(y^2-z^2), y(z^2-x^2),
z(x^2-y^2)$ & 3 & 3  \\
\hline $\Gamma_1^{\prime},H_1^{\prime},M_1^{\prime}$ &
xyz[$x^4(y^2-z^2)$+$y^4(z^2-x^2)$+ $z^4(x^2-y^2)$] & 1 & 9  \\
\hline$\Gamma_2^{\prime},H_2^{\prime},M_2^{\prime}$ & xyz &
1 & 3 \\
\hline $\Gamma_{12^{\prime}},H_{12^{\prime}},M_{12^{\prime}}$ &
(xyz[$z^2-\frac{1}{2}(x^2+y^2)$], xyz$(x^2-y^2)$ & 2 & 5 \\
\hline $\Gamma_{15^{\prime}},H_{15^{\prime}},M_{15^{\prime}}$ &
$xy(x^2-y^2),yz(y^2-z^2),zx(z^2-x^2)$ & 3 & 4 \\
\hline $\Gamma_{25^{\prime}},H_{25^{\prime}},M_{25^{\prime}}$ &
(xy,yz, zx) & 3 & 2 \\
\hline\hline
\end{tabular}%
\end{tabular}%
\newline
$\qquad $Table A1 is copied from Table IV of \cite{Callaway}.
\newline

The wave functions at the point P have the group p symmetry.
Irreducible representations and basis functions appear in Table A2.
The group of P is the tetrahedral group. It is interesting to see
the way in which the representations at $\Gamma$ combine to give
those at P: $P_1$ contains functions belonging to $\Gamma_1$ and
$\Gamma_2'$; $P_2$ contains those belonging to $\Gamma _2$ and
$\Gamma_1'$; $P_3$ contains $\Gamma_{12}$ and $\Gamma_{12}$'; $P_4$,
contains $\Gamma_{15}$ and $\Gamma_{25}'$; $P_5$ contains
$\Gamma_{25}$ and $\Gamma_{15}'$ \cite{Callaway}.

\begin{tabular}{l}
\\
Table A2.\ The irreducible representations and basis
Functions of the Group P \\
\begin{tabular}{|l|l|l|l|l|l|l|l|l}
\hline $T_d$ & Basis function & Deg. & L \\
\hline $P_1$ & 1, xyz & 1 & 0, 3  \\
\hline $P_2$ & $x^4(y^2-z^2)+y^4(z^2-x^2)+z^4(x^2-y^2)$ & 1 & 6\\
\hline $P_3$ & $z^2-\frac{1}{2}(x^2+y^2)$,$(x^2-y^2)$;
(xyz[$z^2-\frac{1}{2}(x^2+y^2)$], xyz$(x^2-y^2)$ & 2 & 2,
5\\
\hline $P_4$ & x, y, z; xy, yz, zx & 3 & 1, 2 \\
\hline $P_5$ &
\begin{tabular}{l}
$x(y^2-z^2), y(z^2-x^2), z(x^2-y^2)$ %
\end{tabular} & 3 & 3 \\
\hline\hline
\end{tabular}%
\end{tabular}%
\newline
$\qquad $ Table A2 is copied from Table V of \cite{Callaway}.
\newline\\
The wave functions at the point N have the symmetries of the group
N\\
\begin{tabular}{l}
\\
Table A3.\ The irreducible representations and basis functions
of the Group N \\
\begin{tabular}{|l|l|l|l|l|l|l|l|l|l|l|l|}
\hline $D_{2h}$ & E & $C_4$ & $C_1$ & $C_2$ & J & J$C_4$ & J$C_1$ &
J$C_2$ & B. function & d & L \\
\hline $N_1$ & 1 & 1 & 1 & 1 & 1 & 1 & 1 & 1 & 1, xy &
1 & 0,2 \\
\hline $N_2$ & 1 & -1 & 1 & -1 & 1 & -1 & 1 & -1 & z(x-y) & 1 &
2\\
\hline $N_3$ & 1 & -1 & -1 & 1 & 1 & -1 & -1 & 1 & z(x+y) & 1
& 2 \\
\hline $N_4$ & 1 & 1 & -1 & -1 & 1 & 1 & -1 & -1 & $x^2-y^2$ & 1 &
2\\
\hline $N^{\prime}_1$ & 1 & -1 & 1 & -1 & -1 & 1 & -1 & 1 & x+y & 1
& 1 \\
\hline $N^{\prime}_2$ & 1 & 1 & 1 & 1 & -1 & -1 & -1 & -1 & z
$(x^2-y^2)$ & 1 & 2 \\
\hline $N^{\prime}_3$ & 1 & 1 & -1 & -1 & -1
& -1 & 1 & 1 & z & 1 & 1\\
\hline $N^{\prime}_4$ & 1 & -1 & -1 & 1 & -1 & 1 & 1 & -1 & x-y &
1 & 1\\
\hline\hline
\end{tabular}%
\end{tabular}%
\newline
$\qquad $Table A3 is copied from Table VI of \cite{Callaway} and
Table 29 of \cite{Energy Band}.\\
\begin{tabular}{l}
\\
Table A4.\ The irreducible representations and basis functions of
the group $\Delta$ \\
\begin{tabular}{|l|l|l|l|l|l|l|l|l|}
\hline $C_{4v}$ & E & C$_4^2$ & C$_4$ & JC$_4^2$ & J$C_2$ & B.
function & Deg. & L \\
\hline $\Delta_1$ & 1 & 1 & 1 & 1 & 1 & 1 & 1& 0 \\
\hline $\Delta_2$ & 1 & 1 & -1 & 1 & -1 & $(x^2-y^2)$ & 1 & 2 \\
\hline $\Delta^{\prime}_2$ & 1 & 1 & -1 & -1 & 1 & xy & 1 & 2 \\
\hline $\Delta^{\prime}_1$ & 1 & 1 & 1 & -1 & -1 & xy$(x^2-y^2)$ & 1
& 4 \\
\hline $\Delta_5$ & 2 & -2 & 0 & 0 & 0 & (x, y) & 1 & 1 \\
\hline\hline
\end{tabular}%
\end{tabular}%
\newline
\, \, \, \, \, \, Table A4 is copied from Table 22 of \cite{Energy
Band}.
\newline\\
\begin{tabular}{l}
\\
Table A5.\ The irreducible representations and basis functions of
the Group $ \Lambda$ \\
\begin{tabular}{|l|l|l|l|l|l|l|l|l|l|}
\hline $C_{3v}$ & E & 2C$_3$ & 3$\sigma_v$ & B. function &
Deg. & L \\
\hline $\Lambda_1 or F_1$ & 1 & 1 & 1 & 1 & 1 & 0 \\
\hline $\Lambda_2 or F_2$ & 1 & 1 & -1 & xy(x-y), yz(y-z),
zx(z-x) & 1 & 4 \\
\hline $\Lambda_3 or F_3$ & 2 & -1 & 0 & [(x - z), (y-z)] &
1 & 1 \\
\hline\hline
\end{tabular}%
\end{tabular}%
\newline
$\qquad $ Table A5 is copied from Table 23 of \cite{Energy Band} and
Table 7.7 of \cite{Joshi}. \newline

\begin{tabular}{l}
\\
Table A6.\ The Irreducible Representations and Basis Functions
of the Groups $\Sigma$, D and G \\
\begin{tabular}{|l|l|l|l|l|l|l|l|l|l|}
\hline $C_{2v}$ & E & C$_2$ & J$C^2_4$ & J$C_2$ & B.
function & Deg. & L \\
\hline $\Sigma_1,D_1,G_1$ & 1 & 1 & 1 & 1 & 1 & 1 & 0 \\
\hline $\Sigma_2,D_2,G_2$ & 1 & 1 & -1 & -1 & z(x-y) & 1 & 2 \\
\hline $\Sigma_3,D_3,G_3$ & 1 & -1 & -1 & 1 & z & 1 & 1 \\
\hline $\Sigma_4,D_4,G_4$ & 1 & -1 & 1 & -1 & x-y & 1 & 1 \\
\hline\hline
\end{tabular}%
\end{tabular}%
\newline
$\qquad $Table A6 is copied from Table 24 of \cite{Energy Band} and
Table
7.7 of \cite {Joshi}. \newline\\
The compatibility relations between states along symmetry axes and
the states at the end-points appear in the following tables:
\newline\\
\begin{tabular}{l}
\\
Table A7. The compatibility relations of the $\Gamma $-$\Delta
$, $\Gamma $-$\Lambda $ and $\Gamma $-$\Sigma $ \\
\begin{tabular}{|l|l|l|l|l|l|l|l|l|l|l|}
\hline $\Gamma$-Point($O_h$) & $\Gamma _{1}$ & $\Gamma _{2}$ &
$\Gamma _{12}$ & $\Gamma _{15}^{^{\prime }}$ &
$\Gamma _{25}^{^{\prime }}$ \\
\hline $\Delta-Axis(C_{4v})$ & $\Delta _{1}$ & $\Delta _{2}$ &
$\Delta _{1} \Delta _{2}$ & $\Delta _{1}^{^{,}}\Delta _{5}$ &
$\Delta _{2}^{^{,}}\Delta _{5}$ \\
\hline $\Lambda-Axis(C_{3v})$ & $\Lambda _{1}$ & $\Lambda _{2}$ &
 $\Lambda _{3}$ & $\Lambda _{2}\Lambda _{3}$ & $\Lambda _{1}
 \Lambda _{3}$ \\
\hline $\Sigma-Axis(C_{2v})$ & $\Sigma _{1}$ & $\Sigma _{4}$ &
$\Sigma _{1}\Sigma _{4}$ & $\Sigma _{2}\Sigma _{3}\Sigma _{4}$ &
$\Sigma _{1}\Sigma _{2}\Sigma _{3}$ \\
\hline \hline $\Gamma-Point(O_h)$ & $\Gamma _{1}^{^{\prime }} $ &
$\Gamma_{2}^{^{\prime }}$ & $\Gamma_{12}^{^{\prime }}$ & $\Gamma
_{15}$ & $\Gamma _{25}$ \\
\hline $\Delta-Axis(C_{4v})$ & $\Delta _{1}^{^{,}}$ & $\Delta
_{2}^{^{,}}$ & $ \Delta _{1}^{^{,}}\Delta _{2}^{^{,}}$ & $\Delta
_{1}\Delta _{5}$ & $ \Delta_{2}\Delta _{5}$\\
\hline $\Lambda-Axis(C_{3v})$ & $\Lambda _{2}$ & $\Lambda _{1}$ &
$\Lambda _{3}$ & $\Lambda _{1}\Lambda _{3}$ & $\Lambda _{2}
\Lambda _{3}$ \\
\hline $\Sigma-Axis(C_{2v})$ & $\Sigma _{2}$ & $\Sigma _{3}$ &
$\Sigma _{2}\Sigma _{3}$ & $\Sigma _{1}\Sigma _{3}\Sigma _{4}$ &
$\Sigma _{1}\Sigma _{2}\Sigma _{4}$\\
\hline\hline
\end{tabular}%
\end{tabular}%
\newline
$\qquad $Table A7 is copied from table 26 of \cite{Energy Band} and
Table 8.5 of \cite{Joshi}.
\newline \\
\begin{tabular}{l}
\\
Table A8. The compatibility relations of the H-$\Delta
$, and H--F-axis\\
\begin{tabular}{|l|l|l|l|l|l|l|l|l|l|l|}
\hline $\Gamma$-Point($O_h$) & $H _{1}$ & $H _{2}$ & $H _{12}$ &%
$H_{15}^{^{\prime }}$ & $H _{25}^{^{\prime }}$ \\
\hline $\Delta-Axis(C_{4v})$ & $\Delta _{1}$ & $\Delta _{2}$ &
$\Delta _{1} \Delta _{2}$ & $\Delta _{1}^{^{,}}\Delta _{5}$ &
$\Delta _{2}^{^{,}}\Delta _{5}$ \\
\hline $F-Axis(C_{3v})$ & $F_{1}$ & $F_{2}$ & $F_{3}$ & $F_{2}F_{3}$
& $F_{1}F_{3}$ \\
\hline \hline $\Gamma-Point(O_h)$ & $H _{1}^{^{\prime }} $ &
$H_{2}^{^{\prime }}$&$H_{12}^{^{\prime }}$&$H_{15}$ &$H_{25}$ \\
\hline $\Delta-Axis(C_{4v})$ & $\Delta _{1}^{^{,}}$ & $\Delta
_{2}^{^{,}}$ & $ \Delta _{1}^{^{,}}\Delta _{2}^{^{,}}$ & $\Delta
_{1}\Delta _{5}$ & $ \Delta_{2}\Delta _{5}$ \\
\hline $F-Axis(C_{3v})$ & $F _{2}$ & $F_{1}$ & $F_{3}$ & %
$F_{1}F_{3}$ & $F _{2}F_{3}$ \\
\hline\hline
\end{tabular}%
\end{tabular}%
\newline
$\qquad $Table A8 is copied from Table 8.5 of \cite{Joshi}.
\newline \\

\begin{tabular}{l}
\\
Table A9. The compatibility relations of the point M--the
G-axis \\
\begin{tabular}{|l|l|l|l|l|l|l|l|l|l|l|}
\hline M-Point($O_h$) & $M _{1}$ & $M_{2}$ & $M_{12}$ & $M
_{15}^{^{\prime }}$ &
$M _{25}^{^{\prime }}$ \\
\hline $G-axis(C_{2v})$ & $G _{1}$ & $G _{4}$ & $G _{1}G_{4}$ & $G
_{2}G _{3}G _{4}$ & $G_{1}G_{2}G_{3}$ \\
\hline \hline $M-Point(O_h)$ & $M_{1}^{^{\prime }} $ &
$M_{2}^{^{\prime }}$
& $M_{12}^{^{\prime }}$ & $M_{15}$ & $M _{25}$ \\
\hline $G-axis(C_{2v})$ & $G _{2}$ & $G_{3}$ & $G_{2}G_{3}$ & $G
_{1}G_{3}G_{4}$ &
$G _{1}G _{2}G _{4}$ \\
\hline\hline
\end{tabular}%
\end{tabular}%
\newline
$\qquad$$\qquad$Table A9 is copied from Table 8.5 of \cite{Joshi}.
\newline \\

\begin{tabular}{l}
\\
Table A10.\ The compatibility relations of P-D, P-F and
P-$\Lambda$ \\
\begin{tabular}{|l|l|l|l|l|l|l|l|l|l|}
\hline $T_d$ & $P_1$ & $P_2$ & $P_3$ & $P_4$ & $P_5$ \\
\hline $C_{3v}$ & $\Lambda_1$ & $\Lambda_2$ & $\Lambda_3$ &
$\Lambda_1$ $\Lambda_3$ & $\Lambda_2$ $\Lambda_3$ \\
\hline $C_{3v}$ & $F_1$ & $F_2$ & $F_3$ & $F_1F_3$ & $F_2F_3$\\
\hline $C_{2v}$ & $D_1$ & $D_2$ & $D_1D_2$ & $D_1D_3D_4$ &
$D_2D_3D_4$\\
\hline\hline
\end{tabular}%
\end{tabular}%
\newline
$\qquad $$\qquad$ Table A10 is copied from Table 8.5 of
\cite{Joshi}.
\newline \\

\begin{tabular}{l}
\\
Table A11.\ The compatibility relations of N-D, N-G and N-$\Sigma$\\
\begin{tabular}{|l|l|l|l|l|l|l|l|l|l|}
\hline $D_{2h}$ & $N_1$ & $N_2$ & $N_3$ & $N_4$ & $N^{\prime}_1$ &
$N^{\prime}_2$ & $N^{\prime}_3$ & $N^{\prime}_4$ \\
\hline $C_{2v}$ & $\Sigma_1$ & $\Sigma_2$ & $\Sigma_3$ & $\Sigma_4$
& $\Sigma_1$ & $\Sigma_2$ & $\Sigma_3$ & $\Sigma_4$ \\
\hline $C_{2v}$ & $D_1$ & $D_3$ & $D_4$ & $D_2$ & $D_4$ & $D_2$ &
$D_1$ & $D_3$ \\
\hline $C_{2v}$ & $G_1$ & $G_3$ & $G_2$ & $G_4$ & $G_4$ & $G_2$ &
$G_3$ & $G_1$ \\
\hline\hline
\end{tabular}%
\end{tabular}%
\newline
$\qquad $ Table A11 is copied from Table 8.5 of \cite{Joshi}.

\section{The Angular Momenta J and Parities P of the Energy Bands}

$\qquad$ From Fig. 2--Fig. 8 and Table A1--Table A11 as well as the
formulae ($\ref{s = 1/2}$)--($\ref{Excited P}$), we can find the
orbit angular momenta L, the total angular momenta J and the
parities P of the energy bands on the $\Delta $-axis, the
$\Lambda$-axis, the $\Sigma$-axis, the D-axis, the F-axis and the
G-axis as appearing in Table B1--Table B6. Since the compatibility
relations (in Table 7, Table 8 and Table 9) cannot completely
determine the representations at a symmetry point correspondent to
an energy band, generally, we choose the representation with lower L
value for the lower energy band; and we choose the representation
with higher L value for higher energy band.
\begin{equation}
choosing\,\,\, the\,\, representation\,\, with\,\,\,
lower\,\,\,L,\,for\,\,\,lower\,\,\, energy\,\,\, band. \label{L-L-L}
\end{equation}
This is a useful formula for choosing a correct representation.
\subsection{The J and P of the energy bands on $\Delta$-axis}

$\qquad $There are many four-fold energy bands in Fig. 2. For a
four-fold energy band, there are always two irreducible
representations in the same band since the highest fold of the
representations is only three.

For these complex cases, in order to get a definite result, we
assume that if the two representations have different orbit angular
momenta (L) and parities (P), the parity (P) of the energy band is
determined by the representations with higher fold (most three
fold). The total angular momentum J is the common J values of the
two representations in ground state. For example, the energy band
with M = 1299 and $\vec{n}$ = (101,011,-101,0-11), has $H_{15}$ (3
fold)+ $\frac{1}{2}H_{12}$(1 fold). The P = ``+" is determined by
$H_{15}$ from (\ref{Excited P}). $H_{15}$ has s = $\frac{1}{2}$ from
(\ref{s = 1/2}) and L = 1 from Table A1. Thus J = $\frac{1}{2},
\frac{3}{2}$. Similarly, $ \frac{1}{2}H_{12}$ has s = $\frac{1}{2}$
and J = $\frac{3}{2}, \frac{5}{2}$. Therefore the two
representations have a common $J^p$ = $\frac{3}{2}^+$. In exited
state, however, the angular momentum L is determined by the
representations with higher fold (most three-fold). Since a
three-fold single group representation correspond to a four-fold
double representation and a two-fold double representation, excited
band may excite the four-fold double representation \cite{Double}.
Such as, the band with M = 2019 and $\vec{n}$ = (112, 1-12, -112,
-1-12), has the representation $H_{15}$ and $H_2^{\prime}$. The L
value and P are determined by $H_{15}$ as shown in Table B1.
Similarly, we can find the J and P for the energy bands with M =
1659 and $\vec{n}$ = (10-1,-10-1,01-1,0-1-1); M = 1659 and $\vec{n}$
= (10-1,-10-1,01-1,0-1-1). The results appearing in Table B1.

For the band with M = 2379 and $\vec{n}$ = (200,-200,020,0-20),
there are two representations $\Gamma_{12}$ and
$\frac{2}{3}\Gamma_{15}$. The two-fold single representation
$\Gamma_{12}$ corresponding to a four-fold double representation,
and the $\frac{2}{3}\Gamma_{15}$ corresponding to a four-fold double
representation also \cite{Double}. From Table A1 and (\ref{Max J}),
for $\frac{2}{3}\Gamma_{15}$, $J^P$ = $\frac{5}{2}^+, \frac{7}{2}^+,
\frac{9}{ 2}^+, \frac{11}{2}^+$ from (\ref{Excited P}); for
$\Gamma_{12}$, $J^P$ = $\frac{5}{2 }^-, \frac{7}{2}^-,
\frac{9}{2}^-, \frac{11}{2}^-$ from (\ref{Excited P}).

There are three single bands: M = 939, $\vec{n}$ = (0,0,0) $\to$
N(939)-ground baryon from (\ref{N(939)}); M = 1299, $\vec{n}$ =
(0,0,2) $ \Lambda(1299)$ is not ground baryons, the representation
is $H_1$, L = 0 from Table A1, J = $\frac{1}{2}$, P = ``-" from
(\ref{Excited P}); M = 2379, $\vec{n}$ = (0, 0, -2), $\Lambda_{c}$
is a ground baryon, the representation $\Delta_1$ $\to$ $J^P$ =
$\frac{1}{2}^+$ from (\ref{Ground P}).

From Fig.2, Table A1, Table A4, Table A7 and Table A8, we can deduce
the total angular momenta J and parities P of the energy bands on
the $\Delta$-axis as appearing in Table B1.

\begin{tabular}{l}
$\qquad$$\qquad$$\qquad$Table B1. The J and P of the energy bands
on the $\Delta$-axis \\
\begin{tabular}{|l|l|l|l|l|l|l|l|l|l|}
\hline M & $(n_1, n_2, n_3)$ & s & Repres. & L & J(1/2) & J(3/2) &
P\\ \hline 939 & (0, 0, 0) & $\frac{1}{2}$ & $\Delta_1$ & 0* &
$\frac{1}{2}$ &  & +  \\
\hline 1299 &
\begin{tabular}{l}
(101,-101, \\
011,0-11)%
\end{tabular}
& $\frac{1}{2}$ &
\begin{tabular}{l}
$H_{15}$ \\
$\frac{1}{2}H_{12}$%
\end{tabular}
& 1 & $\frac{3}{2}$ &  & + \\
\hline
1299 & (0, 0, 2) & $\frac{1}{2}$ & $H_1$ & 0 & $\frac{1}{2}$ &  & - \\
\hline
1659 &
\begin{tabular}{l}
(110,1-10, \\
-110,-1-10)%
\end{tabular}
& $\frac{1}{2}$, $\frac{3}{2}$ &
\begin{tabular}{l}
$\Gamma^{\prime}_{25}$ \\
$\frac{1}{2}\Gamma_{12}$%
\end{tabular}
& 2 & $\frac{3}{2},\frac{5}{2}$ &
$\frac{1}{2},\frac{3}{2},\frac{5}{2},\frac{7}{2}$ & - \\
\hline 1659 &
\begin{tabular}{l}
(10-1,-10-1, \\
01-1,0-1-1)%
\end{tabular}
& $\frac{1}{2}$, $\frac{3}{2}$ &
\begin{tabular}{l}
$\Gamma_{15}$ \\
$\frac{1}{2}\Gamma_{12}$%
\end{tabular}
& 1 & $\frac{1}{2},\frac{3}{2}$ & $\frac{1}{2},\frac{3}{2},
\frac{5}{2}$ & + \\
\hline 2019 &
\begin{tabular}{l}
(112,1-12, \\
-112,-1-12)%
\end{tabular}
& $\frac{1}{2}$, $\frac{3}{2}$ &
\begin{tabular}{l}
$H_{15}$ \\
$H_2'$%
\end{tabular}
&
\begin{tabular}{l}
1 \\
3  %
\end{tabular}&
\begin{tabular}{l}
$\frac{1}{2},\frac{3}{2}$  \\
$\frac{5}{2}$$\frac{7}{2}$; %
\end{tabular}&
\begin{tabular}{l}
$\frac{1}{2}$,$\frac{3}{2}$,$\frac{5}{2}$ \\
$\frac{3}{2}$,$\frac{5}{2}$,$\frac{7}{2}$,$\frac{9}{2}$%
\end{tabular} & + \\

\hline 2379 &
\begin{tabular}{l}
(200,-200, \\
020,0-20)%
\end{tabular}
& $\frac{1}{2}$, $\frac{3}{2}$ &
\begin{tabular}{l}
$\Gamma_{12}$ \\
$\frac{2}{3}\Gamma_{15}$%
\end{tabular}
&
\begin{tabular}{l}
2, \\
4. \\
1, \\
3, \\
5.%
\end{tabular}
&
\begin{tabular}{l}
$\frac{3}{2}$,$\frac{5}{2}$; \\
$\frac{7}{2}$,$\frac{9}{2}$. \\
$\frac{1}{2}$,$\frac{3}{2}$; \\
$\frac{5}{2}$,$\frac{7}{2}$; \\
$\frac{9}{2}$,$\frac{11}{2}$%
\end{tabular}
&
\begin{tabular}{l}
$\frac{1}{2}$,$\frac{3}{2}$,$\frac{5}{2}$,$\frac{7}{2}$; \\
$\frac{5}{2}$ $\frac{7}{2}$,$\frac{9}{2}$,$\frac{11}{2}$. \\
$\frac{1}{2}$,$\frac{3}{2}$,$\frac{5}{2}$; \\
$\frac{3}{2}$,$\frac{5}{2}$,$\frac{7}{2}$,$\frac{9}{2}$; \\
$\frac{7}{2}$,$\frac{9}{2}$,$\frac{11}{2}$,$\frac{13}{2}$.%
\end{tabular}
&
\begin{tabular}{l}
- \\
- \\
+ \\
+ \\
+%
\end{tabular}\\
\hline 2379 & (0, 0, -2) & $\frac{1}{2}$ & $\Delta_1$ & 0*
& $\frac{1}{2}$ &  & + \\
\hline 2739 &
\begin{tabular}{l}
(121,1-21, \\
-121,-1-21, \\
211,2-11, \\
-211,-2-11)%
\end{tabular}
& $\frac{1}{2}$,$\frac{3}{2}$ &
\begin{tabular}{l}
$H_{15}$ \\
$H_{12}$ \\
$H_{25}^{\prime}$%
\end{tabular}
&  &  &  &  \\
\hline 2739 &
\begin{tabular}{l}
(121,1-21, \\
-121,-1-21,%
\end{tabular}
& $\frac{1}{2}$,$\frac{3}{2}$ &
\begin{tabular}{l}
$H_{15}$ \\
$\frac{1}{2}$$H_{12}$%
\end{tabular}
&
\begin{tabular}{l}
1 \\
3  \\
5  %
\end{tabular} &
\begin{tabular}{l}
$\frac{1}{2}$,$\frac{3}{2}$ \\
$\frac{5}{2}$,$\frac{7}{2}$ \\
$\frac{9}{2}$,$\frac{11}{2}$%
\end{tabular}
&
\begin{tabular}{l}
$\frac{1}{2}$,$\frac{3}{2}$,$\frac{5}{2}$ \\
$\frac{3}{2}$,$\frac{5}{2}$,$\frac{7}{2}$,$\frac{9}{2}$ \\
$\frac{7}{2}$,$\frac{9}{2}$,$\frac{11}{2}$,$\frac{13}{2}$%
\end{tabular} & + \\

\hline 2739 &
\begin{tabular}{l}
(211,2-11, \\
-211,-2-11)%
\end{tabular}
& $\frac{1}{2}$,$\frac{3}{2}$ &
\begin{tabular}{l}
$H_{25}^{\prime}$ \\
$\frac{1}{2}$$H_{12}$%
\end{tabular}
&
\begin{tabular}{l}
2 \\
4 \\
6 %
\end{tabular}
&
\begin{tabular}{l}
$\frac{3}{2}$,$\frac{5}{2}$ \\
$\frac{7}{2}$,$\frac{9}{2}$ \\
$\frac{11}{2}$,$\frac{13}{2}$ %
\end{tabular}
&
\begin{tabular}{l}
$\frac{1}{2}$,$\frac{3}{2}$,$\frac{5}{2}$,$\frac{7}{2}$ \\
$\frac{5}{2}$,$\frac{7}{2}$,$\frac{9}{2}$,$\frac{11}{2}$ \\
$\frac{9}{2}$,$\frac{11}{2}$,$\frac{13}{2}$,$\frac{15}{2}$%
\end{tabular}
& - \\

\hline 2739 &
\begin{tabular}{l}
(202,-202, \\
022,0-22)%
\end{tabular}
& $\frac{1}{2}$,$\frac{3}{2}$ &
\begin{tabular}{l}
$H_{25}$ \\
$\frac{1}{2}$$H_{12}$%
\end{tabular}
&
\begin{tabular}{l}
3 \\
5%
\end{tabular}
&
\begin{tabular}{l}
$\frac{5}{2}$,$\frac{7}{2}$ \\
$\frac{9}{2}$,$\frac{11}{2}$%
\end{tabular}
&
\begin{tabular}{l}
$\frac{3}{2}$,$\frac{5}{2}$,$\frac{7}{2}$,$\frac{9}{2}$ \\
$\frac{7}{2}$,$\frac{9}{2}$,$\frac{11}{2}$,$\frac{13}{2}$%
\end{tabular}
& + \\
\hline 2739 &
\begin{tabular}{l}
(013,0-13, \\
103,-103)%
\end{tabular}
& $\frac{1}{2}$,$\frac{3}{2}$ &
\begin{tabular}{l}
$H_{15}^{\prime}$\\
$\frac{1}{2}$$H_{12}$%
\end{tabular}
&
\begin{tabular}{l}
4  \\
6  %
\end{tabular}
&
\begin{tabular}{l}
$\frac{7}{2}$,$\frac{9}{2}$ \\
$\frac{11}{2}$,$\frac{13}{2}$%
\end{tabular}
&
\begin{tabular}{l}
$\frac{5}{2}$,$\frac{7}{2}$,$\frac{9}{2}$,$\frac{11}{2}$ \\
$\frac{9}{2}$,$\frac{11}{2}$,$\frac{13}{2}$,$\frac{15}{2}$%
\end{tabular}
& -\\

\hline 3099 &
\begin{tabular}{l}
(12-1,1-2-1, \\
-12-1,-1-2-1)%
\end{tabular}
& $\frac{1}{2}$,$\frac{3}{2}$ &
\begin{tabular}{l}
$\Gamma_{25}^{\prime}$\\
$\frac{1}{2}$$\Gamma_{12}$%
\end{tabular}
&
\begin{tabular}{l}
2  \\
4  \\
6  %
\end{tabular}
&
\begin{tabular}{l}
$\frac{3}{2}$,$\frac{5}{2}$ \\
$\frac{7}{2}$,$\frac{9}{2}$ \\
$\frac{11}{2}$,$\frac{13}{2}$%
\end{tabular}
&
\begin{tabular}{l}
$\frac{1}{2}$,$\frac{3}{2}$,$\frac{5}{2}$,$\frac{7}{2}$ \\
$\frac{5}{2}$,$\frac{7}{2}$,$\frac{9}{2}$,$\frac{11}{2}$ \\
$\frac{9}{2}$,$\frac{11}{2}$,$\frac{13}{2}$,$\frac{15}{2}$%
\end{tabular}
& -\\

\hline 3099 &
\begin{tabular}{l}
(21-1,2-1-1, \\
-21-1,-2-11)%
\end{tabular}
& $\frac{1}{2}$,$\frac{3}{2}$ &
\begin{tabular}{l}
$\Gamma_{15}^{\prime}$\\
$\frac{1}{2}$$\Gamma_{12}$%
\end{tabular}
&
\begin{tabular}{l}
4  \\
6  %
\end{tabular}
&
\begin{tabular}{l}
$\frac{7}{2}$,$\frac{9}{2}$ \\
$\frac{11}{2}$,$\frac{13}{2}$%
\end{tabular}
&
\begin{tabular}{l}
$\frac{5}{2}$,$\frac{7}{2}$,$\frac{9}{2}$,$\frac{11}{2}$ \\
$\frac{9}{2}$,$\frac{11}{2}$,$\frac{13}{2}$,$\frac{15}{2}$%
\end{tabular}
& -\\

\hline 3099 &
\begin{tabular}{l}
(11-2,1-1-2, \\
-11-2,-1-1-2)%
\end{tabular}
& $\frac{1}{2}$,$\frac{3}{2}$ &
\begin{tabular}{l}
$\Gamma_{25}$\\
$\Gamma_{2}'$%
\end{tabular}
&
\begin{tabular}{l}
3  \\
5 %
\end{tabular}
&
\begin{tabular}{l}
$\frac{5}{2}$,$\frac{7}{2}$ \\
$\frac{9}{2}$,$\frac{11}{2}$%
\end{tabular}
&
\begin{tabular}{l}
$\frac{3}{2}$,$\frac{5}{2}$,$\frac{7}{2}$,$\frac{9}{2}$ \\
$\frac{7}{2}$,$\frac{9}{2}$,$\frac{11}{2}$,$\frac{13}{2}$%
\end{tabular}
& + \\

\hline
\end{tabular}%
\end{tabular}%
\newline

0* means L = 0 from Table A4. \newline
\newline

\subsection{The J and P of the energy bands on $\Lambda$-axis}

$\qquad$From Fig.3, Table A1, Table A2, Table A7 and Table 10, we
can get the irreducible representations of the energy bands at the
lowest energy point. Using Table A1, Table A2 and
$(\ref{J})$--$(\ref{Excited P})$, we can find the J and P of the
energy bands on $\Lambda$-axis as appearing in Table B2. At
$M_{Fig}$ = 1929 and 2649, the band with P-representations of the
P-group (Table A2). $P_4$ contains functions belonging to
$\Gamma_{15}$ and $\Gamma_{25}'$, $P_{5}$ functions belonging to
$\Gamma_{25}$ and $\Gamma_{15}'$ \cite{Callaway}.\ \ \ \ \ \ \ \ \ \
\newline
\newline
\begin{tabular}{l}
$\qquad$$\qquad$$\qquad$$\qquad$Table B2. The J and P of the
energy bands on the $\Lambda$-axis \\
\begin{tabular}{|l|l|l|l|l|l|l|l|l|l|}
\hline
M (Mev) & ($n_1, n_2, n_3$) & s & Repres. & L & J(1/2) & J(3/2) & P\\
\hline 939 & (0, 0, 0) & $\frac{1}{2}$ & $\Lambda_1$ & 0** &
$\frac{1}{2}$ &  & + \\
\hline 1209 & (011,101,110) & $\frac{1}{2}$ & P$_4$ & 1 &
$\frac{1}{2}$,$\frac{3}{2} $ &  & + \\
\hline 1659 &
\begin{tabular}{l}
(1-10,-110,01-1,  \\
0-11,10-1,-101)   %
\end{tabular} &
$\frac{1}{2}$,$\frac{3}{2}$ &
$\Gamma_{1}$$\Gamma_{12}$$\Gamma_{25^{\prime}}$
& &  &  &   \\
\hline 1659 & (1-10,-110,01-1) & $\frac{1}{2}$,$\frac{3}{2}$ &
$\Gamma_{25^{\prime}} $ & 2 & $\frac{3}{2}$,$\frac{5}{2}$ &
$\frac{1}{2}$,$\frac{3}{2}$,$\frac{5}{2 }$,$\frac{7}{2}$ & - \\
\hline 1659 & (0-11,10-1,-101) & $\frac{1}{2}$,$\frac{3}{2}$ &
$\Gamma_{12}\Gamma_{1}$ & 2 &  &
$\frac{1}{2}$,$\frac{3}{2}$, & - \\
\hline 1659 & (-1-10,-10-1,0-1-1) & $\frac{1}{2}$,$\frac{3}{2}$ &
$\Gamma_{15}$ & 1 & $\frac{1}{2}$,$\frac{3}{2}$ &
$\frac{1}{2}$,$\frac{3}{2}$,$\frac{5}{2}$ & + \\
\hline 1929 &
(020,002,200) & $\frac{1}{2}$,$\frac{3}{2}$ & $P_{4}$ &
\begin{tabular}{l}
1, \\
3; \\
2  %
\end{tabular}
&
\begin{tabular}{l}
$\frac{1}{2}$,$\frac{3}{2}$,  \\
$\frac{5}{2}$,$\frac{7}{2}$;  \\
$\frac{3}{2}$,$\frac{5}{2}$   %
\end{tabular}
&
\begin{tabular}{l}
$\frac{1}{2}$,$\frac{3}{2}$,$\frac{5}{2}$, \\
$\frac{3}{2}$,$\frac{5}{2}$,$\frac{7}{2}$,$\frac{9}{2}$; \\
$\frac{1}{2}$,$\frac{3}{2}$,$\frac{5}{2}$,$\frac{7}{2}$%
\end{tabular}
&
\begin{tabular}{l}
+ \\
+  \\
-%
\end{tabular} \\

\hline 1929 & (211,121,112) & $\frac{1}{2}$,$\frac{3}{2}$ & $P_{4}$
&
\begin{tabular}{l}
1, \\
3. \\
2.  %
\end{tabular}
&
\begin{tabular}{l}
$\frac{1}{2}$,$\frac{3}{2}$,  \\
$\frac{5}{2}$,$\frac{7}{2}$. \\
$\frac{3}{2}$,$\frac{5}{2}$.  %
\end{tabular}
&
\begin{tabular}{l}
$\frac{1}{2}$,$\frac{3}{2}$,$\frac{5}{2}$;  \\
$\frac{3}{2}$,$\frac{5}{2}$,$\frac{7}{2}$,$\frac{9}{2}$. \\
$\frac{1}{2}$,$\frac{3}{2}$,$\frac{5}{2}$,$\frac{7}{2}$. %
\end{tabular}
&
\begin{tabular}{l}
+ \\
+ \\
-%
\end{tabular} \\

\hline 2379 & (00-2,0-20,00-2) & $\frac{1}{2}$,$\frac{3}{2}$ &
$\Gamma_{15}$  &
\begin{tabular}{l}
1 \\
3 %
\end{tabular}
&
\begin{tabular}{l}
$\frac{1}{2}$,$\frac{3}{2}$; \\
$\frac{5}{2}$,$\frac{7}{2}$.%
\end{tabular}
&
\begin{tabular}{l}
$\frac{1}{2}$,$\frac{3}{2}$,$\frac{5}{2}$; \\
$\frac{3}{2}$,$\frac{5}{2}$,$\frac{7}{2}$,$\frac{9}{2}$%
\end{tabular} & + \\

\hline 2649 &
\begin{tabular}{l}
(12-1,1-12,21-1  \\
2-11,-121,-112)   %
\end{tabular} & $\frac{1}{2}$,$\frac{3}{2}$ &$P_4P_4$ & & & & \\

\hline 2649 & (12-1,1-12,21-1) & $\frac{1}{2}$,$\frac{3}{2}$ &
$P_{4}$ &
\begin{tabular}{l}
1,  \\
3;  \\
2,  \\
4.  %
\end{tabular}
&
\begin{tabular}{l}
$\frac{1}{2}$,$\frac{3}{2}$, \\
$\frac{5}{2}$,$\frac{7}{2}$;  \\
$\frac{3}{2}$,$\frac{5}{2}$, \\
$\frac{7}{2}$,$\frac{9}{2}$;  %
\end{tabular}&
\begin{tabular}{l}
$\frac{1}{2}$,$\frac{3}{2}$,$\frac{5}{2}$, \\
$\frac{3}{2}$,$\frac{5}{2}$,$\frac{7}{2}$,$\frac{9}{2}$; \\
$\frac{1}{2}$,$\frac{3}{2}$,$\frac{5}{2}$,$\frac{7}{2}$, \\
$\frac{3}{2}$,$\frac{5}{2}$,$\frac{7}{2}$,$\frac{9}{2}$. %
\end{tabular} &
\begin{tabular}{l}
+  \\
+  \\
-  \\
-  %
\end{tabular}\\

\hline 2649 & (2-11,-121,-112) & $\frac{1}{2}$,$\frac{3}{2}$ &
$P_{4}$ &
\begin{tabular}{l}
1,  \\
3;  \\
2,  \\
4.  %
\end{tabular}
&
\begin{tabular}{l}
$\frac{1}{2}$,$\frac{3}{2}$, \\
$\frac{5}{2}$,$\frac{7}{2}$;  \\
$\frac{3}{2}$,$\frac{5}{2}$, \\
$\frac{7}{2}$,$\frac{9}{2}$;  %
\end{tabular}&
\begin{tabular}{l}
$\frac{1}{2}$,$\frac{3}{2}$,$\frac{5}{2}$, \\
$\frac{3}{2}$,$\frac{5}{2}$,$\frac{7}{2}$,$\frac{9}{2}$; \\
$\frac{1}{2}$,$\frac{3}{2}$,$\frac{5}{2}$,$\frac{7}{2}$, \\
$\frac{5}{2}$,$\frac{7}{2}$,$\frac{9}{2}$,$\frac{11}{2}$. %
\end{tabular} &
\begin{tabular}{l}
+  \\
+  \\
-  \\
-  %
\end{tabular}\\

\hline 2649 & (202,220,022) & $\frac{1}{2}$,$\frac{3}{2}$ & $P_{5}$
&
\begin{tabular}{l}
3,  \\
5;  \\
4,   %
\end{tabular}
&
\begin{tabular}{l}
$\frac{5}{2}$,$\frac{7}{2}$;  \\
$\frac{9}{2}$,$\frac{11}{2}$, \\
$\frac{7}{2}$,$\frac{9}{2}$;  %
\end{tabular}&
\begin{tabular}{l}
$\frac{3}{2}$,$\frac{5}{2}$,$\frac{7}{2}$,$\frac{9}{2}$, \\
$\frac{7}{2}$,$\frac{9}{2}$,$\frac{11}{2}$,$\frac{13}{2}$; \\
$\frac{5}{2}$,$\frac{7}{2}$,$\frac{9}{2}$,$\frac{11}{2}$. %
\end{tabular} &
\begin{tabular}{l}
+  \\
+  \\
-  %
\end{tabular} \\

\hline
\end{tabular}
\end{tabular}\\
\,\,\,\,\,\,0* means L = 0 from Table A4
\newline
\newline

\subsection{The J and P of the energy bands on $\Sigma$-axis}

$\qquad$Using Fig. 4, Table A1, Table A3, Table A7 and
$(\ref{J})$--$(\ref{Excited P})$, we can deduce the J and P of the
energy bands on the $\Sigma$-axis as appearing in Table B3:

\begin{tabular}{l}
$\qquad$$\qquad$$\qquad$$\qquad$Table B3.\ The J and P of energy
bands on the $\Sigma$-axis \\
\begin{tabular}{|l|l|l|l|l|l|l|l|}
\hline
M (Mev) & ($n_1, n_2, n_3$) & s & Repres. & L &
J(1/2) & J(3/2) & P \\
\hline 939 & (0, 0, 0) & $\frac{1}{2}$ & $\Sigma_1$ & 0* &
$\frac{1}{2}$ &  & + \\
\hline 1119 & (1, 1, 0) & $\frac{1}{2}$ & $\Sigma_1$ & 0* &
$\frac{1}{2}$ &  & + \\
\hline 1479 & (101,011) & $\frac{1}{2}$,$\frac{3}{2}$ & $N_1,N_2$ &
0,2 & $\frac{1}{2}$ & $\frac{3}{2}$ & - \\
\hline 1479 & (10-1,01-1) & $\frac{1}{2}$,$\frac{3}{2}$ &
$N^{\prime}_3,N^{\prime}_4 $ & 1 & $\frac{1}{2}$,$\frac{3}{2}$ &
$\frac{1}{2}$,$\frac{3}{2}$,$\frac{5}{2 }$ & + \\
\hline 1659 & (1-10,-110) & $\frac{1}{2}$,$\frac{1}{2}$ &
$\frac{2}{3}\Gamma_{15}$ & 1 & $\frac{1}{2}$,$\frac{3}{2}$ &
$\frac{1}{2}$,$\frac{3}{2}$,$\frac{5}{2}$ & + \\
\hline 1659 &
\begin{tabular}{l}
(-101,-10-1, \\
0-11,0-1-1)%
\end{tabular}
& $\frac{1}{2}$,$\frac{3}{2}$ &
$\Gamma^{\prime}_{25}$,$\frac{1}{2}\Gamma_{12}$  &  &  &  & \\
\hline 1659 & (-101,-10-1) & $\frac{1}{2}$,$\frac{3}{2}$ &
$\frac{2}{3} \Gamma^{\prime}_{25}$ & 2 & $\frac{3}{2}$,$\frac{5}{2}$
& $\frac{1}{2}$,$\frac{3}{2}$,$\frac{5}{2}$,$\frac{7}{2}$ & - \\
\hline 1659 & (0-11,0-1-1) & $\frac{1}{2}$,$\frac{3}{2}$ &
$\frac{1}{3} \Gamma^{\prime}_{25},\frac{1}{2}\Gamma_{12}$ & 2 &
$\frac{3}{2}$,$\frac{5}{2} $ &
$\frac{1}{2}$,$\frac{3}{2}$,$\frac{5}{2}$,$\frac{7}{2}$ & - \\
\hline 1659 & (-1,-1,0) & $\frac{1}{2}$,$\frac{3}{2}$ & $\Sigma_1$ &
0* & $\frac{1}{2}$ & $\frac{3}{2}$ & + \\
\hline 1839 & (200,020) & $\frac{1}{2}$,$\frac{3}{2}$ & $N_1,N_4$ &
2 & $\frac{3}{2}$,$\frac{5}{2}$ &
$\frac{1}{2}$,$\frac{3}{2}$,$\frac{5}{2}$,$\frac{7}{2}$ & - \\
\hline 2199 & (121,211) & $\frac{1}{2}$,$\frac{3}{2}$ & $N_2,N_3$ &
\begin{tabular}{l}
2, \\
4.%
\end{tabular}
&
\begin{tabular}{l}
$\frac{3}{2}$,$\frac{5}{2}$; \\
$\frac{7}{2}$,$\frac{9}{2}$. \\
\end{tabular}
&
\begin{tabular}{l}
$\frac{1}{2}$,$\frac{3}{2}$,$\frac{5}{2}$,$\frac{7}{2}$ \\
$\frac{5}{2}$,$\frac{7}{2}$,$\frac{9}{2}$,$\frac{11}{2}$%
\end{tabular} & - \\

\hline 2199 & (12-1,21-1) & $\frac{1}{2}$,$\frac{3}{2}$ &
$N^{\prime}_1,N^{\prime}_4 $ &
\begin{tabular}{l}
1, \\
3.%
\end{tabular}
&
\begin{tabular}{l}
$\frac{1}{2}$,$\frac{3}{2}$; \\
$\frac{5}{2}$,$\frac{7}{2}$. \\
\end{tabular}
&
\begin{tabular}{l}
$\frac{1}{2}$,$\frac{3}{2}$,$\frac{5}{2}$, \\
$\frac{3}{2}$,$\frac{5}{2}$,$\frac{7}{2}$,$\frac{9}{2}$%
\end{tabular}
& + \\

\hline 2379 & (002,00-2) & $\frac{1}{2}$,$\frac{3}{2}$ &
$\frac{2}{3}\Gamma_{15}$ &
\begin{tabular}{l}
1, \\
3.%
\end{tabular}
&
\begin{tabular}{l}
$\frac{1}{2}$,$\frac{3}{2}$; \\
$\frac{5}{2}$,$\frac{7}{2}$. \\
\end{tabular}
&
\begin{tabular}{l}
$\frac{1}{2}$,$\frac{3}{2}$,$\frac{5}{2}$, \\
$\frac{3}{2}$,$\frac{5}{2}$,$\frac{7}{2}$,$\frac{9}{2}$%
\end{tabular}
& + \\

\hline 2379 & (-200,0-20) & $\frac{1}{2}$,$\frac{3}{2}$ &
$\Gamma_{12}$ &
\begin{tabular}{l}
2, \\
4.%
\end{tabular}
&
\begin{tabular}{l}
$\frac{3}{2}$,$\frac{5}{2}$; \\
$\frac{7}{2}$,$\frac{9}{2}$. \\
\end{tabular}
&
\begin{tabular}{l}
$\frac{1}{2}$,$\frac{3}{2}$,$\frac{5}{2}$,$\frac{7}{2}$ \\
$\frac{5}{2}$,$\frac{7}{2}$,$\frac{9}{2}$,$\frac{11}{2}$%
\end{tabular}
& - \\

\hline 2559 & (112,11-2) & $\frac{1}{2}$,$\frac{3}{2}$ & $N_1$,$N_3$
&
\begin{tabular}{l}
2, \\
4.%
\end{tabular}
&
\begin{tabular}{l}
$\frac{3}{2}$,$\frac{5}{2}$; \\
$\frac{7}{2}$,$\frac{9}{2}$. \\
\end{tabular}
&
\begin{tabular}{l}
$\frac{1}{2}$,$\frac{3}{2}$,$\frac{5}{2}$,$\frac{7}{2}$ \\
$\frac{5}{2}$,$\frac{7}{2}$,$\frac{9}{2}$,$\frac{11}{2}$%
\end{tabular}
& - \\

\hline
\end{tabular}%
\end{tabular} \\
\,\,\,\,\,\,0* means L = 0 from Table A4
\newline

\subsection{The J and P of the energy bands on D-axis}

$\qquad$Using Fig.5, Table A2, Table A3, Table A6, Table A8, Table
A9 and $(\ref{J})$--$(\ref{Excited P})$, we can deduce the J and P
of the energy bands on the D-axis as appearing in Table B4. At the
lowest energy $M_{f}$ = 1209, a band with (011,101) will take the
representations with lower L $\frac{1}{3}$$P_4$+$P_1$ from $P-1$ and
$P_4$ from (\ref{L-L-L}. From (\ref{Max J}), $J_{Max}$ $\le$
$\frac{3}{2}$, it cannot get certain L, J and P:

\begin{tabular}{l}
$\qquad$$\qquad$$\qquad$ Table B4.\ The L, J and P of the energy bands on the D-axis \\
\begin{tabular}{|l|l|l|l|l|l|l|l|l|}
\hline
$M_Mev$ & ($n_1, n_2, n_3$) & s & Repres. & L & J(1/2) & J(3/2) & P\\
\hline 1209 & (0,1,1;1,0,1) & $\frac{1}{2}$ & $\frac{1}{3}P_4$+$P_1$
& ? & ? &  & ? \\
\hline 1479 & (10-1,01-1) & $\frac{1}{2}$,$\frac{3}{2}$, & $N_1,N_2$
& 2 & $ \frac{3}{2}$,$\frac{5}{2}$  &
$ \frac{1}{2}$,$\frac{3}{2}$,$ \frac{5}{2}$,$\frac{7}{2}$ & - \\
\hline 1839 & \begin{tabular}{l}
(1-10,-110,  \\
020,200)    %
\end{tabular} &  &  &  &  &  &  \\
\hline 1839 & (1-10,-110) & $\frac{1}{2}$,$\frac{3}{2}$ & $N_1,N_4$
& 2 & $\frac{3}{2},\frac{5}{2}$  &
$\frac{1}{2},\frac{3}{2}$,$\frac{5}{2},\frac{7}{2}$ & - \\
\hline 1839 & (020,200) & $\frac{1}{2}$,$\frac{3}{2}$ &
$N^{\prime}_1,N^{\prime}_4$ & 1 & $\frac{1}{2}$,$\frac{3}{2}$ &
$\frac{1}{2}$,$\frac{3}{2}$,$\frac{5}{2} $ & + \\
\hline 1929 &
\begin{tabular}{l}
(-101,0-11, \\
211,121) %
\end{tabular} &$\frac{1}{2}$,$\frac{3}{2}$ &
$P_4$, $\frac{1}{2}$ $P_{3}$  &
\begin{tabular}{l}
1   \\
2   %
\end{tabular} &
\begin{tabular}{l}
$\frac{1}{2}$,$\frac{3}{2}$ \\
$\frac{3}{2}$,$\frac{5}{2}$  %
\end{tabular}&
\begin{tabular}{l}
$\frac{1}{2}$,$\frac{3}{2}$,$\frac{5}{2}$ \\
$\frac{1}{2}$,$\frac{3}{2}$,$\frac{5}{2}$,$\frac{7}{2}$ %
\end{tabular}&
\begin{tabular}{l}
+    \\
-    %
\end{tabular}\\
\hline 1929 & (002,112) & $\frac{1}{2}$,$\frac{3}{2}$ &
$\frac{1}{2}$$P_{3}$,$\frac{1}{3}P_{4}$  &  & & \\
\hline 1929 & (002) & $\frac{1}{2}$,$\frac{3}{2}$ &
$\frac{1}{2}$$P_{3}$ & 2 & $\frac{3}{2}$,$\frac{5}{2}$&
$\frac{3}{2}$,$\frac{5}{2},\frac{7}{2},\frac{9}{2}$ & - \\
\hline 1929 & (112) & $\frac{1}{2}$,$\frac{3}{2}$ &
$\frac{1}{3}$$P_4$ &
\begin{tabular}{l}
1 \\
2  %
\end{tabular} &
\begin{tabular}{l}
$\frac{1}{2}$,$\frac{3}{2}$\\
$\frac{3}{2}$,$\frac{5}{2}$ %
\end{tabular} &
\begin{tabular}{l}
$\frac{1}{2},\frac{2}{3},\frac{5}{2}$\\
$\frac{1}{2},\frac{2}{3},\frac{5}{2},\frac{7}{2}$  %
\end{tabular} &
\begin{tabular}{l}
+ \\
-  %
\end{tabular} \\
\hline 2199 &
\begin{tabular}{l}
(12-1,21-1,  \\
-10-1,0-1-1)  %
\end{tabular} & $\frac{1}{2},\frac{2}{3}$& $P_4$$P_5$ & & & & \\
\hline 2199 & (1,2,-1;2,1,-1) & $\frac{1}{2}$,$\frac{3}{2}$ &
$N_3,N_4$ &
\begin{tabular}{l}
2, \\
4.%
\end{tabular}
&
\begin{tabular}{l}
$\frac{3}{2}$,$\frac{5}{2}$; \\
$\frac{7}{2}$,$\frac{9}{2}$.%
\end{tabular}
&
\begin{tabular}{l}
$\frac{1}{2},\frac{3}{2},\frac{5}{3},\frac{7}{2}$; \\
$\frac{5}{2},\frac{7}{2},\frac{9}{2},\frac{11}{2}$.%
\end{tabular} & - \\
\hline 2199 & (-1,0,-1;0,-1,-1) & $\frac{1}{2}$,$\frac{3}{2}$ &
$N^{\prime}_3, N^{\prime}_4$ &
\begin{tabular}{l}
1, \\
3.%
\end{tabular}
&
\begin{tabular}{l}
$\frac{1}{2}$,$\frac{3}{2}$ \\
$\frac{5}{2}$,$\frac{7}{2}$%
\end{tabular} &
\begin{tabular}{l}
$\frac{1}{2}\frac{3}{2}\frac{5}{2}$; \\
$\frac{3}{2},\frac{5}{2},\frac{7}{2},\frac{9}{2}$%
\end{tabular} & + \\

\hline 2559 & (220;-1-10) & $\frac{1}{2}$,$\frac{3}{2}$ &
&  &  &  &  \\
\hline 2559 & (2,2,0) & $\frac{1}{2}$,$\frac{3}{2}$ & $N_1$ &
\begin{tabular}{l}
0 \\
2  %
\end{tabular}
& \begin{tabular}{l}
$\frac{1}{2}$ \\
$\frac{3}{2}$,$\frac{5}{2}$%
\end{tabular}
& \begin{tabular}{l}
$\frac{3}{2}$\\
$\frac{1}{2},\frac{3}{2},\frac{5}{2}$,$\frac{7}{2}$ %
\end{tabular} & - \\

\hline 2559 & (-1,-1,0) & $\frac{1}{2}$,$\frac{3}{2}$ & $N_2$ &
\begin{tabular}{l}
2 \\
4  %
\end{tabular}
& \begin{tabular}{l}
$\frac{3}{2}$,$\frac{5}{2}$ \\
$\frac{7}{2}$,$\frac{9}{2}$%
\end{tabular}
& \begin{tabular}{l}
$\frac{1}{2},\frac{3}{2},\frac{5}{2}$,$\frac{7}{2}$\\
$\frac{5}{2},\frac{7}{2},\frac{9}{2}$,$\frac{11}{2}$ %
\end{tabular} & - \\

\hline 2559 & (11-2,00-2) & 2559 &  &  &  &  & \\
\hline 2559 & (1,1,-2) & $\frac{1}{2}$,$\frac{3}{2}$ &
$N_3^{\prime}$ & \begin{tabular}{l}
1 \\
3  %
\end{tabular}
& \begin{tabular}{l}
$\frac{1}{2}$,$\frac{3}{2}$ \\
$\frac{5}{2}$,$\frac{7}{2}$%
\end{tabular}
& \begin{tabular}{l}
$\frac{1}{2},\frac{3}{2},\frac{5}{2}$\\
$\frac{1}{2},\frac{3}{2},\frac{5}{2}$,$\frac{9}{2}$ %
\end{tabular} & + \\
\hline 2559 & (0,0,-2) & $\frac{1}{2}$,$\frac{3}{2}$ &
$N_1^{\prime}$ &
\begin{tabular}{l}
1 \\
3  %
\end{tabular}
& \begin{tabular}{l}
$\frac{1}{2}$,$\frac{3}{2}$ \\
$\frac{5}{2}$,$\frac{7}{2}$%
\end{tabular}
& \begin{tabular}{l}
$\frac{1}{2},\frac{3}{2},\frac{5}{2}$\\
$\frac{3}{2},\frac{5}{2},\frac{7}{2}$,$\frac{9}{2}$ %
\end{tabular} & + \\

\hline 2649 & (-121,2-11) & $\frac{1}{2}$,$\frac{3}{2}$ &
$\frac{2}{3}$$P_4$ &
\begin{tabular}{l}
1, \\
3; \\
2, \\
4.%
\end{tabular}
&
\begin{tabular}{l}
$\frac{1}{2}$,$\frac{3}{2}$; \\
$\frac{5}{2}$,$\frac{7}{2}$. \\
$\frac{3}{2}$,$\frac{5}{2}$; \\
$\frac{7}{2}$,$\frac{9}{2}$.%
\end{tabular}
&
\begin{tabular}{l}
$\frac{1}{2}$,$\frac{3}{2}$,$\frac{5}{2}$; \\
$\frac{3}{2}$,$\frac{5}{2}$,$\frac{7}{2}\frac{9}{2}$. \\
$\frac{1}{2}$,$\frac{3}{2}$,$\frac{5}{2}\frac{7}{2}$; \\
$\frac{5}{2}$,$\frac{7}{2}$,$\frac{9}{2}\frac{11}{2}$.%
\end{tabular}
&
\begin{tabular}{l}
+ \\
+ \\
- \\
-%
\end{tabular} \\
\hline 2649 & \begin{tabular}{l}
(-112,1-12, \\
202,022)   %
\end{tabular}&  &  &  &  &  &  \\
\hline 2649 & (-112,1-12) & $\frac{1}{2}$,$\frac{3}{2}$ &
$\frac{2}{3}P_5$ &
\begin{tabular}{l}
3 \\
5.%
\end{tabular}
&
\begin{tabular}{l}
$\frac{5}{2}$,$\frac{7}{2}$; \\
$\frac{9}{2}$,$\frac{11}{2}$.%
\end{tabular}
&
\begin{tabular}{l}
$\frac{3}{2}$,$\frac{5}{2}$,$\frac{7}{2}\frac{9}{2}$. \\
$\frac{7}{2}$,$\frac{9}{2}$,$\frac{11}{2}\frac{13}{2}$.%
\end{tabular}
&
\begin{tabular}{l}
+ \\
+%
\end{tabular} \\
\hline 2649 & (202,022) & $\frac{1}{2}$,$\frac{3}{2}$ &
$P_1$,$\frac{1}{3}$$P_5$ &
\begin{tabular}{l}
3 \\
5%
\end{tabular}
&
\begin{tabular}{l}
$\frac{5}{2}$,$\frac{7}{2}$; \\
$\frac{9}{2}$,$\frac{11}{2}$.%
\end{tabular}
&
\begin{tabular}{l}
$\frac{3}{2}$,$\frac{5}{2}$,$\frac{7}{2}$,$\frac{9}{2}$; \\
$\frac{7}{2}$,$\frac{9}{2}$,$\frac{11}{2}$,$\frac{13}{2}$%
\end{tabular} & + \\
\hline 2919 & (2-1-1,-12-1) & $\frac{1}{2}$,$\frac{3}{2}$ &
$N_1,N_2$ &
\begin{tabular}{l}
2, \\
4, \\
6.%
\end{tabular}
&
\begin{tabular}{l}
$\frac{3}{2}$,$\frac{5}{2}$; \\
$\frac{7}{2}$,$\frac{9}{2}$; \\
$\frac{11}{2}$,$\frac{13}{2}$.%
\end{tabular}
&
\begin{tabular}{l}
$\frac{1}{2}$,$\frac{3}{2}$,$\frac{5}{2}$,$\frac{7}{2}$. \\
$\frac{5}{2}$,$\frac{7}{2}$,$\frac{9}{2}$,$\frac{11}{2}$. \\
$\frac{9}{2}$,$\frac{11}{2}$,$\frac{13}{2}$,$\frac{15}{2}$.%
\end{tabular}
&
\begin{tabular}{l}
- \\
-%
\end{tabular} \\
\hline
\end{tabular}%
\end{tabular}%
\newline

\subsection{The J and P of the energy bands on F-axis}
\begin{tabular}{l}
$\qquad$$\qquad$$\qquad$Table B5a. The J and P of the energy
bands on F-axis \\
\begin{tabular}{|l|l|l|l|l|l|l|l|l|l|}
\hline $M_Mev$ & ($n_1, n_2, n_3$) & s & Repr. & L & J(1/2) &
J(3/2) & P\\
\hline 1209 & (101,011) & $\frac{1}{2}$ & $\frac{2}{3}P_4$ &
1 & $\frac{1}{2}$,$\frac{3}{2}$ &  & + \\
\hline 1299 & (002,-101,0-11) & $\frac{1}{2}$ &
$H_1 H_{12}$ &  &  &  &   \\
\hline 1299 & (002) & $\frac{1}{2}$ & $H_{1}$ & 0 &
$\frac{1}{2}$ &  & -  \\
\hline 1299 & (-101,0-11) & $\frac{1}{2}$ & $H_{12}$ & 2 &
$\frac{3}{2}$,$\frac{5}{2}$ &  & -  \\
\hline 1929 &(112,-110,1-10) &$\frac{1}{2}$,$\frac{3}{2}$ & & &
& & \\
\hline 1929 & (112) & $\frac{1}{2}$,$\frac{3}{2}$ & $P_1$ & 0 &
$\frac{1}{2}$ & $\frac{3}{2}$ & - \\
\hline 1929 & (-110,1-10)& $\frac{1}{2}$,$\frac{3}{2}$ & $P_3$ & 2 &
$\frac{3}{2}$, $\frac{5}{2}$ &
$\frac{1}{2}$,$\frac{3}{2}$,$\frac{5}{2}$,$\frac{7}{2}$ & - \\

\hline 1929 &
\begin{tabular}{l}
(01-1,10-1,020 \\
200,121,211)%
\end{tabular}
& $\frac{1}{2}$,$\frac{3}{2}$ & $P_4P_5$ &  &  &  &  \\
\hline 1929 & (01-1,10-1,020) & $\frac{1}{2}$,$\frac{3}{2}$ & $P_4$&
&  &  &  \\
\hline 1929 & (020) & $\frac{1}{2}$,$\frac{3}{2}$ & $\frac{1}{3}P_4$
& \begin{tabular} {l}
1  \\
2   %
\end{tabular} &
\begin{tabular} {l}
$\frac{1}{2}$,$\frac{3}{2}$ \\
$\frac{3}{2}$,$\frac{5}{2}$ %
\end{tabular}  &
\begin{tabular} {l}
$\frac{1}{2}$,$\frac{3}{2}$,$\frac{5}{2}$  \\
$\frac{1}{2}$,$\frac{3}{2}$,$\frac{5}{2}$,$\frac{7}{2}$   %
\end{tabular} &
\begin{tabular} {l}
+  \\
-   %
\end{tabular} \\

\hline 1929 & (01-1,10-1) & $\frac{1}{2}$,$\frac{3}{2}$ &
$\frac{2}{3}P_4$ &
\begin{tabular} {l}
1  \\
2   %
\end{tabular} &
\begin{tabular} {l}
$\frac{1}{2}$,$\frac{3}{2}$ \\
$\frac{3}{2}$,$\frac{5}{2}$ %
\end{tabular}  &
\begin{tabular} {l}
$\frac{1}{2}$,$\frac{3}{2}$,$\frac{5}{2}$  \\
$\frac{1}{2}$,$\frac{3}{2}$,$\frac{5}{2}$,$\frac{7}{2}$   %
\end{tabular} &
\begin{tabular} {l}
+  \\
-   %
\end{tabular} \\

\hline 1929 & (200,121,211) & $\frac{1}{2}$,$\frac{3}{2}$ &
$P_5$ & &  &  & \\
\hline 1929 & (200) & $\frac{1}{2}$,$\frac{3}{2}$ & $\frac{1}{3}P_5$
& 3 & $\frac{5 }{2}$,$\frac{7}{2}$ & $\frac{3}{2}$,$\frac{5}{2}$,
$\frac{7}{2}$,$\frac{9}{2}$ & + \\
\hline 1929 & (121,211) & $\frac{1}{2}$,$\frac{3}{2}$ &
$\frac{2}{3}P_5$ & 3 & $ \frac{5}{2}$,$\frac{7}{2}$ &
$\frac{3}{2}$,$\frac{5}{2}$,$\frac{7}{2}$,$\frac{9}{2}$ & +\\
\hline 2019 & (-1-10,-112,1-12) & $\frac{1}{2}$,$\frac{3}{2}$ &
$H_{25}'$ &  &  & \\

\hline 2019 & (-112,1-12) & $\frac{1}{2}$,$\frac{3}{2}$ &
$\frac{2}{3} H_{25}'$ &
\begin{tabular} {l}
2  \\
4  %
\end{tabular} &
\begin{tabular} {l}
$\frac{3}{2}$,$\frac{5}{2}$;\\
$\frac{7}{2}$,$\frac{9}{2}$; %
\end{tabular} &
\begin{tabular} {l}
$\frac{1}{2}$,$\frac{3}{2}$,$\frac{5}{2}$,$\frac{7}{2}$  \\
$\frac{5}{2}$,$\frac{7}{2}$,$\frac{9}{2}$,$\frac{11}{2}$ %
\end{tabular} & - \\

\hline 2019 & (-1,-1,0) & $\frac{1}{2}$,$\frac{3}{2}$ &
$\frac{1}{3}H_{25}'$ &
\begin{tabular} {l}
2  \\
4  %
\end{tabular} &
\begin{tabular} {l}
$\frac{3}{2}$,$\frac{5}{2}$;\\
$\frac{7}{2}$,$\frac{9}{2}$; %
\end{tabular} &
\begin{tabular} {l}
$\frac{1}{2}$,$\frac{3}{2}$,$\frac{5}{2}$,$\frac{7}{2}$  \\
$\frac{5}{2}$,$\frac{7}{2}$,$\frac{9}{2}$,$\frac{11}{2}$ %
\end{tabular} & - \\

\hline 2019 & (-1,-1,2) & $\frac{1}{2}$,$\frac{3}{2}$ &
$H^{\prime}_{2}$ & 3 & $\frac{5}{2}$,$\frac{7}{2}$ &
$\frac{3}{2}$,$\frac{5}{2}$,$\frac{7}{2}$,$\frac{9}{2}$ & + \\

\hline 2649 &
\begin{tabular}{l}
(202,022,-121, \\
2-11,0-1-1,-10-1)%
\end{tabular} & $\frac{1}{2}$,$\frac{3}{2}$ &$P_4P_5$ & &  &  & \\
\hline 2649 & (202,022,-121) & $\frac{1}{2}$,$\frac{3}{2}$ & $P_4$ &
1 & $\frac{1}{2}$,$\frac{3}{2}$ &
$\frac{1}{2}$,$\frac{3}{2}$,$\frac{5}{2}$ & + \\
\hline
2649 & (202,022) & $\frac{1}{2}$,$\frac{3}{2}$ & $P_4$ &
\begin{tabular}{l}
1,3 \\
2.%
\end{tabular}
&
\begin{tabular}{l}
$\frac{1}{2}$,$\frac{3}{2}$;$\frac{5}{2}$,$\frac{7}{2}$ \\
$\frac{3}{2}$,$\frac{5}{2}$.%
\end{tabular}
&
\begin{tabular}{l}
$\frac{1}{2}$,$\frac{3}{2}$,$\frac{5}{2}$;
$\frac{3}{2}$,$\frac{5}{2}$,$\frac{7}{2}$,$\frac{9}{2}$. \\
$\frac{1}{2}$,$\frac{3}{2}$,$\frac{5}{2}$,$\frac{7}{2}$.%
\end{tabular} &
\begin{tabular}{l}
+ \\
-  %
\end{tabular} \\
\hline 2649 & (-121) & $\frac{1}{2}$,$\frac{3}{2}$ & $P_4$ &
\begin{tabular}{l}
1,3 \\
2.%
\end{tabular}
&
\begin{tabular}{l}
$\frac{1}{2}$,$\frac{3}{2}$;$\frac{5}{2}$,$\frac{7}{2}$ \\
$\frac{3}{2}$,$\frac{5}{2}$.%
\end{tabular}
&
\begin{tabular}{l}
$\frac{1}{2}$,$\frac{3}{2}$,$\frac{5}{2}$;
$\frac{3}{2}$,$\frac{5}{2}$,$\frac{7}{2}$,$\frac{9}{2}$. \\
$\frac{1}{2}$,$\frac{3}{2}$,$\frac{5}{2}$,$\frac{7}{2}$.%
\end{tabular}
&
\begin{tabular}{l}
+ \\
-%
\end{tabular} \\
\hline 2649 & (2-11,0-1-1,-10-1) & $\frac{1}{2}$,$\frac{3}{2}$ &
$P_5$ &  &  &  &  \\
\hline 2649 & (0-1-1,-10-1) & $\frac{1}{2}$,$\frac{3}{2}$ & $P_5$ &3
& $\frac{5}{2} $,$\frac{7}{2}$ &
$\frac{3}{2}$$\frac{5}{2}$$\frac{7}{2}$ $\frac{9}{2}$ & + \\
\hline 2649 & (2,-1,-1) & $\frac{1}{2}$,$\frac{3}{2}$ & $P_5$ & 3 &
$\frac{5}{2}$,$ \frac{7}{2}$ &
$\frac{3}{2}$$\frac{5}{2}$$\frac{7}{2}$ $\frac{9}{2}$ & + \\
\hline 2649 & (21-1,12-1,220) & $\frac{1}{2}$,$\frac{3}{2}$ & $P_4$
&
\begin{tabular}{l}
1,3 \\
2.%
\end{tabular}
&
\begin{tabular}{l}
$\frac{1}{2}$,$\frac{3}{2}$;$\frac{5}{2}$,$\frac{7}{2}$ \\
$\frac{3}{2}$,$\frac{5}{2}$.%
\end{tabular}
&
\begin{tabular}{l}
$\frac{1}{2}$,$\frac{3}{2}$,$\frac{5}{2}$;
$\frac{3}{2}$,$\frac{5}{2}$,$\frac{7}{2}$,$\frac{9}{2}$. \\
$\frac{1}{2}$,$\frac{3}{2}$,$\frac{5}{2}$,$\frac{7}{2}$.%
\end{tabular}
&
\begin{tabular}{l}
+ \\
-%
\end{tabular} \\
\hline
2649 & (21-1,12-1) & $\frac{1}{2}$,$\frac{3}{2}$ & $P_4$ &
\begin{tabular}{l}
1,3 \\
2.%
\end{tabular}
&
\begin{tabular}{l}
$\frac{1}{2}$,$\frac{3}{2}$;$\frac{5}{2}$,$\frac{7}{2}$ \\
$\frac{3}{2}$,$\frac{5}{2}$.%
\end{tabular}
&
\begin{tabular}{l}
$\frac{1}{2}$,$\frac{3}{2}$,$\frac{5}{2}$;
$\frac{3}{2}$,$\frac{5}{2}$,$\frac{7}{2}$,$\frac{9}{2}$. \\
$\frac{1}{2}$,$\frac{3}{2}$,$\frac{5}{2}$,$\frac{7}{2}$.%
\end{tabular}
&
\begin{tabular}{l}
+ \\
-%
\end{tabular} \\
\hline
2649 & (2,2,0) & $\frac{1}{2}$,$\frac{3}{2}$ & $P_4$ &
\begin{tabular}{l}
1,3 \\
2.%
\end{tabular}
&
\begin{tabular}{l}
$\frac{1}{2}$,$\frac{3}{2}$;$\frac{5}{2}$,$\frac{7}{2}$ \\
$\frac{3}{2}$,$\frac{5}{2}$.%
\end{tabular}
&
\begin{tabular}{l}
$\frac{1}{2}$,$\frac{3}{2}$,$\frac{5}{2}$;
$\frac{3}{2}$,$\frac{5}{2}$,$\frac{7}{2}$,$\frac{9}{2}$. \\
$\frac{1}{2}$,$\frac{3}{2}$,$\frac{5}{2}$,$\frac{7}{2}$.%
\end{tabular}
&
\begin{tabular}{l}
+ \\
-%
\end{tabular}\\
\hline
\end{tabular}%
\end{tabular}%
\newline
$\qquad $ $\qquad $ $\qquad $ \newline
\newline
\begin{tabular}{l}

\,\,\,Table B5b. The J and P of the energy bands on F-axis
(continued from B5a) \\
\begin{tabular}{|l|l|l|l|l|l|l|l|l|l|}
\hline
$M_Mev$ & ($n_1, n_2, n_3$) & s & Repr. & L & J(1/2) & J(3/2) & P  \\
\hline
2739 &
\begin{tabular}{l}
(013,103,1-21 \\
-211,0-20,-200)%
\end{tabular}&$\frac{1}{2}$,$\frac{3}{2}$ & $H_{15}$,
$H_{25^{\prime}}$ &  &  &  & \\
\hline 2739 & (013,103,1-21) & $\frac{1}{2}$,$\frac{3}{2}$ &
$H_{15}$ &
\begin{tabular}{l}
1, \\
3.%
\end{tabular}
&
\begin{tabular}{l}
$\frac{1}{2}$,$\frac{3}{2}$; \\
$\frac{5}{2}$,$\frac{7}{2}$.%
\end{tabular}
&
\begin{tabular}{l}
$\frac{1}{2}$,$\frac{3}{2}$,$\frac{5}{2}$; \\
$\frac{3}{2}$,$\frac{5}{2}$,$\frac{7}{2}$,$\frac{9}{2}$.%
\end{tabular}
&
\begin{tabular}{l}
+ \\
+%
\end{tabular} \\
\hline 2739 & (013,103) & $\frac{1}{2}$,$\frac{3}{2}$ & $H_{15}$ &
\begin{tabular}{l}
1, \\
3.%
\end{tabular}
&
\begin{tabular}{l}
$\frac{1}{2}$,$\frac{3}{2}$; \\
$\frac{5}{2}$,$\frac{7}{2}$.%
\end{tabular}
&
\begin{tabular}{l}
$\frac{1}{2}$,$\frac{3}{2}$,$\frac{5}{2}$; \\
$\frac{3}{2}$,$\frac{5}{2}$,$\frac{7}{2}$,$\frac{9}{2}$.%
\end{tabular}
&
\begin{tabular}{l}
+ \\
+ %
\end{tabular} \\
\hline 2739 & (1-21) & $\frac{1}{2}$,$\frac{3}{2}$ &
$\frac{1}{3}$$H_{15}$ &
\begin{tabular}{l}
1, \\
3.%
\end{tabular}
&
\begin{tabular}{l}
$\frac{1}{2}$,$\frac{3}{2}$; \\
$\frac{5}{2}$,$\frac{7}{2}$.%
\end{tabular}
&
\begin{tabular}{l}
$\frac{1}{2}$,$\frac{3}{2}$,$\frac{5}{2}$; \\
$\frac{3}{2}$,$\frac{5}{2}$,$\frac{7}{2}$,$\frac{9}{2}$.%
\end{tabular}
&
\begin{tabular}{l}
+ \\
+ %
\end{tabular} \\
\hline 2739 & (-211,0-20,-200) & $\frac{1}{2}$,$\frac{3}{2}$ &
$H_{25^{\prime}}$ & & &   \\
\hline 2739 & (0-20,-200) & $\frac{1}{2}$,$\frac{3}{2}$ &
$\frac{2}{3}$$H_{25^{\prime}}$ &
\begin{tabular}{l}
2, \\
4.%
\end{tabular}
&
\begin{tabular}{l}
$\frac{3}{2}$,$\frac{5}{2}$; \\
$\frac{7}{2}$,$\frac{9}{2}$.%
\end{tabular}
&
\begin{tabular}{l}
$\frac{1}{2}$,$\frac{3}{2}$,$\frac{5}{2}$,$\frac{7}{2}$; \\
$\frac{5}{2}$,$\frac{7}{2}$,$\frac{9}{2}$,$\frac{11}{2}$.%
\end{tabular}
&
\begin{tabular}{l}
- \\
- %
\end{tabular} \\
\hline 2739 & (-2,1,1) & $\frac{1}{2}$,$\frac{3}{2}$ &
$\frac{1}{3}$$H_{25^{\prime}}$ &
\begin{tabular}{l}
2, \\
4.%
\end{tabular}
&
\begin{tabular}{l}
$\frac{3}{2}$,$\frac{5}{2}$; \\
$\frac{7}{2}$,$\frac{9}{2}$.%
\end{tabular}
&
\begin{tabular}{l}
$\frac{1}{2}$,$\frac{3}{2}$,$\frac{5}{2}$,$\frac{7}{2}$; \\
$\frac{5}{2}$,$\frac{7}{2}$,$\frac{9}{2}$,$\frac{11}{2}$.%
\end{tabular}
&
\begin{tabular}{l}
- \\
- %
\end{tabular} \\

\hline 2739 &
\begin{tabular}{l}
(0-13,-103,0-22, \\
-202,0-2-1,-1-21)%
\end{tabular}
& $\frac{1}{2}$,$\frac{3}{2}$ & $H_{25}$$H_{15^{\prime}}$
& & &  & \\
\hline 2739 & (0-13,-103,0-22) & $\frac{1}{2}$,$\frac{3}{2}$ &
$H_{25}$ &
\begin{tabular}{l}
3, \\
5.%
\end{tabular}
&
\begin{tabular}{l}
$\frac{5}{2}$,$\frac{7}{2}$; \\
$\frac{9}{2}$,$\frac{11}{2}$.%
\end{tabular}
&
\begin{tabular}{l}
$\frac{3}{2}$,$\frac{5}{2}$,$\frac{7}{2}$,$\frac{9}{2}$; \\
$\frac{7}{2}$,$\frac{9}{2}$,$\frac{11}{2}$,$\frac{13}{2}$.%
\end{tabular}
&
\begin{tabular}{l}
+ \\
+  %
\end{tabular} \\
\hline 2739 & (0-13,-103) & $\frac{1}{2}$,$\frac{3}{2}$ &
$\frac{2}{3}$$H_{25}$ &
\begin{tabular}{l}
3, \\
5.%
\end{tabular}
&
\begin{tabular}{l}
$\frac{5}{2}$,$\frac{7}{2}$; \\
$\frac{9}{2}$,$\frac{11}{2}$.%
\end{tabular}
&
\begin{tabular}{l}
$\frac{3}{2}$,$\frac{5}{2}$,$\frac{7}{2}$,$\frac{9}{2}$; \\
$\frac{7}{2}$,$\frac{9}{2}$,$\frac{11}{2}$,$\frac{13}{2}$.%
\end{tabular}
&
\begin{tabular}{l}
+ \\
+  %
\end{tabular} \\
\hline 2739 & (1-21) & $\frac{1}{2}$,$\frac{3}{2}$ & $H_{25}$ &
\begin{tabular}{l}
3, \\
5.%
\end{tabular}
&
\begin{tabular}{l}
$\frac{5}{2}$,$\frac{7}{2}$; \\
$\frac{9}{2}$,$\frac{11}{2}$.%
\end{tabular}
&
\begin{tabular}{l}
$\frac{3}{2}$,$\frac{5}{2}$,$\frac{7}{2}$,$\frac{9}{2}$; \\
$\frac{7}{2}$,$\frac{9}{2}$,$\frac{11}{2}$,$\frac{13}{2}$.%
\end{tabular}
&
\begin{tabular}{l}
+ \\
+ %
\end{tabular} \\
\hline 2739 & (-202,-2-11,-1-21) & $\frac{1}{2}$,$\frac{3}{2}$ &
$H_{15^{\prime}}$ & 4 &  &  & \\
\hline 2739 & (-2-11,-1-21) & $\frac{1}{2}$,$\frac{3}{2}$ &
$\frac{2}{3}$$H_{15^{\prime}}$ & 4 & $\frac{7}{2}$,$\frac{9}{2}$ &
$\frac{5}{2}$$\frac{7}{2}$$\frac{9}{2}$ $\frac{11}{2}$ & -\\
\hline 2739 & (-2,0,2) & $\frac{1}{2}$,$\frac{3}{2}$ &
$H_{15^{\prime}}$ & 4 & $\frac{7}{2}$,$\frac{9}{2}$ &
$\frac{5}{2}$$\frac{7}{2}$$\frac{9}{2}$ $\frac{11}{2}$ & - \\

\hline
\end{tabular}%
\end{tabular}%
\newline\\

\subsection{The J and P of the energy bands on G-axis}

\begin{tabular}{l}
$\qquad$$\qquad$$\qquad$$\qquad$$\qquad$Table B6.\ The J and P
of the energy bands on the G-axis \\
\begin{tabular}{|l|l|l|l|l|l|l|l|l|l|l|}
\hline M (Mev) & ($n_1, n_2, n_3$) & s & Repres. & L &
J(1/2) & J(3/2) & P\\
\hline 1299 & (1,0,1;0,1,-1) & $\frac{1}{2}$ & $\frac{2}{3}M_{15}$ &
1 & $\frac{1}{2 }$,$\frac{3}{2}$ &  &+ \\
\hline 1299 & (2,0,0;1,-1,0) & $\frac{1}{2}$
&$M_1$$\ref{1}{3}M_{15}$
&  &  & & \\
\hline 1299 & (2,0,0,) & $\frac{1}{2}$ & $\frac{1}{3}M_{15}$ & 1 &
$\frac{1}{2},\frac{3}{2}$ &  &  +  \\
\hline 1299 & (1,-1,0,) & $\frac{1}{2}$ & $M_{1}$ & 0 &
$\frac{1}{2}$  & $\frac{1}{2}$ & -  \\
\hline 1479 & (0,1,1;0,1,-1) & $\frac{1}{2}$,$\frac{3}{2}$ &
$N_3',N_4'$ & 1 & $\frac{1}{2}$,$\frac{3}{2}$ &
$\frac{1}{2}$,$\frac{3}{2}$,$\frac{5}{2}$ & + \\
\hline 1839 & (0,2,0;-1,1,0) & $\frac{1}{2}$,$\frac{3}{2}$ &
$N_1N_4$ &  &  &  & \\
\hline 1839 & (0,2,0,) & $\frac{1}{2}$,$\frac{3}{2}$ & $N_1$ &
\begin{tabular}{l}
0  \\
2   %
\end{tabular}&
\begin{tabular}{l}
$\frac{1}{2}$ \\
$\frac{3}{2}$, $\frac{5}{2}$ %
\end{tabular} &
\begin{tabular}{l}
$\frac{3}{2}$ \\
$\frac{1}{2}$,$\frac{3}{2}$,$\frac{5}{2}$,$\frac{7}{2}$ %
\end{tabular} & - \\
\hline 1839 & (-1,1,0) & $\frac{1}{2}$,$\frac{3}{2}$ & $N_4$ & 2 &
$\frac{3}{2}$,$\frac{5}{2}$ &
$\frac{1}{2}$,$\frac{3}{2}$,$\frac{5}{2}$,$\frac{5}{2}$ &- \\
\hline 2019 & (0-11,0-1-1,21-1,211) & $\frac{1}{2}$,$\frac{3}{2}$
&$M_{25}^{\prime},\frac{1}{3}M_{15}$
&  &  &  & \\
\hline 2019 & (0-11,0-1-1) & $\frac{1}{2}$,$\frac{3}{2}$ &
$\frac{1}{3} M_{25}^{\prime},\frac{1}{3}M_{15}$ & 2,1 &
$\frac{3}{2}$ & $\frac{1}{2}$,$\frac{3}{2}$,$\frac{5}{2}$ & ? \\
\hline 2019 & (21-1,211) & $\frac{1}{2}$,$\frac{3}{2}$ &
$\frac{2}{3}M^{\prime}_{25}$ &
\begin{tabular}{l}
2 \\
4 %
\end{tabular} &
\begin{tabular}{l}
$\frac{3}{2},\frac{5}{2}$ \\
$\frac{7}{2},\frac{9}{2}$ %
\end{tabular} &
\begin{tabular}{l}
$\frac{1}{2},\frac{3}{2},\frac{5}{2},\frac{7}{2}$\\
$\frac{5}{2},\frac{7}{2},\frac{9}{2},\frac{11}{2}$  %
\end{tabular} & - \\

\hline 2019 & (2,-1,1;2,-1,-1) & $\frac{1}{2}$,$\frac{3}{2}$ &
$\frac{2}{3}M_{15}$ &
\begin{tabular}{l}
1 \\
3 %
\end{tabular} &
\begin{tabular}{l}
$\frac{1}{2},\frac{3}{2}$ \\
$\frac{5}{2},\frac{7}{2}$ %
\end{tabular} &
\begin{tabular}{l}
$\frac{1}{2},\frac{3}{2},\frac{5}{2}$\\
$\frac{3}{2},\frac{5}{2},\frac{7}{2},\frac{9}{2}$  %
\end{tabular} & + \\

\hline 2199 & (-101,-10-1,121,12-1) & $\frac{1}{2}$,$\frac{3}{2}$&
&  &
&  &  \\
\hline 2199 & (-101,-10-1) & $\frac{1}{2}$,$\frac{3}{2}$ &
$N_3^{\prime}$$ N_4^{\prime}$ &
\begin{tabular}{l}
1 \\
3 %
\end{tabular} &
\begin{tabular}{l}
$\frac{1}{2},\frac{3}{2}$ \\
$\frac{5}{2},\frac{7}{2}$ %
\end{tabular} &
\begin{tabular}{l}
$\frac{1}{2},\frac{3}{2},\frac{5}{2}$\\
$\frac{1}{2},\frac{3}{2},\frac{5}{2},\frac{7}{2}$ %
\end{tabular} & + \\

\hline 2199 & (121,12-1) & $\frac{1}{2}$,$\frac{3}{2}$ &
$N_3$ $N_4$
&
\begin{tabular}{l}
2 \\
4 %
\end{tabular} &
\begin{tabular}{l}
$\frac{3}{2},\frac{5}{2}$ \\
$\frac{7}{2},\frac{9}{2}$ %
\end{tabular} &
\begin{tabular}{l}
$\frac{1}{2},\frac{3}{2},\frac{5}{2},\frac{7}{2}$\\
$\frac{5}{2},\frac{7}{2},\frac{9}{2},\frac{11}{2}$ %
\end{tabular} & - \\

\hline 2559 & (002,00-2) & $\frac{1}{2}$,$\frac{3}{2}$ & $N_1, N_3$&
\begin{tabular}{l}
2 \\
4 %
\end{tabular} &
\begin{tabular}{l}
$\frac{3}{2},\frac{5}{2}$ \\
$\frac{7}{2},\frac{9}{2}$ %
\end{tabular} &
\begin{tabular}{l}
$\frac{1}{2},\frac{3}{2},\frac{5}{2},\frac{7}{2}$  \\
$\frac{5}{2},\frac{7}{2},\frac{9}{2},\frac{11}{2}$  %
\end{tabular} & - \\

\hline 2559 & (112,11-2) & $\frac{1}{2}$,$\frac{3}{2}$ &
$N^{\prime}_1, N^{\prime}_3 $ &
\begin{tabular}{l}
1 \\
3 %
\end{tabular} &
\begin{tabular}{l}
$\frac{1}{2},\frac{3}{2}$ \\
$\frac{5}{2},\frac{7}{2}$ %
\end{tabular} &
\begin{tabular}{l}
$\frac{1}{2},\frac{3}{2},\frac{5}{2}$\\
$\frac{3}{2},\frac{3}{2},\frac{5}{2},\frac{7}{2}$  %
\end{tabular} & + \\

\hline 2559 & (220,-1-10) & $\frac{1}{2}$,$\frac{3}{2}$
&  &  &  &  & \\
\hline 2559 & (2,2,0 ) & $\frac{1}{2}$,$\frac{3}{2}$ &
$N^{\prime}_1$ &
\begin{tabular}{l}
1 \\
3 %
\end{tabular} &
\begin{tabular}{l}
$\frac{1}{2},\frac{3}{2}$ \\
$\frac{5}{2},\frac{7}{2}$ %
\end{tabular} &
\begin{tabular}{l}
$\frac{1}{2},\frac{3}{2},\frac{5}{2}$\\
$\frac{3}{2},\frac{3}{2},\frac{5}{2},,\frac{7}{2}$  %
\end{tabular} & + \\

\hline 2559 & (-1,-1,0) & $\frac{1}{2}$,$\frac{3}{2}$ & $N_1$ &
\begin{tabular}{l}
0  \\
2  \\
4   %
\end{tabular} &
\begin{tabular}{l}
$\frac{1}{2}$  \\
$\frac{3}{2}$,$\frac{5}{2}$ \\
$\frac{7}{2}$,$\frac{9}{2}$ %
\end{tabular} &
\begin{tabular}{l}
$\frac{3}{2}$  \\
$\frac{1}{2}$,$\frac{3}{2}$,$\frac{5}{2}$,$\frac{7}{2}$ \\
$\frac{5}{2}$,$\frac{7}{2}$,$\frac{9}{2}$,$\frac{11}{2}$ %
\end{tabular} &  - \\

\hline 2739 & (202,20-2) & $\frac{1}{2}$,$\frac{3}{2}$ &
$\frac{2}{3}M_{15}$ &
\begin{tabular}{l}
1  \\
3  %
\end{tabular} &
\begin{tabular}{l}
$\frac{1}{2}$,$\frac{3}{2}$  \\
$\frac{5}{2}$,$\frac{7}{2}$  %
\end{tabular} &
\begin{tabular}{l}
$\frac{1}{2}$,$\frac{3}{2}$,$\frac{5}{2}$  \\
$\frac{3}{2}$,$\frac{5}{2}$,$\frac{7}{2}$,$\frac{9}{2}$ %
\end{tabular} &  + \\

\hline 2739 & (1-12,1-1-2) & $\frac{1}{2}$,$\frac{3}{2}$
&$\frac{2}{3}M_{25}$  &
\begin{tabular}{l}
3  \\
5   %
\end{tabular} &
\begin{tabular}{l}
$\frac{5}{2}$,$\frac{7}{2}$ \\
$\frac{9}{2}$,$\frac{11}{2}$ %
\end{tabular} &
\begin{tabular}{l}
$\frac{3}{2}$,$\frac{5}{2}$,$\frac{7}{2}$,$\frac{9}{2}$ \\
$\frac{7}{2}$,$\frac{9}{2}$,$\frac{11}{2}$,$\frac{13}{2}$ %
\end{tabular} &  + \\

\hline 2739 & (3, 1, 0) & $\frac{1}{2}$,$\frac{3}{2}$ &
$\frac{1}{3}M_{15}$ &
\begin{tabular}{l}
1  \\
3  %
\end{tabular} &
\begin{tabular}{l}
$\frac{1}{2}$,$\frac{3}{2}$  \\
$\frac{5}{2}$,$\frac{7}{2}$  %
\end{tabular} &
\begin{tabular}{l}
$\frac{1}{2}$,$\frac{3}{2}$,$\frac{5}{2}$  \\
$\frac{3}{2}$,$\frac{5}{2}$,$\frac{7}{2}$,$\frac{9}{2}$ %
\end{tabular} &  + \\

\hline 2739 & (0,-2,0) & $\frac{1}{2}$,$\frac{3}{2}$
&$\frac{1}{3}M_{25}$  & 3  & $\frac{5}{2}$,$\frac{7}{2}$ &
$\frac{3}{2}$,$\frac{5}{2}$$\frac{7}{2}$,$\frac{9}{2}$ &  + \\

\hline 2739 & (301,30-1) & $\frac{1}{2}$,$\frac{3}{2}$ & $M_{12}$ &
\begin{tabular}{l}
2  \\
4   %
\end{tabular} &
\begin{tabular}{l}
$\frac{3}{2}$,$\frac{5}{2}$ \\
$\frac{7}{2}$,$\frac{9}{2}$ %
\end{tabular} &
\begin{tabular}{l}
$\frac{1}{2}$,$\frac{3}{2}$,$\frac{5}{2}$,$\frac{7}{2}$ \\
$\frac{5}{2}$,$\frac{7}{2}$,$\frac{9}{2}$,$\frac{11}{2}$ %
\end{tabular} &  - \\

\hline 2739 & (1-21,1-2-1) & $\frac{1}{2}$,$\frac{3}{2}$ &
$\frac{2}{3}$$M_{25'}$ &
\begin{tabular}{l}
2  \\
4   %
\end{tabular} &
\begin{tabular}{l}
$\frac{3}{2}$,$\frac{5}{2}$ \\
$\frac{7}{2}$,$\frac{9}{2}$ %
\end{tabular} &
\begin{tabular}{l}
$\frac{1}{2}$,$\frac{3}{2}$,$\frac{5}{2}$,$\frac{7}{2}$ \\
$\frac{5}{2}$,$\frac{7}{2}$,$\frac{9}{2}$,$\frac{11}{2}$ %
\end{tabular} &  - \\

\hline 2739 & (3,-1,0) & $\frac{1}{2}$,$\frac{3}{2}$ &
$\frac{1}{2}$$M_{12}$ &
\begin{tabular}{l}
2  \\
4   %
\end{tabular} &
\begin{tabular}{l}
$\frac{3}{2}$,$\frac{5}{2}$ \\
$\frac{7}{2}$,$\frac{9}{2}$ %
\end{tabular} &
\begin{tabular}{l}
$\frac{1}{2}$,$\frac{3}{2}$,$\frac{5}{2}$,$\frac{7}{2}$ \\
$\frac{5}{2}$,$\frac{7}{2}$,$\frac{9}{2}$,$\frac{11}{2}$ %
\end{tabular} &  - \\

\hline 2739 & (2,-2,0) & $\frac{1}{2}$,$\frac{3}{2}$ &
$\frac{1}{2}$$M_{12}$ &
\begin{tabular}{l}
2  \\
4   %
\end{tabular} &
\begin{tabular}{l}
$\frac{3}{2}$,$\frac{5}{2}$ \\
$\frac{7}{2}$,$\frac{9}{2}$ %
\end{tabular} &
\begin{tabular}{l}
$\frac{1}{2}$,$\frac{3}{2}$,$\frac{5}{2}$,$\frac{7}{2}$ \\
$\frac{5}{2}$,$\frac{7}{2}$,$\frac{9}{2}$,$\frac{11}{2}$ %
\end{tabular} &  - \\

\hline
\end{tabular}
\end{tabular}

\end{document}